\documentclass[twocolumn]{aastex63} 

\usepackage{graphicx} 
\usepackage{float} 
\usepackage{wrapfig} 
\usepackage{lipsum} 
\usepackage{hyperref}
\usepackage{times}
\usepackage{amsmath}
\usepackage{graphicx}
\usepackage{subfigure}
\usepackage{placeins}
\usepackage{hyperref}
\usepackage{gensymb}
\usepackage{upgreek}
\usepackage{natbib}

\usepackage{graphicx}
\usepackage{subfigure}
\usepackage{multirow}
\usepackage{comment}
\usepackage{natbib}
\usepackage{hyperref}
\usepackage{mathtools}
\usepackage{mathrsfs}
\usepackage{fontenc}
\usepackage{color}
\usepackage{url}
\usepackage{hyperref}
\usepackage{gensymb}
\usepackage{pifont}
\graphicspath{{./}}

\newcommand       \K            {\,{\rm K}}
\newcommand       \mum          {{\rm \mu m}}

\shorttitle{Theoretical Diagnostics for AGN Environments under the View of JWST}
\shortauthors{Zhang et al.}

\begin{document}

\title{Theoretical Diagnostics for the Physical Conditions in Active Galactic Nuclei under the View of JWST}

\author[0000-0003-4937-9077]{Lulu Zhang}
\affiliation{The University of Texas at San Antonio, One UTSA Circle, San Antonio, TX 78249, USA; lulu.zhang@utsa.edu, l.l.zhangastro@gmail.com}

\author[0000-0003-4949-7217]{Ric I. Davies}
\affiliation{Max-Planck-Institut für extraterrestrische Physik, Postfach 1312, D-85741, Garching, Germany}

\author[0000-0001-7827-5758]{Chris Packham}
\affiliation{The University of Texas at San Antonio, One UTSA Circle, San Antonio, TX 78249, USA; lulu.zhang@utsa.edu, l.l.zhangastro@gmail.com}
\affiliation{National Astronomical Observatory of Japan, National Institutes of Natural Sciences (NINS), 2-21-1 Osawa, Mitaka, Tokyo 181-8588, Japan}

\author[0000-0002-4457-5733]{Erin K. S. Hicks}
\affiliation{Department of Physics and Astronomy, University of Alaska Anchorage, Anchorage, AK 99508-4664, USA}
\affiliation{The University of Texas at San Antonio, One UTSA Circle, San Antonio, TX 78249, USA; lulu.zhang@utsa.edu, l.l.zhangastro@gmail.com}
\affiliation{Department of Physics, University of Alaska, Fairbanks, Alaska 99775-5920, USA}

\author[0009-0007-6992-2555]{Daniel E. Delaney}
\affiliation{Department of Physics, University of Alaska, Fairbanks, Alaska 99775-5920, USA}
\affiliation{Department of Physics and Astronomy, University of Alaska Anchorage, Anchorage, AK 99508-4664, USA}

\author[0000-0002-4005-9619]{Miguel Pereira-Santaella}
\affiliation{Instituto de F{\'i}sica Fundamental, CSIC, Calle Serrano 123, 28006 Madrid, Spain}

\author[0000-0002-9610-0123]{Laura Hermosa Mu{\~n}oz}
\affiliation{Centro de Astrobiolog\'{\i}a (CAB), CSIC-INTA, Camino Bajo del Castillo s/n, E-28692 Villanueva de la Ca\~nada, Madrid, Spain}

\author[0000-0002-9627-5281]{Ismael Garc{\'i}a-Bernete}
\affiliation{Centro de Astrobiolog\'{\i}a (CAB), CSIC-INTA, Camino Bajo del Castillo s/n, E-28692 Villanueva de la Ca\~nada, Madrid, Spain}

\author[0000-0001-5231-2645]{Claudio Ricci}
\affiliation{Instituto de Estudios Astrof{\'i}sicos, Facultad de Ingenier{\'i}a y Ciencias, Universidad Diego Portales, Av. Ej{\'e}rcito Libertador 441, Santiago, Chile}
\affiliation{Kavli Institute for Astronomy and Astrophysics, Peking University, Beijing 100871, People’s Republic of China}

\author[0000-0001-6854-7545]{Dimitra Rigopoulou}
\affiliation{Department of Physics, University of Oxford, Keble Road, Oxford OX1 3RH, UK}
\affiliation{School of Sciences, European University Cyprus, Diogenes street, Engomi, 1516 Nicosia, Cyprus}

\author[0000-0001-6794-2519]{Almudena Alonso-Herrero}
\affiliation{Centro de Astrobiolog\'{\i}a (CAB), CSIC-INTA, Camino Bajo del Castillo s/n, E-28692 Villanueva de la Ca\~nada, Madrid, Spain}

\author[0000-0003-1810-0889]{Martin J. Ward}
\affiliation{Centre for Extragalactic Astronomy, Durham University, South Road, Durham DH1 3LE, UK}

\author[0000-0001-9791-4228]{Enrica Bellocchi}
\affiliation{Departamento de F\'isica de la Tierra y Astrof\'isica, Fac. de CC. F\'isicas, Universidad Complutense de Madrid, 28040 Madrid, Spain}
\affiliation{Instituto de F\'isica de Part\'iculas y del Cosmos IPARCOS, Fac. CC. F\'isicas, Universidad Complutense de Madrid, 28040 Madrid, Spain}

\author[0000-0001-8353-649X]{Cristina Ramos Almeida}
\affiliation{Instituto de Astrof{\'i}sica de Canarias, Calle V{\'i}a L{\'a}ctea, s/n, E-38205, La Laguna, Tenerife, Spain}
\affiliation{Departamento de Astrof{\'i}sica, Universidad de La Laguna, E-38206, La Laguna, Tenerife, Spain}

\author[0000-0003-2658-7893]{Francoise Combes}
\affiliation{LERMA, Observatoire de Paris, Coll{\`e}ge de France, PSL University, CNRS, Sorbonne University, Paris}

\author[0000-0001-6186-8792]{Masatoshi Imanishi}
\affiliation{National Astronomical Observatory of Japan, National Institutes of Natural Sciences, 2-21-1 Osawa, Mitaka, Tokyo 181-8588, Japan}
\affiliation{Department of Astronomy, School of Science, Graduate University for Advanced Studies (SOKENDAI), Mitaka, Tokyo 181-8588, Japan}

\author[0000-0002-2356-8358]{Omaira Gonz{\'a}lez-Mart{\'i}n}
\affiliation{Instituto de Radioastronom{\'i}a and Astrof{\'i}sica (IRyA-UNAM), 3-72 (Xangari), 8701, Morelia, Mexico}

\author[0000-0003-0699-6083]{Tanio D{\'i}az-Santos}
\affiliation{Institute of Astrophysics, Foundation for Research and Technology-Hellas (FORTH), Heraklion 70013, Greece}
\affiliation{School of Sciences, European University Cyprus, Diogenes street, Engomi, 1516 Nicosia, Cyprus}

\author[0000-0003-3589-3294]{Anelise Audibert}
\affiliation{Instituto de Astrof{\'i}sica de Canarias, Calle V{\'i}a L{\'a}ctea, s/n, E-38205, La Laguna, Tenerife, Spain}
\affiliation{Departamento de Astrof{\'i}sica, Universidad de La Laguna, E-38206, La Laguna, Tenerife, Spain}

\author[0000-0002-0690-8824]{\'{A}lvaro Labiano}
\affiliation{Telespazio UK for the European Space Agency (ESA), ESAC, Camino Bajo del Castillo s/n, 28692 Villanueva de la Ca{\~n}ada, Spain}

\author[0000-0003-4209-639X]{Nancy A. Levenson}
\affiliation{Space Telescope Science Institute, 3700 San Martin Drive Baltimore, Maryland 21218, USA}

\author[0000-0003-0444-6897]{Santiago Garc{\'i}a-Burillo}
\affiliation{Observatorio Astron{\'o}mico Nacional (OAN-IGN)-Observatorio de Madrid, Alfonso XII, 3, 28014, Madrid, Spain}

\author[0000-0003-4809-6147]{Lindsay Fuller}
\affiliation{The University of Texas at San Antonio, One UTSA Circle, San Antonio, TX 78249, USA; lulu.zhang@utsa.edu, l.l.zhangastro@gmail.com}



\begin{abstract}
With excellent spectral and angular resolutions and, especially, sensitivity, the JWST allows us to observe infrared emission lines that were previously inaccessible or barely accessible. These emission lines are promising for evaluating the physical conditions in different galaxies. Based on {\sc MAPPINGS V} photoionization models, we systematically analyze the dependence of over 20 mid-infrared (mid-IR) emission lines covered by the Mid-Infrared Instrument (MIRI) onboard JWST on the physical conditions of different galactic environments, in particular narrow line regions (NLRs) in active galactic nuclei (AGN). We find that mid-IR emission lines of highly ionized argon (i.e., [Ar~{\small V}]7.90,13.10) and neon (i.e., [Ne~{\small V}]14.32,24.32, [Ne~{\small VI}]7.65) are effective in diagnosing the physical conditions in AGN. We accordingly propose new prescriptions to constrain the ionization parameter ($U$), peak energy of the AGN spectrum ($E_{\rm peak}$), metallicity ($\rm 12+log (O/H)$), and gas pressure ($P/k$) in AGN. These new calibrations are applied to the central regions of six Seyfert galaxies included in the Galaxy Activity, Torus, and Outflow Survey (GATOS) as a proof of concept. We also discuss the similarity and difference in the calibrations of these diagnostics in AGN of different luminosities, highlighting the impact of hard X-ray emission and particularly radiative shocks, as well as the different diagnostics in star-forming regions. Finally, we propose diagnostic diagrams involving [Ar~{\small V}]7.90 and [Ne~{\small VI}]7.65 to demonstrate the feasibility of using the results of this study to distinguish galactic regions governed by different excitation sources.
\end{abstract}


\section{Introduction}

Supermassive black holes reside in most, if not all, massive galaxies and up to $\sim 40\,\%$ of the black holes in emission-line galaxies manifest as AGN (i.e., \citealt{Kauffmann.etal.2003}; \citealt{Miller.etal.2003}; and see review \citealt{Ho2008}). A growing body of evidence has revealed that AGN play an important role in determining the evolutionary pathways and final fate of galaxies through different modes of feedback processes (e.g., \citealt{DiMatteo.etal.2005, Hopkins.etal.2008, Weinberger.etal.2017, Dave.etal.2019}; and see review \citealt{McNamara&Nulsen2007, Fabian2012, Harrison&RamosAlmeida2024}). The electromagnetic spectrum of a galaxy contains a wealth of information on the fundamental physical conditions contributing to the recurrent feedback processes (see review by \citealt{Padovani.etal.2017, Kewley.etal.2019, Sajina.etal.2022}). In this regard, theoretical photoionization modeling plays a crucial role in interpreting observed AGN spectra, and thus using observations to constrain the physical conditions in AGN for a comprehensive understanding of the galaxy evolution.

AGN photoionization modeling has been developed for over a half century from the early calculations of quasars and X-ray nebulae (e.g., \citealt{Burbidge.etal.1966, Osterbrock&Parker1966, Tarter&Salpeter1969, Tarter.etal.1969}; see review \citealt{Kewley.etal.2019}). During this period, two large photoionization codes (i.e., {\sc MAPPINGS} and {\sc CLOUDY}) gradually developed into the most commonly used codes for photoionization modeling. Specifically, both {\sc MAPPINGS} and {\sc CLOUDY} include self-consistent treatment of nebular and dust physics with different atomic data sets and calculation methods. In addition, {\sc MAPPINGS} includes radiative shock physics (\citealt{Sutherland&Dopita1993, Sutherland&Dopita2017}), while {\sc CLOUDY} includes molecular physics (\citealt{Ferland.etal.1998, Chatzikos.etal.2023}). Therefore, {\sc MAPPINGS} can be applied to H~{\small II} regions, AGN, and shocked regions, while {\sc CLOUDY} can be applied to H~{\small II} regions, AGN, and photodissociation regions. Despite the difference, the theoretical AGN photoionization models calculated by {\sc MAPPINGS} and {\sc CLOUDY} with the same inputs present similar correlations between emission line ratios, gas metallicity and ionization parameter (\citealt{Zhu.etal.2023}).

Among key parameters of the theoretical photoionization modeling, the effect of different metallicities on the predicted emission line ratios has been extensively studied for the direct calibration of metallicity in star-forming galaxies, as well as in AGN (e.g., \citealt{Nagao.etal.2011, Pereira-Santaella.etal.2017, Kewley.etal.2019, Perez-Diaz.etal.2022, Martinez-Paredes.etal.2023, Zhu.etal.2024}; and references therein). The effect of another key parameter, the ionization parameter ($U$), commonly defined as the ratio of the number density of incident ionizing photons ($\Phi$) and the number density of hydrogen atoms ($n_{\rm H}$) multiplied by the speed of light ($U=\frac{\Phi}{c~n_{\rm H}}$ where $\Phi = \frac{\int^{\infty}_{\nu0}L_{\nu}d\nu/h\nu}{4\pi r^2}$), has been also considered in the above studies to better constrain the calibration of metallicity. However, an accurate calibration of the ionization parameter $U$ itself, especially in AGN, is still lacking. $U$ is an important parameter determined together by the spectral energy distribution (SED) of the ionizing radiation field, the gas density, and the distance of the ionized gas relative to the ionizing source. Variations in $U$ shed important light on underlying evolutionary processes in galaxies at different redshifts (e.g., \citealt{Abel.etal.2009, Sanders.etal.2016, Kashino&Inoue2019, Reddy.etal.2023}).

More specifically, measurements of the ionization parameter in AGN can also be used to help quantify gas outflow rates (e.g., \citealt{Baron&Netzer2019, Davies.etal.2020}), as well as help decode the feedback effects responsible for observations such as, but not limited to, the unique characteristics of polycyclic aromatic hydrocarbon (PAH) emission (e.g., \citealt{Egorov.etal.2023, Garcia-Bernete.etal.2024b, Zhang.etal.2024b}) in AGN. Empirical calibrations of $U$ can be achieved using any paired lines with different dependences on $U$ of their emissivities (which are intrinsically dependent on electron density and temperature). JWST spectral observations provide the access to study a variety of infrared emission lines covering a wide range of critical densities and ionization potentials (e.g., \citealt{Pereira-Santaella.etal.2022, Pereira-Santaella.etal.2024, Armus.etal.2023, Goold.etal.2024, Zhang.etal.2024a, AlonsoHerrero.etal.2025, Buiten.etal.2025, Ceci.etal.2025, HermosaMunoz.etal.2025, Marconcini.etal.2025, Ogle.etal.2025, RamosAlmeida.etal.2025}). Most of these emission lines were barely available in the past due to the limited spectral resolution and/or sensitivity of previous instruments such as spectrographs onboard Infrared Space Observatory and Spitzer space telescope. Thus, we now have an unprecedented opportunity to explore in depth and fully utilize infrared emission lines detected in AGN with JWST to evaluate the effect of the ionization parameter in diagnostic diagrams.

This paper aims to explore the theoretical dependence of the mid-IR fine structure emission line ratios on the physical conditions of AGN, in particular the ionization parameter, and as the complementarity, the $E_{\rm peak}$, metallicity, and gas pressure, which are also important physical parameters in AGN. This study can provide a more comprehensive explanation for JWST observations, thereby further advancing the analysis of future observational work. Specifically, we focus on mid-IR emission lines accessible by the Medium Resolution Spectrograph (MRS; \citealt{Wells.etal.2015, Labiano.etal.2021, Argyriou.etal.2023}) on the MIRI (\citealt{Rieke.etal.2015, Wright.etal.2015, Wright.etal.2023}) of JWST (\citealt{Gardner.etal.2023, Rigby.etal.2023}). The near-IR ($\sim 1-5\,\mum$) spectra of galaxies, although covering some weak coronal lines, are mainly composed of emission lines from hydrogen, helium, and iron and are not considered here.

This paper is structured as follows. After introducing the adopted theoretical models (Section~\ref{sec2}), we elaborate on the calibration of $U$ and propose some optimal prescriptions according to a feature extraction method based on the Pearson correlation coefficient combining some practical criteria (Section~\ref{sec3.1}). The following sections discuss the calibrations of $E_{\rm peak}$ (Section~\ref{sec3.2}), metallicity (Section~\ref{sec3.3}), and gas pressure (Section~\ref{sec3.4}), as well as the application of these calibrations in observed AGN (Section~\ref{sec3.5}). In addition, we discuss the calibrations in low-luminosity AGN (Section~\ref{sec4.1}), highlighting the effect of hard X-ray radiation (Section~\ref{sec4.1.1}) and potential shocks (Section~\ref{sec4.1.2}). Section~\ref{sec4} also discusses the different calibrations in star-forming regions (Section~\ref{sec4.2}) and the potential application of the above results in future studies of distinguishing galactic regions governed by different excitation sources (Section~\ref{sec4.3}). A summary of the main results of this paper is provided in Section~\ref{sec5}.


\section{Theoretical Models}\label{sec2}

As mentioned in the introduction section, the two most popular codes for photoionization modeling are {\sc MAPPINGS} and {\sc CLOUDY}, which are implemented independently with self-consistent atomic data and physical processes (\citealt{Sutherland&Dopita1993, Ferland.etal.1998, Sutherland&Dopita2017, Chatzikos.etal.2023}). Specifically, {\sc MAPPINGS} can be applied to H~{\small II} regions, AGN, and shocked regions, while {\sc CLOUDY} can be applied to H~{\small II} regions, AGN, and photodissociation regions. Overall, the theoretical AGN photoionization models calculated by the two codes provide similar predictions when using the same inputs, with most of the predicted emission line ratios agreeing to within 0.1 dex (\citealt{Zhu.etal.2023}). The existing discrepancy is mainly due to the difference in the implementation of inherent physical processes and the adopted atomic data sets. Since this paper requires theoretical models for shocked regions without attempting to model the photodissociation regions, all theoretical models presented in the following sections are calculated using {\sc MAPPINGS}.

\begin{figure}[!ht]
\center{\includegraphics[width=1\linewidth]{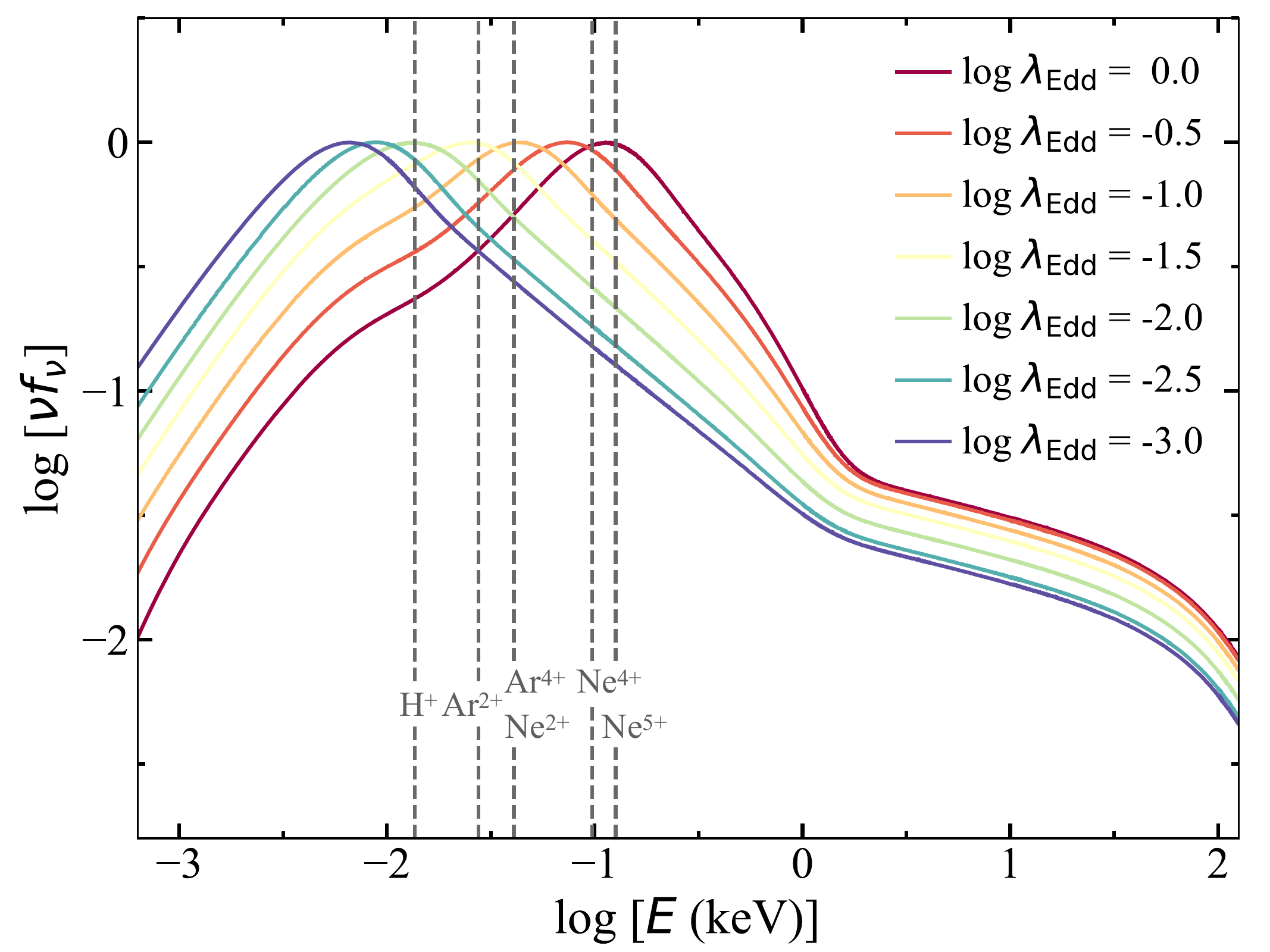}}
\caption{Examples of the AGN ionizing spectra calculated by {\sc OPTXAGNF} for AGN photoionization models with the black hole mass as $10^7\,M_{\odot}$ and the Eddington ratio (i.e., $\lambda_{\rm Edd}$) ranging from $10^{-3}$ to $1$. The vertical lines indicate the required ionization energy of marked ions. All AGN ionizing spectra are normalized by the peak flux for illustration. Note that the energy corresponding to the peak flux (i.e., $E_{\rm peak}$, in unit of keV) increases along with the increase of $\lambda_{\rm Edd}$ and the decrease of black hole mass.}\label{SpecAGN}
\end{figure}

We use the latest version (5.21) {\sc MAPPINGS~V}\footnote{\url{https://mappings.anu.edu.au/}} code to model the interstellar medium (ISM) in the NLRs surrounding the AGN. The latest available atomic data from {\sc CHIANTI} version 10 (\citealt{DelZanna.etal.2021}) is included in {\sc MAPPINGS~V}. For the theoretical AGN photoionization models, we adopt the realistic SEDs derived by the AGN radiation model {\sc OPTXAGNF}\footnote{\url{https://heasarc.gsfc.nasa.gov/xanadu/xspec/manual/node211.html}} as the incident ionizing radiation sources (\citealt{Done.etal.2012, Jin.etal.2012}). The adopted AGN radiation sources include the soft X-ray excess with energy between $\sim 0.2 - 2$ keV, which can be important for the excitation of infrared forbidden lines. To model various types of AGN, we sample the black hole mass (from $10^6\,M_{\odot}$ to $10^9\,M_{\odot}$) and Eddington ratio (from 10$^{-5}$ to 1) both in increments of 0.5 dex step size as the inputs to the embedded {\sc OPTXAGNF} in {\sc MAPPINGS~V}. The black hole mass and Eddington ratio together contribute to a shift of the peak energy, i.e., $E_{\rm peak}$, of the AGN spectrum as described by Equations~4~--~6 in \citeauthor{Thomas.etal.2016} (2016, see also Figure~3 therein). Other parameters of {\sc OPTXAGNF} are set to be the recommended values\footnote{$a_{*}=0$, $r_{\rm cor} = 40\,R_{g}$, $r_{\rm out} = 10^{4}\,R_{g}$, $kT_{e} = 0.2\,{\rm keV}$, $\tau = 15$, $\Gamma = 2.2$, and $f_{\rm pl} = 0.2$.} for NLRs in {\sc MAPPINGS~V} (see Figure~\ref{SpecAGN} for examples of the AGN ionizing radiation sources). Potential caveats of the adopted AGN SEDs are discussed in Section~\ref{sec4.1}.

For each AGN photoionization model, we adopt the abundance scaling relations provided by \cite{Nicholls.etal.2017} and consider the metallicity variation from 12+log (O/H) = 8.150, 8.427, 8.632, 8.760, 8.850, 8.943 to 8.997 (i.e., 0.2, 0.4, 0.7, 1.0, 1.3, 1.7, and 2.0 $Z_{\odot}$). The adopted abundance scaling relations applied a more realistic cosmic abundance standard (\citealt{Nieva&Przybilla2012}), instead of a solar abundance standard. Dust depletion effect, an important factor in ensuring reliable models in high metallicity regions (\citealt{Zhu.etal.2023}), is included in the model with the depletion factors derived by \cite{Thomas.etal.2018} according to the results of \cite{Jenkins2009}. We adopt a plane-parallel geometry in the AGN photoionization modeling with the ionization parameter log~$U$ at the inner edge of the cloud varying from $-4.8$ to $-1.3$ by a 0.5 dex step size.\footnote{The log~$U$ values already include the systematic differences in definition from the values calculated using spherical geometry models and can therefore be directly compared with them.} Furthermore, we adopt an isobaric structure in the calculation, with the gas pressure log~$(P/k)$ varying from 5.5 to 8.5, unless specifically noted, by a 0.5 dex step size. An isochoric structure can be selected as well, while the model results based on these two structures do not change much in terms of the line ratio predictions (e.g., \citealt{Pereira-Santaella.etal.2010,Pereira-Santaella.etal.2017,Pereira-Santaella.etal.2024, Nagao.etal.2011}). The detailed calculation of the AGN model proceeds at a 2~\% photon absorption step and stops when the hydrogen becomes 99~\% neutral (with the $N_{\rm H}$ of $\sim 10^{21} - 10^{22}\,\rm cm^{-2}$). The model results calculated using the different models involved in this paper are available in Appendix~\ref{secA}.

\section{Modeling Results and Applications}\label{sec3}

The primary goal of this work is to extract the optimal line ratio diagnostics accessible by JWST/MRS spectral observations to quantify the ionization parameter (i.e., $U$) in AGN for science goals as exampled in the introduction section. We are also interested in diagnostics of the hardness of ionizing radiation field (i.e., $E_{\rm peak}$), metallicity, and gas pressure in AGN, as these parameters play important roles in determining the physical conditions and hence corresponding evolutionary processes therein. 

To this end, we first select 23 sufficiently strong lines (overall $\gtrsim$ 0.1\% of H$\beta$ emission) within JWST/MRS bands based on their line strengths predicted by the theoretical AGN models in order to analyze their dependence on the physical conditions of AGN (see Table~\ref{tablines} for the emission line list and Table~\ref{tabRes1} for the modeling results). We confirm that these selected lines are all detectable with JWST/MRS according to recent studies of nearby Seyferts (e..g, \citealt{Pereira-Santaella.etal.2022, Armus.etal.2023, Zhang.etal.2024a, AlonsoHerrero.etal.2025, Ceci.etal.2025, HermosaMunoz.etal.2025, Marconcini.etal.2025}). The only exception is the [Fe~{\small II}]~25.99~$\mum$ emission line, which was not reported in all of the observations, especially for the nuclear regions. The lacking of [Fe~{\small II}]~25.99~$\mum$ emission is likely because that most Fe$^{+}$ are in high level populations (corresponding to stronger [Fe~{\small II}]~5.34~$\mum$ emission) in the hot environments adjacent to AGN, or simply due to the sensitivity issue of MRS channel 4 observations. All combinations of any two lines in Table~\ref{tablines} are under consideration for the selection of the optimal diagnostics.

\startlongtable
\setlength{\tabcolsep}{3pt}
\begin{deluxetable}{lccc||lccc}
\tabletypesize{\footnotesize}
\tablecolumns{8}
\tablecaption{Basic Information of Selected Emission Lines}
\tablehead{
\colhead{Line} & \colhead{$\lambda$} & \colhead{IP} & \colhead{log~$n_{c}$} &\colhead{Line} & \colhead{$\lambda$} & \colhead{IP} & \colhead{log~$n_{c}$} \\
\colhead{(-)} & \colhead{($\mum$)} & \colhead{(eV)} & \colhead{[${\rm cm^{-3}}$]} &\colhead{(-)} & \colhead{($\mum$)} & \colhead{(eV)} & \colhead{[${\rm cm^{-3}}$]} \\
\colhead{(1)} & \colhead{(2)} & \colhead{(3)} & \colhead{(4)} & \colhead{(1)} & \colhead{(2)} & \colhead{(3)} & \colhead{(4)}}
\startdata
$\rm [Fe~II]$ & 5.34 & 7.9 & 2.48 & $\rm[S~IV]$ & 10.51 & 34.8 & 4.75 \\
$\rm[Fe~VIII]$ & 5.45 & 124 & - & $\rm Hu\alpha$ & 12.37 & 13.6 & - \\
$\rm[Mg~VII]$ & 5.50 & 186.5 & 6.53 & $\rm[Ne~II]$ & 12.81 & 21.6 & 5.80 \\
$\rm[Mg~V]$ & 5.61 & 109.2 & 6.60 & $\rm[Ar~V]$ & 13.10 & 40.7 & 4.47 \\
$\rm[Ar~II]$ & 6.99 & 15.8 & 5.62 & $\rm[Ne~V]$ & 14.32 & 97.1 & 4.51 \\
$\rm[Na~III]$ & 7.32 & 47.3 & - &  $\rm[Ne~III]$ & 15.56 & 40.1 & 5.32 \\
$\rm Pf\alpha$ & 7.46 & 13.6 & - & $\rm[S~III]$ & 18.71 & 23.3 & 4.08 \\
$\rm[Ne~VI]$ & 7.65 & 126.2 & 5.80 & $\rm[Ar~III]$ & 21.83 & 27.6 & 5.38 \\
$\rm[Fe~VII]$ & 7.82 & 99.1 & 6.10 & $\rm[Ne~V]$ & 24.32 & 97.1 & 3.77 \\
$\rm[Ar~V]$ & 7.90 & 40.7 & 5.20 & $\rm[O~IV]$ & 25.89 & 54.9 & 4.00 \\
$\rm[Ar~III]$ & 8.99 & 27.6 & 5.28 & $\rm[Fe~II]$ & 25.99 & 7.9 & 4.11 \\
$\rm[Fe~VII]$ & 9.53 & 99.1 & 5.74 & $$ & $$ & $$ & $$ \\
\enddata
\tablecomments{Column (1): line name. Column (2): rest wavelength. Column (3): \\ionization potential. Column (4): critical electron density for collisional \\excitation calculated by {\sc PyNeb} (\citealt{Luridiana.etal.2015}) at $T_{e} = 10^{4}\,\rm K$.}
\label{tablines}
\end{deluxetable}

\subsection{Ionization Parameter Diagnostics}\label{sec3.1}

\begin{figure*}[!ht]
\center{\includegraphics[width=1\textwidth]{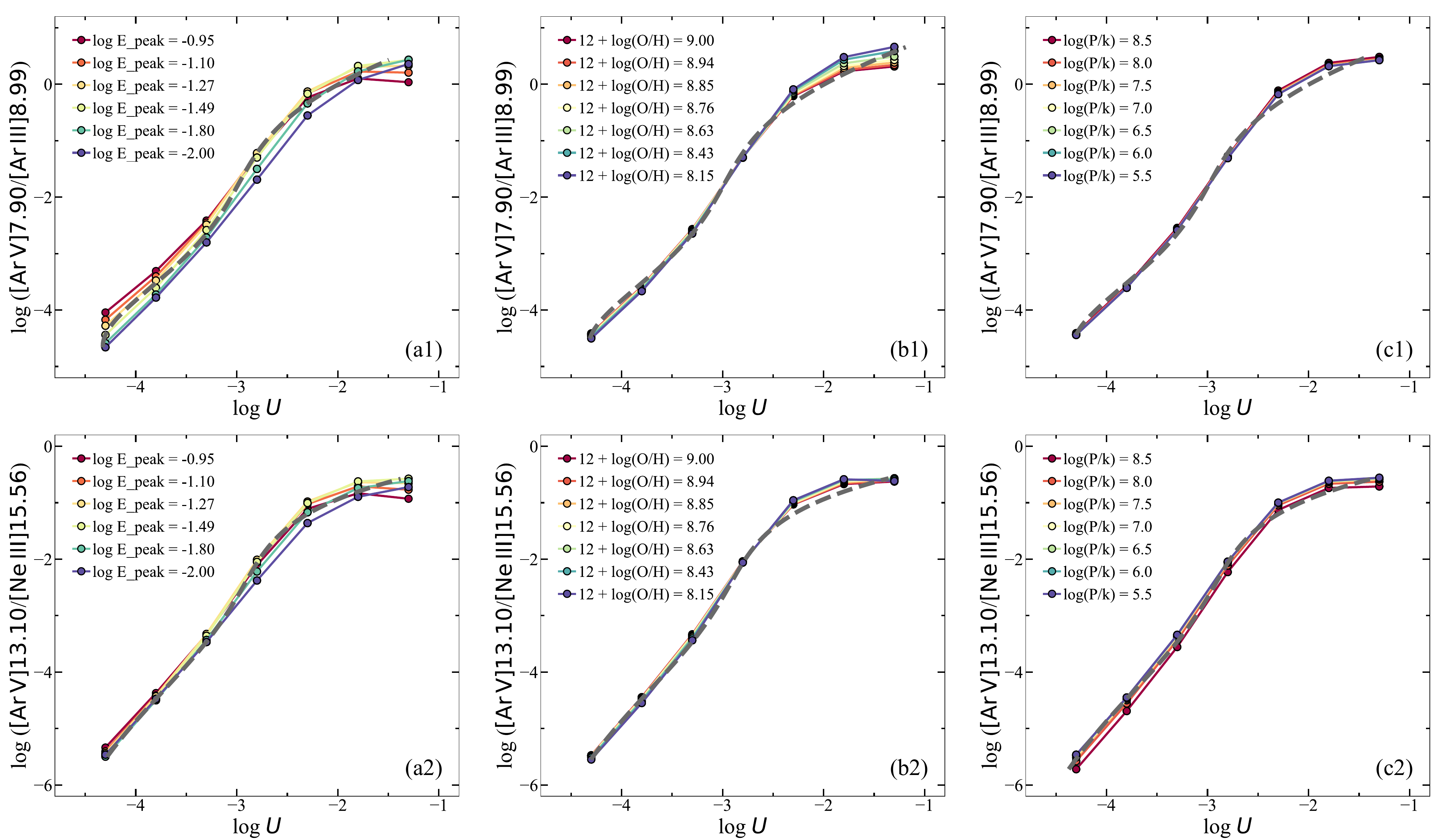}}
\caption{Dependence of (top) [Ar~{\small V}]7.90/[Ar~{\small III}]8.99 and (bottom) [Ar~{\small V}]13.10/[Ne~{\small III}]15.56 on the ionization parameter ($U$), as predicted by AGN photoionization models (colored points). Each panel displays model results under specific parameter constraints: (a1\&a2) 12 + log (O/H) = 8.76 and log $(P/k)$ = 7.0; (b1\&b2) log $E_{\rm peak}$ = $-1.49$ and log $(P/k)$ = 7.0; (c1\&c2) log $E_{\rm peak}$ = $-1.49$ and 12 + log (O/H) = 8.76, while the gray-dashed line represents the best-fit correlation (Equation~\ref{equ1}) derived from all AGN models in Section~\ref{sec3}.}\label{Fig_U_AGN}
\end{figure*}

To efficiently extract the optimal $U$ diagnostics, we use a feature extraction method based on the Pearson correlation coefficient.\footnote{Pearson correlation coefficient measures the linear correlation between two variables and it is essentially a normalized measurement of the covariance. Its value lies between $-1$ and 1, where $-1$ and 1 indicate ideal negative and positive linear correlation, respectively.} Specifically, for all combinations of any two lines in Table~\ref{tablines}, we first calculate the Pearson correlation coefficient between the logarithmic line ratio and log~$U$ of theoretical AGN models. All line ratios are then ranked according to the absolute values of their corresponding Pearson correlation coefficients. As an additional quality control check, we visually inspect each correlation between the logarithmic line ratios and log~$U$, from the correlations with high to low absolute Pearson correlation coefficients. Overall, line ratios involving [Ar~{\small V}] work the best as $U$ diagnostics, followed by line ratios involving [Ne~{\small V}], and finally line ratios involving [O~{\small IV}] with larger scatters. This result indicates that the [Ar~{\small V}] emissivity has the strongest dependance on $U$.

Considering that [Ar~{\small V}] has the best performance in $U$ diagnostics, we then combine some practical criteria to select the optimal $U$ diagnostics from the line ratios involving [Ar~{\small V}]. Specifically, we first consider the line ratios of paired lines with close wavelengths to better avoid the influence of dust extinction in observations (even if relatively small in mid-IR). Moreover, the line ratios of paired lines close in wavelengths falling in the same JWST/MRS channel are also more time efficient to be observed. Secondly, when given the comparable robustness, we prefer the line ratios of paired lines with shorter wavelengths, since shorter wavelengths correspond to higher spatial and spectral resolutions in JWST observations, and are applicable to a wider redshift range. Based on the result of the feature extraction process and these practical criteria, Figure~\ref{Fig_U_AGN} demonstrates two optimal $U$ diagnostics (i.e., [Ar~{\small V}]7.90/[Ar~{\small III}]8.99, and [Ar~{\small V}]13.10/[Ne~{\small III}]15.56). Besides the tight correlations, the two $U$ diagnostics also have their own pros and cons.

\begin{figure*}[!ht]
\center{\includegraphics[width=1\textwidth]{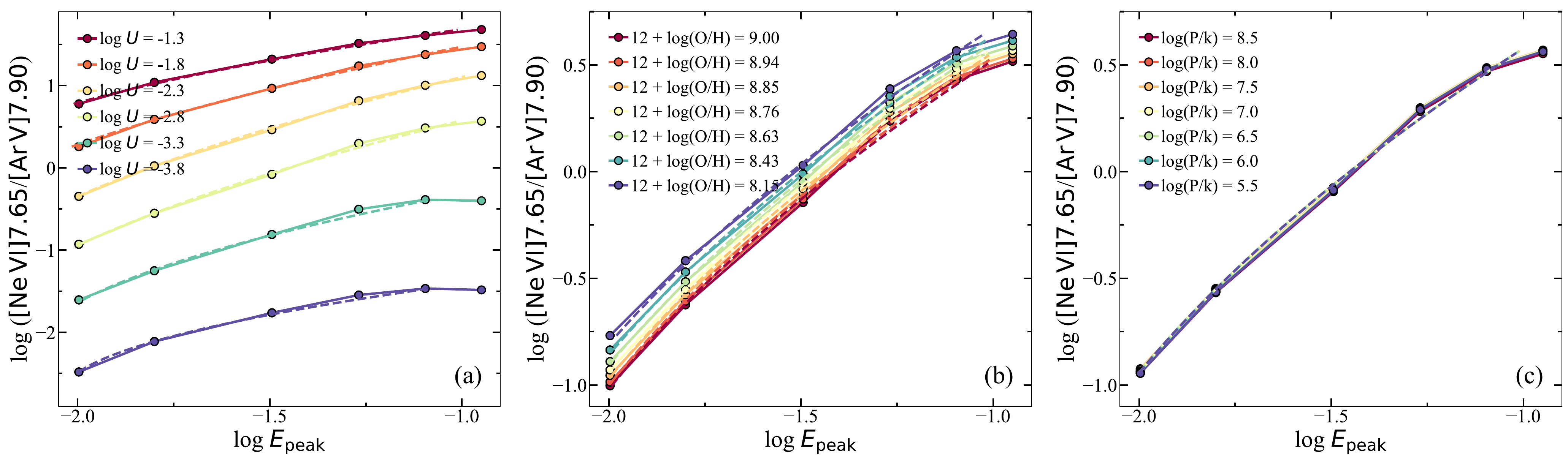}}
\caption{Dependence of [Ne~{\small VI}]7.65/[Ar~{\small V}]7.90 on the input AGN spectrum peak energy ($E_{\rm peak}$, in keV), as predicted by AGN photoionization models (colored points). Each panel displays model results under specific parameter constraints: (a) 12 + log (O/H) = 8.76 and log $(P/k)$ = 7; (b) log $U$ = $-2.8$ and log $(P/k)$ = 7; (c) log $U$ = $-2.8$ and 12 + log (O/H) = 8.76. The dashed lines represent the best-fit correlations (Equation~\ref{equ2}--\ref{equ2_3}) derived from all AGN models with the given log $U$ and 12 + log (O/H).}\label{Fig_Ep_AGN}
\end{figure*}

The line ratio [Ar~{\small V}]7.90/[Ar~{\small III}]8.99 best follows the practical criteria as discussed, but it has a slightly secondary dependence on $E_{\rm peak}$ due to the different ionization potentials of [Ar~{\small V}] and [Ar~{\small III}]. The line ratio [Ar~{\small V}]13.10/[Ne~{\small III}]15.56 is relatively less sensitive to $E_{\rm peak}$ given the similar ionization potentials of [Ar~{\small V}] and [Ne~{\small III}]. However, this line ratio has a slight dependence on gas pressure due to the different critical densities of [Ar~{\small V}]13.10 and [Ne~{\small III}]15.56 and this line ratio falls in longer wavelength MRS band. Nevertheless, these two line ratios, considering the tight correlations as shown in Figure~\ref{Fig_U_AGN}, remain numerically (and physically; see Section~\ref{sec4.1.1}) the best diagnostics for $U$, especially when compared to line ratio diagnostics for $U$ in AGN based on relatively low ionization lines (e.g., \citealt{Pereira-Santaella.etal.2017, Zhu.etal.2024}). An alternative $U$ diagnostic is [Ar~{\small V}]7.90/[Ne~{\small III}]15.56, which has even weaker dependence on both $E_{\rm peak}$ and gas pressure given the comparable ionization potentials and critical densities of [Ar~{\small V}]7.90 and [Ne~{\small III}]15.56, whereas this line ratio is not the best in observations considering their significantly separate wavelengths. This line ratio might be in turn used to constrain different dust extinction models with known $U$.

To calibrate $U$ with these line ratios, we adopt a quintic function, i.e., Equation~\ref{equ1}, 
\noindent
\begin{align}\label{equ1}
\begin{aligned}
{\rm log}\,U = \,\,&c_5\times({\rm log}\,R)^{5} + c_4\times({\rm log}\,R)^{4} + c_3\times({\rm log}\,R)^{3} + \\&c_2\times({\rm log}\,R)^{2} + c_1\times{\rm log}\,R + c_0
\end{aligned}
\end{align}
\noindent
and perform a least-square fit of the curved correlations between the input log~$U$ and predicted line ratios using the Levenberg-Marquardt algorithm. Note that we selected the quintic function (as well as other functions adopted in following subsections) from among different functions simply because they can provide numerically accurate calibrations for the model results studied here, and the functions adopted in other specific studies may differ.

The best-fit coefficients for $R =$ [Ar~{\small V}]7.90/[Ar~{\small III}]8.99, [Ar~{\small V}]13.10/[Ne~{\small III}]15.56, and also [Ar~{\small V}]7.90/[Ne~{\small III}]15.56, are listed in Table~\ref{tabEqu1} in Appendix~\ref{secA}. As a quantification for the robustness of the quintic function fitting, we calculate the root-mean-square (RMS) scatters of the best-fit correlations relative to the model results, which are 0.16 dex and 0.17 dex for [Ar~{\small V}]7.90/[Ar~{\small III}]8.99 and [Ar~{\small V}]13.10/[Ne~{\small III}]15.56, respectively. The best-fit correlations in this section are all for AGN models with relatively high $E_{\rm peak}$ as illustrated in Figure~\ref{Fig_U_AGN}(a). The best-fit results for AGN models with low $E_{\rm peak}$ (namely low-luminosity AGN) are specifically discussed in Section~\ref{sec4} given the bi-valuation dependence of predicted line ratios on $E_{\rm peak}$. In fact, since the line ratio diagnostics of $U$ have basically no dependence on $E_{\rm peak}$, the best-fit results for different $E_{\rm peak}$ ranges are essentially the same.

\subsection{Peak Energy Diagnostics}\label{sec3.2}

As illustrated in Figure~\ref{SpecAGN}, the energy corresponding to the peak flux (i.e., $E_{\rm peak}$, in keV) of the AGN ionizing spectrum increases along with the increase of $\lambda_{\rm Edd}$. Higher $E_{\rm peak}$ of the AGN ionizing spectrum with relatively more high energy ionizing photons essentially corresponds to harder radiation field in AGN. Therefore, in the following discussion, we use the radiation filed hardness in AGN synonymously with the $E_{\rm peak}$. Variations in radiation filed hardness and the corresponding line ratio diagnostics in observations are critical to constrain theoretical models (e.g., \citealt{Rigopoulou.etal.1996, Thornley.etal.2000, Giveon.etal.2002, Verma.etal.2003,Draine.etal.2021}) and to explain observational ISM properties of different galaxies (e.g., \citealt{Genzel.etal.1998, Sturm.etal.2002, Hunt.etal.2010, Perez-Montero.etal.2024, Perez-Montero.etal.2025, Baron.etal.2025}). The ratios of mid-IR emission lines from the same element but of different excitation levels (e.g., [Ne~{\small V}]14.32/[Ne~{\small II}]12.81, [Ne~{\small III}]15.56/[Ne~{\small II}]12.81, [Ar~{\small III}]8.99/[Ar~{\small II}]6.99, and [S~{\small IV}]10.51/[S~{\small III}]18.71) have long been proposed to be good diagnostics of the ionizing radiation field hardness (e.g., \citealt{Genzel.etal.1998, Thornley.etal.2000, Giveon.etal.2002, Verma.etal.2003}). The ratios of some UV/optical emission lines (e.g., He{\small II}$\lambda$1640/He{\small I}$\lambda$3187, He{\small II}$\lambda$1640/H$\beta$, and [He{\small II}]$\lambda$4686/H$\beta$) have also been recently proposed to have the same function (\citealt{Zhu.etal.2024}). 

\begin{figure*}[!ht]
\center{\includegraphics[width=1\textwidth]{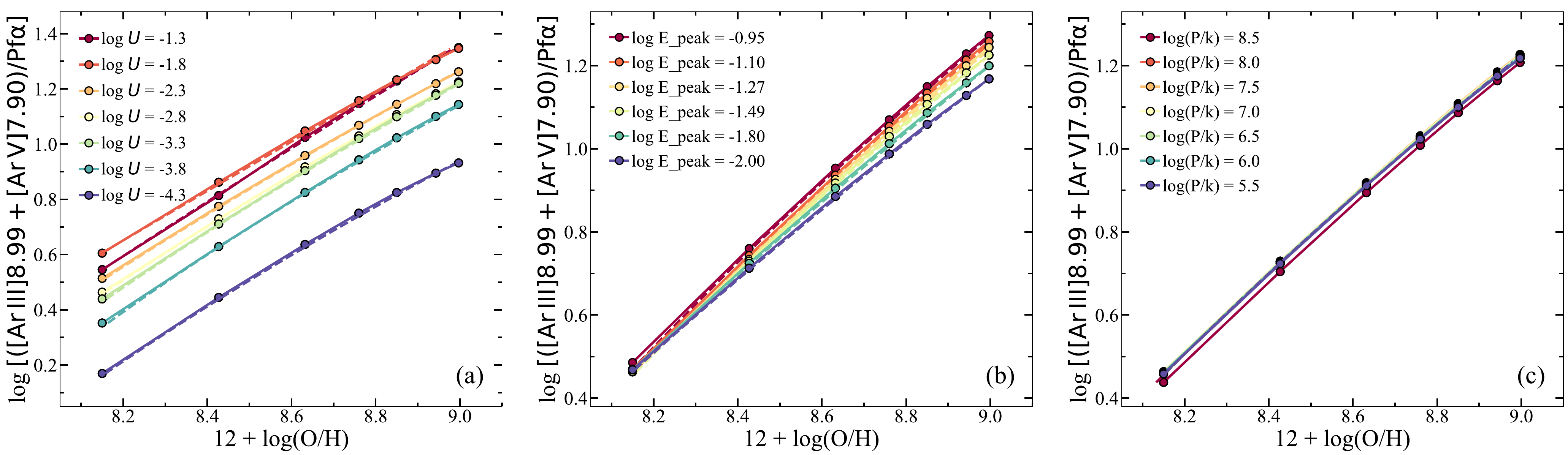}}
\caption{Dependence of ([Ar~{\small III}]8.99+[Ar~{\small V}]7.90)/Pf$\alpha$ on the metallicity (12 + log (O/H)), as predicted by AGN photoionization models (colored points). Each panel displays model results under specific parameter constraints: (a) log $E_{\rm peak}$ = $-1.49$ and log $(P/k)$ = 7.0; (b) log $U$ = $-2.8$ and log $(P/k)$ = 7.0; (c) log $U$ = $-2.8$ and log $E_{\rm peak}$ = $-1.49$. The dashed lines represent the best-fit correlations (Equation~\ref{equ3}--\ref{equ3_3}) derived from all AGN models with the given log $U$ and log $E_{\rm peak}$.}\label{Fig_A_AGN}
\end{figure*}

In principle, the optimal $E_{\rm peak}$ diagnostics of AGN within JWST/MRS bands are those line ratios involving the infrared lines of very high ionization potentials (e.g., [Ne~{\small VI}]7.65, [Fe~{\small VIII}]5.45, and [Mg~{\small V}]5.61 with IP $>$ 100 eV). These high ionization potential lines have several advantages in diagnosing $E_{\rm peak}$ in AGN. First, these lines can only be produced in regions with very high energy photons and are hence more sensitive to the variation of the peak energy of AGN ionizing radiation field. Moreover, these high ionization potential lines can barely be produced by starbursts, which can contribute significantly to the excitation of the relatively low ionization potential lines such as [Ne~{\small III}]15.56, [Ar~{\small III}]8.99, and [S~{\small IV}]10.51 (see Section~\ref{sec4.2}). We noted that [Ne~{\small V}]14.32 and [Ne~{\small VI}]7.65 have been recently detected using JWST/MIRI observations in the nuclear region of M83, previously known as a purely starburst system, but such detection are still more likely due to excitation of fast radiative shocks or weakly accreting AGN rather than starbursts (i.e., \citealt{Hernandez.etal.2025}). These lines are also less affected by dust extinction than the UV/optical diagnostics. We therefore expect that these infrared lines of very high ionization potentials will be more widely used in diagnosing the radiation field hardness in the era of JWST.

Applying the same feature extraction process as detailed in Section~\ref{sec3.1}, we find that none of the line ratios works as effectively in directly diagnosing $E_{\rm peak}$ as in diagnosing $U$. Some line ratios investigated here do have a secondary dependence on $E_{\rm peak}$, but still with a primary dependence on $U$. We therefore adopt a different strategy for the calibration of $E_{\rm peak}$ as described by Equation~\ref{equ2}.
\noindent
\begin{align}\label{equ2}
\begin{aligned}
{\rm log}\,E_{\rm peak} = a\times({\rm log}\,R - b)^{2} + c,
\end{aligned}
\end{align}
\noindent
where
\noindent
\begin{align}\label{equ2_1}
\begin{aligned}
a = \,\,&a_{Z,2}\times(12+{\rm log\,(O/H)})^{2} + \\&a_{Z,1}\times(12+{\rm log\,(O/H)}) + \\&a_{U,1}\times{\rm log}\,U + a_0,
\end{aligned}
\end{align}
\noindent
\begin{align}\label{equ2_2}
\begin{aligned}
b = \,\,&b_{Z,2}\times(12+{\rm log\,(O/H)})^{2} + \\&b_{Z,1}\times(12+{\rm log\,(O/H)}) + \\&b_{U,1}\times{\rm log}\,U + b_0,
\end{aligned}
\end{align}
\noindent
\begin{align}\label{equ2_3}
\begin{aligned}
c = \,\,&c_{Z,2}\times(12+{\rm log\,(O/H)})^{2} + \\&c_{Z,1}\times(12+{\rm log\,(O/H)}) + \\&c_{U,1}\times{\rm log}\,U + c_0.
\end{aligned}
\end{align}
\noindent
The prerequisites of this prescription are the measurements of ionization parameter, which can be estimated as discussed above, and metallicity, which can also be constrained using the infrared line ratios as detailed in Section~\ref{sec3.3}.

Following the strategy in Equations~\ref{equ2}$-$\ref{equ2_3}, we examined the predicted dependence of all the line ratios involving high ionization potential lines on $E_{\rm peak}$ and found that the one composed of emission lines from two noble gases [Ne~{\small VI}]7.65/[Ar~{\small V}]7.90 is the best among other emission line ratios within JWST/MRS bands in diagnosing $E_{\rm peak}$ in AGN (see Figure~\ref{Fig_Ep_AGN}). This result is consistent with the expectation for the high ionization potential lines in diagnosing $E_{\rm peak}$ in AGN. For any value of log~$U$ between $-3.8$ and $-1.3$, the increase in [Ne~{\small VI}]7.65/[Ar~{\small V}]7.90 exceeds 1 dex as log~$E_{\rm peak}$ changes from $-2.0$ to $-1.0$. Accurate calibrations of log~$U$ values are available as discussed in Section~\ref{sec3.1}, and the scatter of [Ne~{\small VI}]7.65/[Ar~{\small V}]7.90 due to variations in metallicity remains within 0.30 dex at the low end and 0.15 dex at the high end of log$E_{\rm peak}$.

[Ne~{\small VI}]7.65/[Ar~{\small V}]7.90 follows the same practical criteria applied to the $U$ diagnostics. The best-fit coefficients in Equation~\ref{equ2} for $R =$ [Ne~{\small VI}]7.65/[Ar~{\small V}]7.90 in terms of different $U$ ranges are listed in Table~\ref{tabEqu2}. The RMS scatters of the piecewise fitting correlations relative to the model results range from 0.02 to 0.05 dex. Possible alternative $E_{\rm peak}$ diagnostics in theory are [Fe~{\small VIII}]5.45/[Mg~{\small V}]5.61 and [Mg~{\small V}]5.61/[Fe~{\small VII}]7.82. However, their diagnostic reliability is greatly undermined in observations due to the significant dust depletion, which can be up to $\sim$ 99\%, of iron and magnesium in galaxies. The complex and inhomogeneous extinction due to dirty water ice in the $5.5-7.5\,\mum$ band (e.g., \citealt{Garcia-Bernete.etal.2024a}) can also complicate the application of these two line ratios in diagnosing $E_{\rm peak}$.

\subsection{Metallicity Diagnostics}\label{sec3.3}

\begin{figure*}[!ht]
\center{\includegraphics[width=1\textwidth]{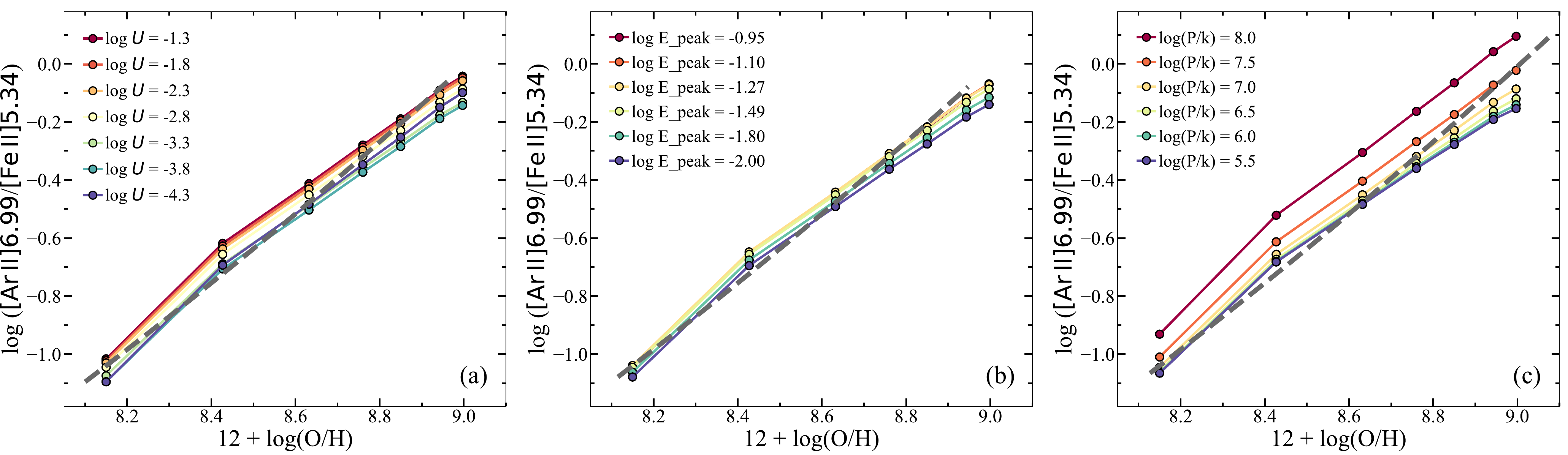}}
\caption{Dependence of [Ar~{\small II}]6.99/[Fe~{\small II}]5.34 on the metallicity (12 + log (O/H)), as predicted by AGN photoionization models (colored points). Each panel displays model results under specific parameter constraints: (a) log $E_{\rm peak}$ = $-1.49$ and log $(P/k)$ = 7.0; (b) log $U$ = $-2.8$ and log $(P/k)$ = 7.0; (c) log $U$ = $-2.8$ and log $E_{\rm peak}$ = $-1.49$, while the gray-dashed line represents the best-fit correlation (Equation~\ref{equ4}) derived from all AGN models with $(P/k) \leq 8.0$.}\label{Fig_A_II_AGN}
\end{figure*}

Following the traditional definition of metallicity, we first study the feasibility of the relative intensities of infrared neon and argon emission to ionized hydrogen emission (i.e., Pf$\alpha$ here, the strongest ionized hydrogen line within MIRI/MRS band) in estimating metallicity. After visually inspecting all such correlations, we find that the high excitation lines from neon and argon (i.e., [Ne~{\small V}], and [Ar~{\small V}]) are better than the relatively low excitation lines (i.e., [Ne~{\small III}], [Ne~{\small II}], [Ar~{\small III}], and [Ar~{\small II}]) in diagnosing metallicity in AGN, although all lines show a primary dependence on $U$ and a weak dependence on $E_{\rm peak}$. The better performance of the highly excited lines is within expectation given their enhanced emissivities in AGN, i.e., higher sensitivities to the radiation field with varying metallicity. Meanwhile, [Ar~{\small V}] emission lines, especially [Ar~{\small V}]7.90, correlate better than [Ne~{\small V}] emission lines, as [Ar~{\small V}] emission lines have much higher critical densities and are therefore not dependent on gas pressure. 

We further find that the addition of [Ar~{\small III}]8.99 eliminates most of the dependence of [Ar~{\small V}]7.90 on $E_{\rm peak}$. Moreover, the addition of [Ar~{\small III}]8.99 mitigates the contamination effect of potential nuclear starbursts in AGN as nuclear starbursts contribute to both the [Ar~{\small III}]8.99 and Pf$\alpha$ emission and hence to a certain degree cancel the contamination effect. We therefore propose an optimal prescription in estimating metallicity with $R =$ ([Ar~{\small III}]8.99+[Ar~{\small V}]7.90)/Pf$\alpha$ (see Figure~\ref{Fig_A_AGN}). For any value of log~$U$ between $-4.3$ and $-1.3$, the increase in ([Ar~{\small III}]8.99+[Ar~{\small V}]7.90)/Pf$\alpha$ is of 0.7 dex as the metallicity changes from 0.2 to 2.0 $Z_{\odot}$. Accurate log~$U$ estimates are available as discussed in Section~\ref{sec3.1}, and the scatter of ([Ar~{\small III}]8.99+[Ar~{\small V}]7.90)/Pf$\alpha$ due to variations in log~$E_{\rm peak}$ ranges from 0.01 to 0.17 dex for different log~$U$ values.

Similar to the calibration of $E_{\rm peak}$, we adopt the strategy as described by Equation~\ref{equ3} to calibrate metallicity in AGN, but with different coefficients.

\noindent
\begin{align}\label{equ3}
\begin{aligned}
12 + {\rm log\,(O/H)} = a\times({\rm log}\,R - b)^{2} + c.
\end{aligned}
\end{align}
\noindent
where

\noindent
\begin{align}\label{equ3_1}
\begin{aligned}
a = \,\,&a_{E,3}\times({\rm log}\,E_{\rm peak})^{3} + a_{E,2}\times({\rm log}\,E_{\rm peak})^{2} + \\&a_{E,1}\times{\rm log}\,E_{\rm peak} + a_{U,1}\times{\rm log}\,U + a_0,
\end{aligned}
\end{align}
\noindent
\begin{align}\label{equ3_2}
\begin{aligned}
b = \,\,&b_{E,3}\times({\rm log}\,E_{\rm peak})^{3} + b_{E,2}\times({\rm log}\,E_{\rm peak})^{2} + \\&b_{E,1}\times{\rm log}\,E_{\rm peak} + b_{U,1}\times{\rm log}\,U + b_0,
\end{aligned}
\end{align}
\noindent
\begin{align}\label{equ3_3}
\begin{aligned}
c = \,\,&c_{E,3}\times({\rm log}\,E_{\rm peak})^{3} + c_{E,2}\times({\rm log}\,E_{\rm peak})^{2} + \\&c_{E,1}\times{\rm log}\,E_{\rm peak} + c_{U,1}\times{\rm log}\,U + c_0.
\end{aligned}
\end{align}
\noindent

For the above calibration, we do not need to estimate the ionization correction factors as in previous metallicity calibration studies (e.g., \citealt{Martin-Hernandez.etal.2002, Verma.etal.2003, Nagao.etal.2011}), because the intrinsic parameters that determine the ionization correction factors are already included in the calibration. The best-fit coefficients for the metallicity calibration with $R =$ ([Ar~{\small III}]8.99+[Ar~{\small V}]7.90)/Pf$\alpha$ in terms of different $U$ ranges are listed in Table~\ref{tabEqu3}. The RMS scatters of the piecewise fitting correlations relative to the model results are of $\sim 0.01 - 0.02$ dex. The accurate metallicity calibration here depends on reliable estimates of $U$ and $E_{\rm peak}$. The former can be accurately estimated as discussed in Section~\ref{sec3.1}, while the latter needs to be estimated iteratively together with the metallicity.

\begin{figure*}[!ht]
\center{\includegraphics[width=1\textwidth]{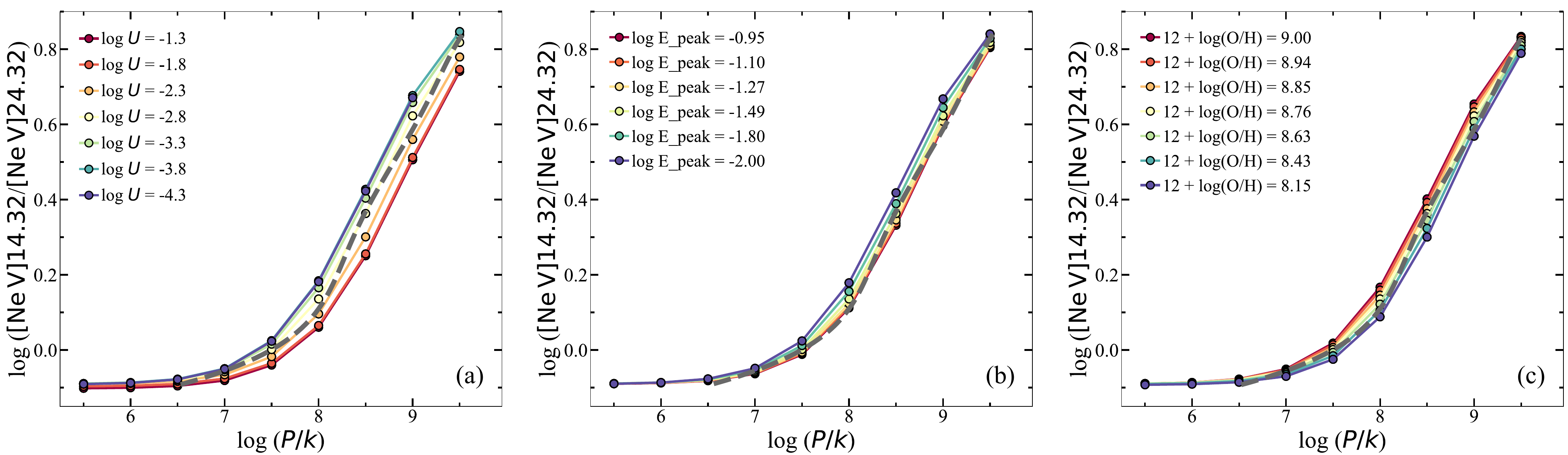}}
\caption{Dependence of [Ne~{\small V}]14.32/[Ne~{\small V}]24.32 on the gas pressure ($P/k$, in $\rm K\,cm^{-3}$), as predicted by AGN photoionization models (colored points). Each panel displays model results under specific parameter constraints: (a) log $E_{\rm peak}$ = $-1.49$ and 12 + log (O/H) = 8.76; (b) log $U$ = $-2.8$ and 12 + log (O/H) = 8.76; (c) log $U$ = $-2.8$ and log $E_{\rm peak}$ = $-1.49$, while the gray-dashed line represents the best-fit correlation (Equation~\ref{equ4}) derived from all AGN models with log $(P/k) \gtrsim 6.5$.}\label{Fig_P_AGN}
\end{figure*}

In addition, through the same feature extraction process as detailed in Section~\ref{sec3.1}, we find that besides the relative intensities of infrared neon and argon emission to ionized hydrogen emission, the line ratios of infrared neon and argon emission to singly ionized iron emission can, in theory, also be used to calibrate metallicity in AGN. In particular, the line ratios involving singly ionized neon or argon (e.g., [Ar~{\small II}]6.99/[Fe~{\small II}]5.34) have an almost linear dependence on metallicity in AGN models. This result is plausibly due to the strong dependence of iron depletion by dust on metallicity (e.g., \citealt{OHalloran.etal.2008}). As illustrated in Figure~\ref{Fig_A_II_AGN} and described by Equation~\ref{equ4}, 
\noindent
\begin{align}\label{equ4}
\begin{aligned}
12 + {\rm log}({\rm O/H}) = c_2\times({\rm log}\,R)^{2} + c_1\times{\rm log}\,R + c_0.
\end{aligned}
\end{align}
\noindent
we simply adopt a quadratic function to calibrate the metallicity with $R =$ [Ar~{\small II}]6.99/[Fe~{\small II}]5.34 (also for [Ne~{\small II}]12.81/[Fe~{\small II}]5.34, [Ar~{\small II}]6.99/[Fe~{\small II}]25.99, and [Ne~{\small II}]12.81/[Fe~{\small II}]25.99, which are more susceptible to dust extinction correction), and the best-fit coefficients are listed in Table~\ref{tabEqu4}.\footnote{The calibration of metallicity using [Ar~{\scriptsize II}]6.99/[Fe~{\scriptsize II}]5.34 and [Ne~{\scriptsize II}]12.81/[Fe~{\scriptsize II}]5.34 presented here is only for log $(P/k) \leq 8.0$ given the strong dependence of [Fe~{\scriptsize II}]5.34 on gas pressure.}

The metallicity calibrated based on [Ar~{\small II}]6.99/[Fe~{\small II}]5.34 has a larger uncertainty than that calibrated based on ([Ar~{\small III}]8.99+[Ar~{\small V}]7.90)/Pf$\alpha$. This larger uncertainty is mainly due to the dependence of [Ar~{\small II}]6.99/[Fe~{\small II}]5.34 (the same for [Ne~{\small II}]12.81/[Fe~{\small II}]5.34) on gas pressure (see Figure~\ref{Fig_A_II_AGN}c). Specifically, the collisional de-excitation of [Fe~{\small II}] is enhanced relative to the radiative de-excitation at high gas pressure (equally high gas density), given its relatively low critical electron density log $(n_{c}/{\rm cm^{-3}})$ = 2.48. Line ratios involving [Fe~{\small II}]25.99 are less sensitive to gas pressure but are more susceptible to dust extinction correction. Moreover, the metallicity calibrations involving iron emission lines are susceptible to dust depletion variations (e.g., \citealt{Konstantopoulou.etal.2022, Konstantopoulou.etal.2024, Hamanowicz.etal.2024}). [Fe~{\small II}] emission is also sensitive to shock excitation that is prevalent in AGN, especially in low-luminosity AGN (see Section~\ref{sec4.1.2}). Another caveat regarding the use of the low excitation infrared line ratios to calibrate metallicity, as well as other parameters, in AGN will be discussed in Section~\ref{sec4.1.1}. Considering these caveats, we propose that [Ar~{\small II}]6.99/[Fe~{\small II}]5.34 (the same for [Ne~{\small II}]12.81/[Fe~{\small II}]5.34, [Ar~{\small II}]6.99/[Fe~{\small II}]25.99, and [Ne~{\small II}]12.81/[Fe~{\small II}]25.99) can only be used for prior estimates in relatively low density environments, while the high-excitation-line diagnostic, i.e., ([Ar~{\small III}]8.99+[Ar~{\small V}]7.90)/Pf$\alpha$, should be used for more robust iterative estimates of metallicity in AGN.

\subsection{Pressure Diagnostics}\label{sec3.4}

The ratios of emission lines of the same ion species but with different critical densities have been widely used as electron density or equivalently gas pressure diagnostics in AGN (e.g., \citealt{Alexander.etal.1999, Dopita.etal.2002, Dudik.etal.2007, Zhu.etal.2024}). Several such emission line ratios are available with JWST/MRS spectral observations (i.e., [Fe~{\small VII}]7.82/[Fe~{\small VII}]9.53, [Ne~{\small V}]14.32/[Ne~{\small V}]24.32, [Ar~{\small V}]7.90/[Ar~{\small V}]13.10, [Ar~{\small III}]8.99/[Ar~{\small III}]21.83, and  [Fe~{\small II}]25.99/[Fe~{\small II}]5.34). We have examined and found that [Ne~{\small V}]14.32/[Ne~{\small V}]24.32 is the best among these infrared line pairs for diagnosing gas pressure in AGN. [Ne~{\small V}]14.32/[Ne~{\small V}]24.32 can be used to diagnose gas pressure in AGN with log~$(P/k) \gtrsim 6.5$, while the other infrared line ratios involved here are insensitive to gas pressure until log~$(P/k) \gtrsim 7.5$ (with the $n_{\rm H}$ of $\sim 10^{3}\,\rm cm^{-3}$). [Ne~{\small V}]14.32/[Ne~{\small V}]24.32 also has the advantage of being unaffected by starburst activity and dust depletion.

As shown in Figure~\ref{Fig_P_AGN}, we include more AGN models with higher gas pressure to better depict the dependence of [Ne~{\small V}]14.32/[Ne~{\small V}]24.32 on gas pressure. We again adopt a quintic function, i.e., Equation~\ref{equ5},
\noindent
\begin{align}\label{equ5}
\begin{aligned}
{\rm log}\,(P/k) = \,\,&c_5\times({\rm log}\,R)^{5} + c_4\times({\rm log}\,R)^{4} + c_3\times({\rm log}\,R)^{3} + \\&c_2\times({\rm log}\,R)^{2} + c_1\times{\rm log}\,R + c_0
\end{aligned}
\end{align}
\noindent
with $R =$ [Ne~{\small V}]14.32/[Ne~{\small V}]24.32 to calibrate the gas pressure and the best-fit coefficients are listed in Table~\ref{tabEqu5}. Note that the adopted calibration is fitted based on the data points with log~$(P/k) \gtrsim 6.5$ and the corresponding RMS uncertainty is 0.19 dex.

\subsection{Applications and Caveats of the Diagnostics}\label{sec3.5}

\subsubsection{Example Applications of the Diagnostics}\label{sec3.5.1}

In the above subsections, we scrutinized the correlations between different line ratios and physical parameters and proposed optimal diagnostics for these physical parameters in a self-consistent manner. In this subsection, we apply these diagnostics to the central $1\farcs5\times1\farcs5$ ($\sim 200 - 300$ pc in physical scales) regions of a sample of six type 1.9/2 Seyferts as a proof of concept. The six Seyferts are included in the Galaxy Activity, Torus, and Outflow Survey (GATOS)\footnote{\url{https://gatos.myportfolio.com}} and were observed by MRS integral field unit (IFU) via the JWST cycle 1 GO program (\#1670; PI: Shimizu, T. Taro). A series of papers from the GATOS collaboration provided discussion in depth of different aspects of one or more of the six targets, including their molecular/ionic emission lines, PAH features, ice/dust properties, and AGN feedback processes such as shocks and outflows (\citealt{AlonsoHerrero.etal.2023, Davies.etal.2024, Esposito.etal.2024, Garcia-Bernete.etal.2024a, Garcia-Bernete.etal.2024b, Garcia-Burillo.etal.2024, HermosaMunoz.etal.2024, Zhang.etal.2024a, Zhang.etal.2024b, Delaney.etal.2025, Esparza-Arredondo.etal.2025, Gonzalez-Martin.etal.2025}).

According to the measured line ratios listed in Table~\ref{tabTest} and the calibrations proposed in Section~\ref{sec3.1}, the six Seyferts have log~$U$ values of $\sim -2.6$ to $-2.4$. As a comparison, the log~$U$ values measured here are consistent with the measurements by \cite{Davies.etal.2020} within 0.2 dex for ESO137$-$G034, MCG$-$05$-$23$-$016, NGC\,3081, and NGC\,5728, and are larger by $\sim 0.4$ dex and $\sim 0.7$ dex for NGC\,5506 and NGC\,7172, respectively. The $U$ values in \cite{Davies.etal.2020} were derived with optical [O~{\small III}]$\lambda5007$/H$\beta$ and [N~{\small II}]$\lambda6584$/H$\alpha$ line ratios measured for a slightly larger ($1\farcs8\times1\farcs8$) aperture from VLT/X-shooter IFU mode observations, and adopted the $U$ calibration by \cite{Baron&Netzer2019} based on theoretical calculations by {\sc CLOUDY}. The different $U$ values measured for the latter two Seyferts can be attributed to their edge-on gas disks blocking most of the optical emission lines emanating from the nuclear region, as discussed at the end of this subsection.

As discussed in Section~\ref{sec3.2}, the [Ne~{\small VI}]7.65/[Ar~{\small V}]7.90 is in principle the best $E_{\rm peak}$ diagnostic within JWST/MRS bands. The observed [Ne~{\small VI}]7.65/[Ar~{\small V}]7.90 line ratios indicate equivalent log~$(E_{\rm peak}/{\rm keV})$ values of $\sim -1.0$ to $-0.5$ in the central $\sim 200-300$ pc regions of the six Seyferts. These values are larger by of $\sim$ 1 dex than what would be expected from the correlation between $E_{\rm peak}$ and $\lambda_{\rm Edd}$ (\citealt{Thomas.etal.2016}) given their $\lambda_{\rm Edd}$ of $\sim 0.01 - 0.06$ (\citealt{Caglar.etal.2020}). On the other hand, with the log~$(E_{\rm peak}/{\rm keV})$ values of $\sim -1.0$ to $-0.5$, the bolometric luminosities of the input AGN ionizing spectra will be over $10^{45}\,\rm erg\,s^{-1}$. This value exceeds the measured bolometric luminosities of $\sim 10^{43.4} - 10^{44.3}\,\rm erg\,s^{-1}$ for the six targets reported by \citet{Davies.etal.2015}, indicating the need for additional heating sources beyond the AGN. Significantly higher abundance of neon relative to argon in these targets than that adopted in the modeling can also result in an overestimation of $E_{\rm peak}$. However, this is not the case here because the two $U$ values calibrated with [Ar~{\small V}]7.90/[Ar~{\small III}]8.99 and [Ar~{\small V}]13.10/[Ne~{\small III}]15.56 are consistent within 0.1 dex for all of the six targets.

A more plausible explanation for the elevated $E_{\rm peak}$ is the presence of an extra heating of soft ($\sim 0.2 - 2$ keV) X-ray to the coronal line [Ne~{\small VI}]7.65 in the nuclear regions of the six Seyferts. Although the adopted AGN ionizing radiation sources in our AGN models have included the soft X-ray excess associated with the accreting black hole, the emission from circumnuclear hot gas or other potential sources can further contribute to the soft X-ray radiation in AGN (see review by \citealt{McNamara&Nulsen2007}). Specifically, there is growing evidence that fast radiative shocks are required to explain the extension of coronal emission in AGN on scales of hundreds of parsecs, which cannot be explained by pure AGN photoionization (e.g., \citealt{Rodriguez-Ardila.etal.2006, Rodriguez-Ardila.etal.2011, Rodriguez-Ardila.etal.2017, Rodriguez-Ardila.etal.2025, Mazzalay.etal.2010, Mazzalay.etal.2013, Durre&Mould2018, May.etal.2018}). We examined and found that the observed [Fe~{\small VIII}]5.45/[Mg~{\small V}]5.61, a less reliable diagnostic of $E_{\rm peak}$ (see Section~\ref{sec3.2}), also indicates an elevated $E_{\rm peak}$, albeit with larger scatter, in the nuclear regions of all the six targets.

The derived metallicities in the central regions of the six Seyferts based on ([Ar~{\small III}]8.99+[Ar~{\small V}]7.90)/Pf$\alpha$ and using the above log~$(E_{\rm peak}/{\rm keV})$ values range from $\sim 0.2 - 0.9\,\,Z_{\odot}$, consistent in statistics with the metallicity distribution of type 2 Seyferts (e.g., \citealt{Nagao.etal.2006, Dors.etal.2019, Armah.etal.2023}), especially within their central regions (\citealt{Armah.etal.2024}). The corresponding gas pressure log~$(P/k)$ based on [Ne~{\small V}]14.32/[Ne~{\small V}]24.32 are of $\sim 7.8 - 8.3$. Since the calibrations of metallicity based on ([Ar~{\small III}]8.99+[Ar~{\small V}]7.90)/Pf$\alpha$ has a weak dependence on $E_{\rm peak}$ (see Figure~\ref{Fig_A_AGN}b), we artificially reduce the above $E_{\rm peak}$ values derived from [Ne~{\small VI}]7.65/[Ar~{\small V}]7.90 by 1 dex considering the potential contribution of radiative shocks to coronal line [Ne~{\small VI}]7.65, and repeat the metallicity calibration to quantify the uncertainty associated with the elevated $E_{\rm peak}$. Specifically, the reduced $E_{\rm peak}$ leads to an increase in the metallicity of $\sim 0.1 - 0.3\,Z_{\odot}$ for the six targets except for NGC\,3081. For the nuclear region of NGC\,3081, which has the most elevated log~$(E_{\rm peak}/{\rm keV})$ value of $-0.5$, the corresponding metallicity increases from $0.9\,Z_{\odot}$ to $2.5\,Z_{\odot}$. The relatively higher metallicity in the nuclear region of NGC\,3081 is consistent with its nature as a massive, early-type barred galaxy, which tends to accumulate more stellar mass and enrich its central region due to a deeper gravitational potential.

The subsolar to solar metallicity in the central regions of the six Seyferts is itself an interesting result. A subsolar metallicity in a galaxy's central region is generally associated with circumnuclear gas inflows, as observed in part of the six Seyferts (e.g., \citealt{Schnorr-Muller.etal.2016, Shimizu.etal.2019, Esparza-Arredondo.etal.2025}). Circumnuclear gas inflows can result in black hole growth as well as nuclear starburst. We therefore examined the contamination effect of potential nuclear starbursts in estimating the metallicity, and found it is negligible or within an increase of $0.2\,Z_{\odot}$, when their contribution to the [Ar~{\small III}]8.99 and Pf$\alpha$ emission of the six targets is less than 30\% or not exceeding 50\%. Another possibility that deserves further specific study is that the mid-IR-emission-line-based metallicity calibration, as well as the calibrations of other parameters discussed here, are more sensitive to the physical condition of the ISM immediately adjacent to the central ionizing source, because of the stronger penetrating power of infrared emission lines. On the contrary, optical emission lines from the inner region can be highly blocked. That is, infrared and optical diagnostics are tracking the physical conditions at different depths around ionized sources. As will be discussed in Section~\ref{sec4.1.1} and references therein, [Ar~{\small III}] and [Ar~{\small V}] emission lines are all emanating from the H~{\small II} regions (i.e., where H atoms are fully ionized) adjacent to the ionizing sources.

The hypothesis that infrared and optical diagnostics essentially trace different layers around ionizing sources can to a certain degree reconcile the larger log~$U$ estimated based on mid-IR emission lines than those estimated based on optical emission lines for NGC\,5506 and NGC\,7172. These two Seyferts, especially the latter, have a edge-on gas disk that largely blocks optical emission lines from the nuclear region (\citealt{Smajic.etal.2012, AlonsoHerrero.etal.2023, Esposito.etal.2024, Garcia-Bernete.etal.2024b}), leaving only the portion from the periphery, and therefore decrease the log~$U$ value estimated based on optical emission line ratios. The correction for dust extinction of the optical emission lines applying the same extinction strength (i.e., $A_{V}$) can not eliminate the differences in $U$ values estimated based on optical and infrared emission line ratios for the following reasons. Firstly, [O~{\small III}]$\lambda5007$/H$\beta$, [N~{\small II}]$\lambda6584$/H$\alpha$, [Ar~{\small V}]7.90/[Ar~{\small III}]8.99, and [Ar~{\small V}]13.10/[Ne~{\small III}]15.56 are not sensitive to dust extinction correction given the close wavelengths of these line ratios and the weak dust extinction in the mid-IR. Secondly, the differences in $U$ values are more likely due to the absolute extinction of the optical emission lines in the inner region rather than partial attenuation, whereas the former effect can not be corrected by a single extinction strength for all lines.

Furthermore, based on the optical R(N2, S2, O3)\footnote{log~R(N2, S2, O3) = 0.9738$\times$log([N II]$\lambda6584$/[S II]$\lambda\lambda6717, 31$) $+$ 0.047$\times$log([N II]$\lambda6584$/H$\alpha$) $-$ 0.183$\times$log([O III]$\lambda5007$/H$\beta$)} prescription of \cite{Zhu.etal.2024} and available emission line ratios extracted from $1\farcs8\times1\farcs8$ apertures of VLT/X-shooter IFU mode observations by \cite{Burtscher.etal.2021}, the derived metallicities for five of the six Seyferts range from $\sim 1.7 - 3.5\,\,Z_{\odot}$ assuming the same abundance scaling relations as adopted in this paper (i.e., \citealt{Nicholls.etal.2017}). Moreover, based on the R([S~{\small II}]$\lambda6717$/[S~{\small II}]$\lambda6731$) prescription of \cite{Zhu.etal.2024} and [S~{\small II}] emission line ratios extracted from $1\farcs8\times1\farcs8$ apertures of VLT/X-shooter IFU mode observations by \cite{Davies.etal.2020}, the derived log~$(P/k)$ values are of $\sim 6.9 - 7.6$ for the six targets. The overall higher metallicity and correspondingly lower gas pressure derived based on optical emission lines again suggest that infrared and optical diagnostics essentially trace different layers around ionizing sources, albeit more convincing conclusions require further dedicated, especially spatially resolved, studies.

\startlongtable
\setlength{\tabcolsep}{3pt}
\begin{deluxetable*}{lccccccc}
\tablecolumns{8}
\tablecaption{Nuclear Emission Line Ratios of Sample Seyferts}
\tablehead{
\colhead{Target} & \colhead{$\rm log\frac{[Ar~V]7.90}{[Ar~III]8.99}$} & \colhead{$\rm log\frac{[Ar~V]13.10}{[Ne~III]15.56}$} & \colhead{$\rm log\frac{[Ne~VI]7.65}{[Ar~V]7.90}$} & \colhead{$\rm log\frac{([Ar~III]8.99+[Ar~V]7.90)}{Pf\alpha}$} & \colhead{$\rm log\frac{[Ne~V]14.32}{[Ne~V]24.32}$} & \colhead{$\rm log\frac{[Ne~VI]7.65}{[Fe~II]5.34}$} & \colhead{$\rm log\frac{[Ar~V]7.90}{[Fe~II]5.34}$} \\
\colhead{(1)} & \colhead{(2)} & \colhead{(3)} & \colhead{(4)} & \colhead{(5)} & \colhead{(6)} & \colhead{(7)} & \colhead{(8)}}
\startdata
ESO137$-$G034 & $-$0.70 $\pm$ 0.11 & $-$1.45 $\pm$ 0.07 & 1.15 $\pm$ 0.20 & 1.00 $\pm$ 0.11 & 0.05 $\pm$ 0.08 & 0.76 $\pm$ 0.17 & $-$0.39 $\pm$ 0.10 \\
MCG$-$05$-$23$-$016 & $-$0.65 $\pm$ 0.07 & $-$1.55 $\pm$ 0.18 & 1.41 $\pm$ 0.04 & 0.69 $\pm$ 0.07 & 0.17 $\pm$ 0.13 & 1.84 $\pm$ 0.03 & 0.43 $\pm$ 0.05 \\
NGC\,3081 & $-$0.75 $\pm$ 0.03 & $-$1.44 $\pm$ 0.01 & 1.33 $\pm$ 0.03 & 1.37 $\pm$ 0.03 & 0.09 $\pm$ 0.01 & 1.79 $\pm$ 0.02 & 0.46 $\pm$ 0.04 \\
NGC\,5506 & $-$0.60 $\pm$ 0.03 & $-$1.32 $\pm$ 0.09 & 1.04 $\pm$ 0.04 & 0.74 $\pm$ 0.03 & 0.24 $\pm$ 0.05 & 0.71 $\pm$ 0.02 & $-$0.33 $\pm$ 0.03 \\
NGC\,5728 & $-$0.52 $\pm$ 0.14 & $-$1.43 $\pm$ 0.06 & 1.22 $\pm$ 0.15 & 1.05 $\pm$ 0.14 & 0.12 $\pm$ 0.03 & 1.09 $\pm$ 0.06 & $-$0.13 $\pm$ 0.14 \\
NGC\,7172 & $-$0.54 $\pm$ 0.18 & $-$1.50 $\pm$ 0.24 & 1.38 $\pm$ 0.03 & 0.65 $\pm$ 0.18 & 0.11 $\pm$ 0.19 & 1.02 $\pm$ 0.03 & $-$0.36 $\pm$ 0.04 \\
\enddata
\tablecomments{Based on the measurements from the central $r = 0\farcs75$ ($\sim 100 - 150$ pc) regions of the sample Seyferts by \cite{Zhang.etal.2024a}.}
\label{tabTest}
\end{deluxetable*}

\begin{figure*}[!ht]
\center{\includegraphics[width=1\textwidth]{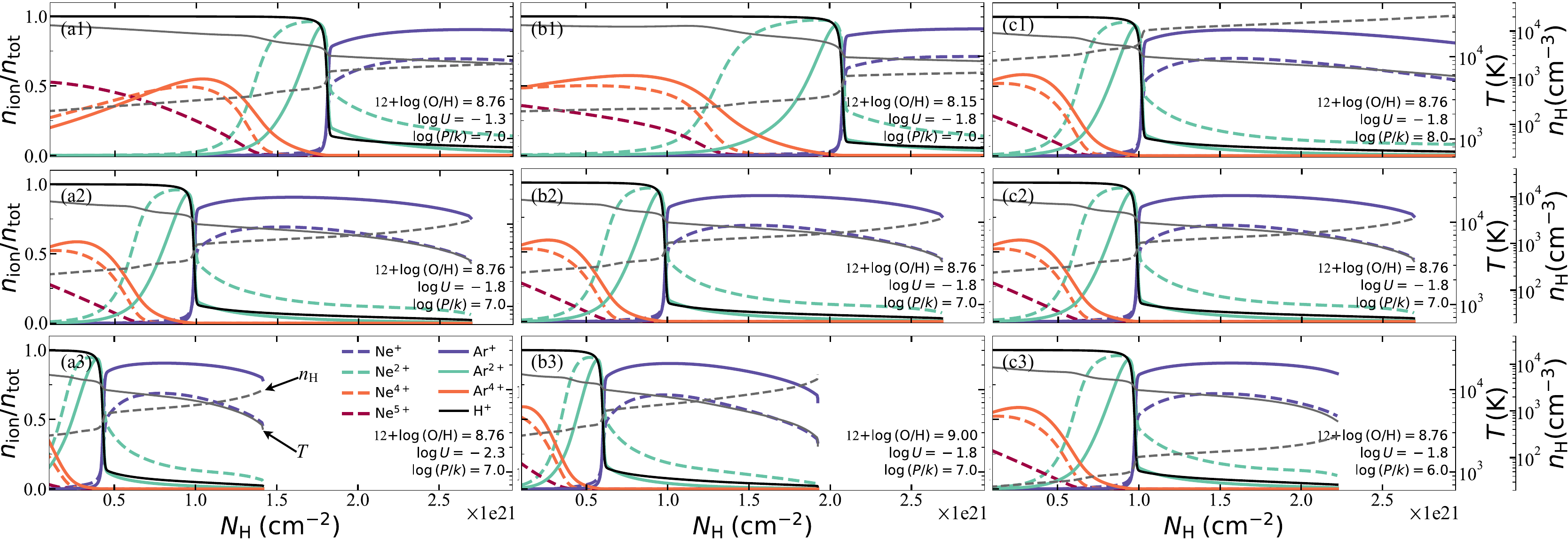}}
\caption{Examples of the ionization structures at different depths into the ionized clouds, as predicted by AGN photoionization models with the black hole mass as $10^{7}\,M_{\odot}$ and log~$\lambda_{\rm Edd}=-0.5$. See the lower right of each panel for the values of other input parameters, i.e., log~$U$, log~$(P/k)$, and 12 + log~(O/H). The colored solid and dashed lines in each panel represent the abundances of different argon and neon ions relative to the total argon or neon abundance, respectively. As a reference, the black-solid line in each panel represents the hydrogen ion abundance relative to the total hydrogen abundance. In addition, the gray-solid and -dashed lines in each panel, as labelled in panel (a3), represent the temperature $T$ and column density ($n_{\rm H}$) as a function of the number density ($N_{\rm H}$), respectively (see the two y-axes on the right).}\label{Fig_IonStruc}
\end{figure*}

\subsubsection{General Applications of the Diagnostics}\label{sec3.5.2}

We discussed above example applications of the proposed diagnostics to six Seyfets, and for more general applications in future work, we suggest following the steps summarized below. First, the ionization parameter can be derived using Equation~\ref{equ1} with $R = \rm[Ar~{\small V}]7.90/[Ar~{\small III}]8.99$ or $\rm[Ar~{\small V}]13.10/[Ne~{\small III}]15.56$ (see Section~\ref{sec3.1}). Then, with the derived ionization parameter, the peak energy and metallicity can be iteratively derived using Equations~\ref{equ2} with $R = \rm[Ne~{\small VI}]7.65/[Ar~{\small V}]7.90$ (see Section~\ref{sec3.2}) and Equations~\ref{equ3} with $R = \rm([Ar~{\small III}]8.99+[Ar~{\small V}]7.90)/Pf\alpha$ (see Section~\ref{sec3.3}). We assume a solar metallicity in the first round iteration and stop the iterations when the peak energy and metallicity converge to final values. At this step, it is important to account for the contribution of radiative shocks to coronal emission lines in AGN; otherwise, these contributions could bias the diagnostics of the AGN spectral peak energy and, consequently, the inferred metallicity (see further discussion in Section~\ref{sec4.3}). Specifically, \cite{Rodriguez-Ardila.etal.2025} found that radiative shocks may contribute up to 80\% of the highly ionized coronal emission lines [Fe~{\small VII}] and [Fe~{\small X}] in AGN. Finally, the gas pressure can be derived using Equation~\ref{equ5} with $R = \rm[Ne~{\small V}]14.32/[Ne~{\small V}]24.32$ (see Section~\ref{sec3.4}), which is related to the gas density via the equation of state. Although the above steps are recommended, Equations~\ref{equ2} and \ref{equ3} can also be used individually if the required secondary parameters (i.e., log~$U$ and 12 + log(O/H) or log~$E_{\rm peak}$) are already known. To obtain more robust statistics, we suggest perturbing the observed diagnostic emission-line fluxes with random noise at their uncertainty levels, and repeating this e.g., 100 times as a Markov Chain Monte Carlo (MCMC) sampling. The resulting 100 sets of line ratios can then be input into Equations~\ref{equ1}--\ref{equ5}, and adopt the corresponding median and standard deviation of the 100 calculations as the final estimate and uncertainty of each physical parameter.

\section{Discussion}\label{sec4}

\subsection{Diagnostics in Low Luminosity AGN}\label{sec4.1}
As noted in Section~\ref{sec3.1}, the AGN photoionization models produce [Ne~{\small VI}]7.65/[Ar~{\small V}]7.90 values with a bi-valuation dependence on $E_{\rm peak}$, which is checked to be mainly due to the non-linear dependance of [Ne~{\small VI}]7.65 emissivity on $E_{\rm peak}$. Taking this into consideration, we perform independent calibrations for AGN models with log$(E_{\rm peak}/{\rm keV})\leq -2.25$ (empirically log~$\lambda_{\rm Edd} \leq -3$) and therefore low luminosity (low-luminosity AGN hereafter), using the same strategies as described by Equations~\ref{equ1}~--~\ref{equ5}. As shown by Figures~\ref{Fig_U_AGN_II}~--~\ref{Fig_P_AGN_II} in Appendix~\ref{secA}, the strategies adopted for AGN models with relatively high $E_{\rm peak}$ are also applicable to AGN models with relatively low $E_{\rm peak}$, but with numerically different best-fit coefficients and RMS scatters as listed in Tables~\ref{tabEqu1}~--~\ref{tabEqu5} in Appendix~\ref{secA}. 

Nevertheless, we note here that the calibrations for low-luminosity AGN in Appendix~\ref{secA} should be used with caution, since Advection-Dominated Accretion Flow (ADAF) model (see review \citealt{Narayan&McClintock2008, Yuan&Narayan2014}) is in principle better than {\sc OPTXAGNF} in calculating the AGN ionizing spectra of these low-luminosity AGN. We keep these calibrations for the sake of completeness and in order to highlight the two factors deserving specific discussion given their potential impact on the proposed diagnostics in low-luminosity AGN. While the first factor does not significantly affect the proposed diagnostics, the second factor may have a large impact, as discussed below.

\subsubsection{The Effect of Hard X-ray Photons}\label{sec4.1.1}

\begin{figure*}[!ht]
\center{\includegraphics[width=0.85\textwidth]{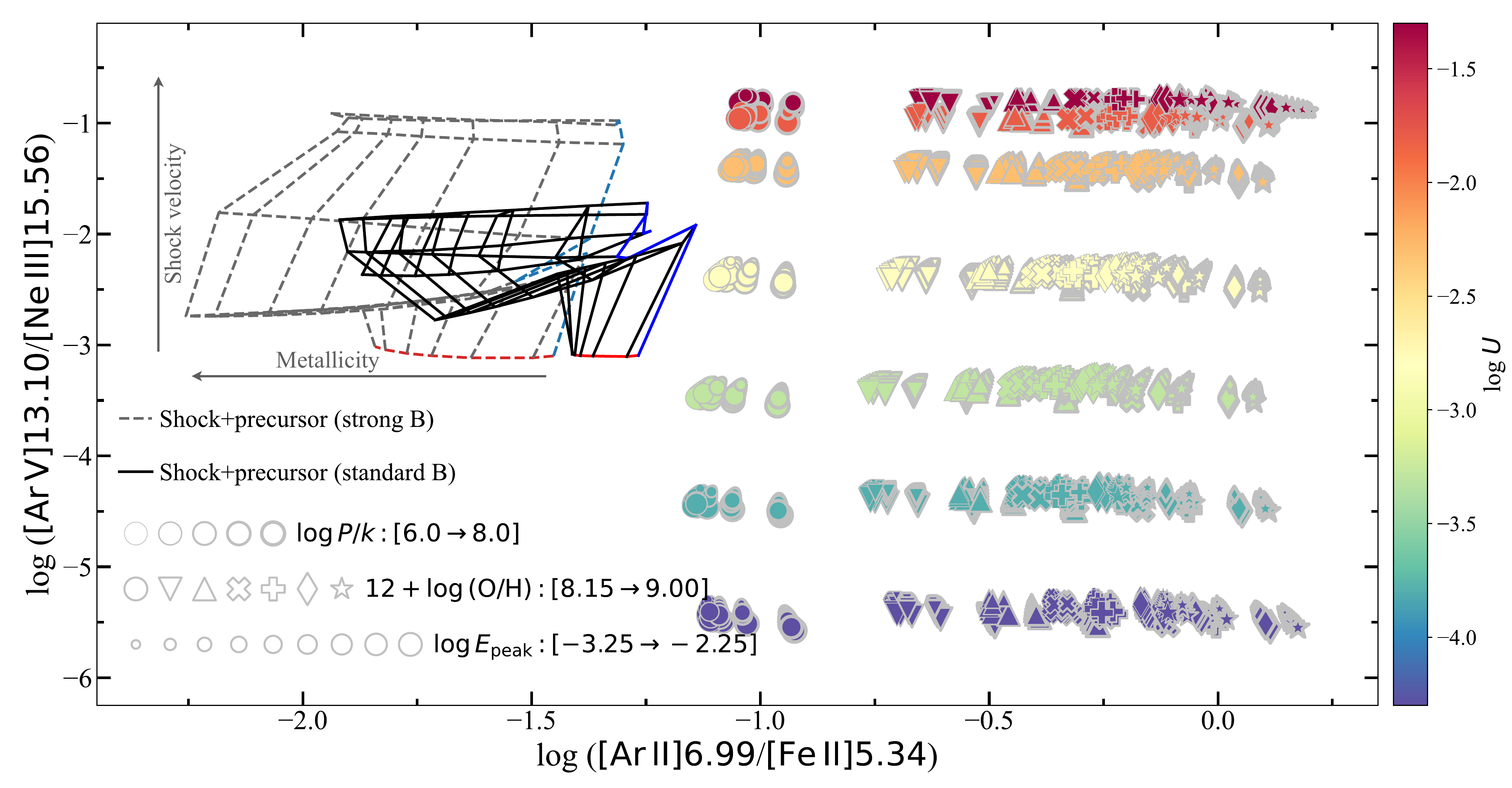}}
\caption{Diagram consisting of [Ar~{\small V}]13.10/[Ne~{\small III}]15.56 versus [Ar~{\small II}]6.99/[Fe~{\small II}]5.34 to illustrate the contamination effects of shock excitation on emission line ratio diagnostics in AGN. For AGN model results in the right, different colors, marker sizes, marker types, and edge widths correspond to different log $U$, log $E_{\rm peak}$, 12 + log (O/H), and log $(P/k)$, respectively. For shock+precursor model results in the left, black-solid and gray-dashed grids correspond to model results of standard and strong magnetic fields, respectively, where the two gray arrows indicate the trend of model results as metallicity and shock velocity increase, and the red and blue lines mark the lower boundaries.}\label{Fig_Shock}
\end{figure*}

Theoretical AGN photoionization models used here adopt an essentially unchanged shape (i.e., $\Gamma$) and contribution (i.e., $f_{\rm pl}$) of the hard ($\sim 2 - 10$ keV) X-ray Comptonization in AGN ionizing spectra. This is a simplified assumption, since $\Gamma$ tends to decrease as the Eddington ratio of AGN decreases, resulting in higher fraction ($f_{\rm pl}$) of hard X-ray photons in low-luminosity AGN (e.g., \citealt{Shemmer.etal.2006, Done.etal.2012, Jin.etal.2012}). Hard X-ray photons are able to penetrate deeper to further irradiate atomic/molecular gas clouds than UV photons, given the greatly reduced photoionization cross sections of H and He in X-ray regime. This effect of X-ray photons leads to an {\it extra} excitation of heavy elements in so called X-ray dissociation regions (or X-ray dominated regions, XDRs, \citealt{Maloney.etal.1996}), which contributes independently to specific infrared emission lines (e.g., \citealt{Meijerink.etal.2007, Ferland.etal.2013}). Theoretical AGN photoionization models used here do not stop until the hydrogen becomes 99~\% neutral and therefore include the XDR contribution. In practical applications, line ratio diagnostics involving emission lines with undetermined XDR contributions have larger uncertainty than those involving only NLR emission lines (e.g., \citealt{Pereira-Santaella.etal.2017}).

Specifically, hard X-ray photons affect the heavy elements (e.g., C, N, O, Ne, Na, Mg, Si, S, Ar, Fe) mainly through the K- and L-shell photoionization, which results in multiply charged ions by the Auger effect. These highly charged ions are rapidly destroyed through the charge exchange reactions with atomic and molecular hydrogen, and other species (\citealt{Abel.etal.2009, Adamkovics.etal.2011}). However, the contribution of XDRs in the excitation of infrared emission lines is significant for singly and doubly charged ions due to their limited charge exchange reactions with atomic and molecular hydrogen (\citealt{Glassgold.etal.2007, Adamkovics.etal.2011, Ferland.etal.2013}). \cite{Pereira-Santaella.etal.2017} found that the XDR excitation can contribute, respectively, to $\sim 80$~\% and $\sim 20$~\% of [Ne~{\small II}]12.81 and [Ne~{\small III}]15.56 emission lines in AGN models with solar metallicity, and these values decrease as metallicity increases. Relevant to this, \cite{Garcia-Bernete.etal.2017} found that [Ne~{\small II}]12.81 and [Ne~{\small III}]15.56 emission show considerable correlations with X-ray continuum from $3-40$ keV bands.

The line ratio diagnostics proposed in this study for AGN are essentially based on the emission lines of highly charged ions (i.e., [Ar~{\small III}],  [Ar~{\small V}], [Ne~{\small V}], and [Ne~{\small VI}]), which are basically only radiated from H~{\small II} regions with the contribution outside H~{\small II} regions less than $\sim 5\%$ (see Figure~\ref{Fig_IonStruc}). Therefore, the line ratio diagnostics proposed here are overall immune to the impact of XDRs, except for the line ratio diagnostics involving singly and some doubly charged ions. For example, the [Ar~{\small II}]6.99/[Fe~{\small II}]5.34 diagnostic discussed in Section~\ref{sec3.3}, and such diagnostics should be used with additional caution. We want to highlight again that the proposed optimal diagnostics (excluding the gas pressure diagnostic in AGN) are all based on emission lines with log~$n_{\rm c}\geq 4.5$ (see Table~\ref{tablines}). Such high critical electron density relative to the available electron density in NLRs largely eliminate the dependence of these proposed diagnostics on the gas density, equivalently the gas pressure (see the right column of Figure~\ref{Fig_IonStruc}). Moreover, the ions of these noble gases, i.e., argon and neon, are immune to the dust depletion effect and other metallicity relevant effects.

\subsubsection{The Effect of Shocks}\label{sec4.1.2}

An important heating mechanism in AGN, especially in low-luminosity AGN, is shock heating (see review \citealt{McNamara&Nulsen2007}), also known as the kinetic mode AGN feedback and widely embraced in modern cosmological simulations (e.g., \citealt{Weinberger.etal.2017, Dave.etal.2019}). Since shocks are prevalent in low-luminosity AGN due to their very low radiative efficiency (e.g., \citealt{Ho.etal.2003, Ho2009}), the shock excitation may have a large impact on the NLR emission line diagnostics. To study the infrared emission line excitation in shocked regions, we use the same {\sc MAPPINGS V} code to create a grid of radiative shock models as detailed by \cite{Sutherland&Dopita2017}. The corresponding radiative shock precursor, which can play an important role in gas excitation in AGN, is also included in the shock modeling.

Following \cite{Sutherland&Dopita2017}, the four varying parameter of the radiative shock models are the metallicity from $Z$ = 0.2, 0.4, 0.7, 1.0, 1.3, 1.7, and 2.0 $Z_{\odot}$; the shock velocity $v_s$ = 100, 150, 200, 300, 400, 600, 800, and 1000 $\rm km~s^{-1}$; the preshock density $n_{\rm H}$ = 1, 10, 100, 1000, and 10000 $\rm cm^{-3}$; and the magnetic to ram pressure ratio $\eta_{\rm M}$ = $B^{2}/(4\pi\rho v_{s}^{2})$ = 0.0, 0.0001, 0.001, 0.01, and 0.1, where 0.0001, 0.01 , and 0.1 corresponding to standard, moderate, and strong magnetic fields, respectively (see Table~\ref{tabRes2} for the modeling results). In accordance with \cite{Allen.etal.2008} and \cite{Alarie&Morisset2019}, the adopted radiative shock precursors are computed separately. Specifically, they are calculated by the {\sc MAPPINGS V} photoionization module using the ionizing radiation emanating from the shock as the input ionizing source. The dust depletion as used in AGN models is included in the shock precursor calculations as dust can be present in radiative shock precursors albeit not in shocks (\citealt{Pereira-Santaella.etal.2024}).

As shown in Figure~\ref{Fig_Shock}, the typical radiative shock models with $n_{\rm H}$ = 100 $\rm cm^{-3}$, $Z$ = 0.2 -- 2 $Z_{\odot}$, and $v_{s}$ = 100 -- 1000 $\rm km~s^{-1}$ for standard and especially strong magnetic fields produce a range of [Ar~{\small V}]13.10/[Ne~{\small III}]15.56 values that are within the range of those produced by the AGN models. Such overlapping exists in not only $U$ diagnostics like [Ar~{\small V}]13.10/[Ne~{\small III}]15.56 but also diagnostics of other physical parameters studied here as further discussed in Section~\ref{sec4.3}. This conclusion does not change with different pre-shock densities not shown in Figure~\ref{Fig_Shock} and holds for high luminosity AGN discussed in Section~\ref{sec3}. In other words, shocked regions can produce diagnostic emission line ratios the same as some AGN-excited regions. Therefore, it is important to identify and exclude the contribution of shock-dominated regions when diagnosing the physical parameters of NLRs in AGN. The identification of shocked regions in AGN has been studied over decades (e.g., \citealt{Dopita&Sutherland1995, Groves.etal.2004, Allen.etal.2008, Alarie&Morisset2019, Kewley.etal.2019, Feltre.etal.2023}), but still deserves further specific studies under the view of JWST. Line ratios involving [Fe~{\small II}] emission lines as illustrated in Figure~\ref{Fig_Shock} are useful here because enhanced [Fe~{\small II}] emission in AGN is always associated with shocked regions (e.g., \citealt{Forbes&Ward1993, Mouri.etal.2000, Storchi-Bergmann.etal.2009, Koo.etal.2016, Rodriguez-Ardila.etal.2017, Durre&Mould2018}). Before scrutinizing them in further specific studies, we briefly discuss the diagrams that distinguish shocked regions in AGN, and candidate strategies for quantifying the contribution of radiative shocks to gas heating in AGN later in Section~\ref{sec4.3}.

\begin{figure}[!ht]
\center{\includegraphics[width=1\linewidth]{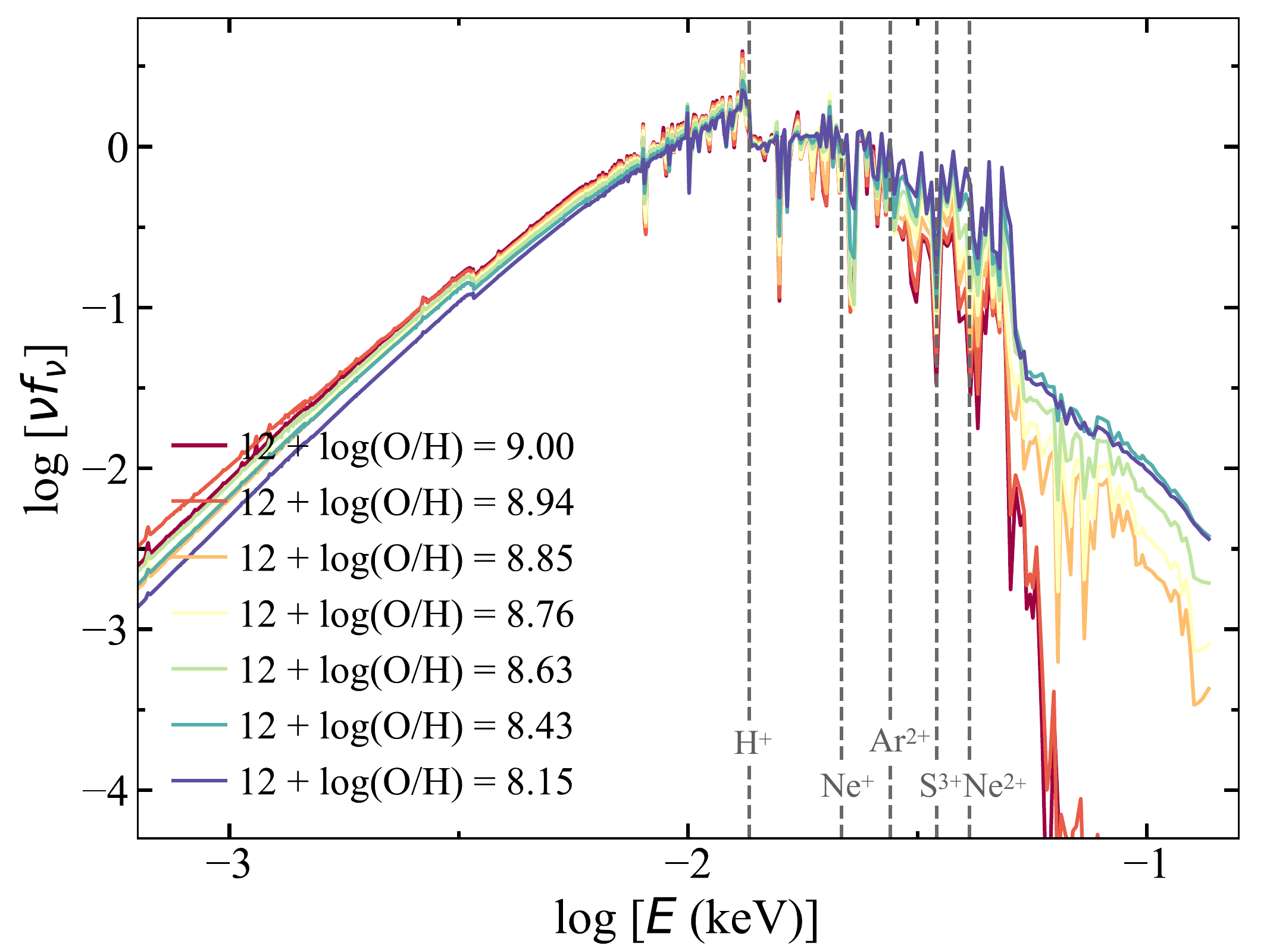}}
\caption{The ionizing spectra calculated by \cite{Thomas.etal.2018} for the SF photoionization models, where the vertical lines indicate the required ionization energy of marked ions. All ionizing spectra are normalized by the flux at the ionization potential of hydrogen atom (i.e., 13.6 eV).}\label{SpecSF}
\end{figure}

\subsection{Diagnostics in Star Forming Regions}\label{sec4.2}

\begin{figure*}[!ht]
\center{\includegraphics[width=1\textwidth]{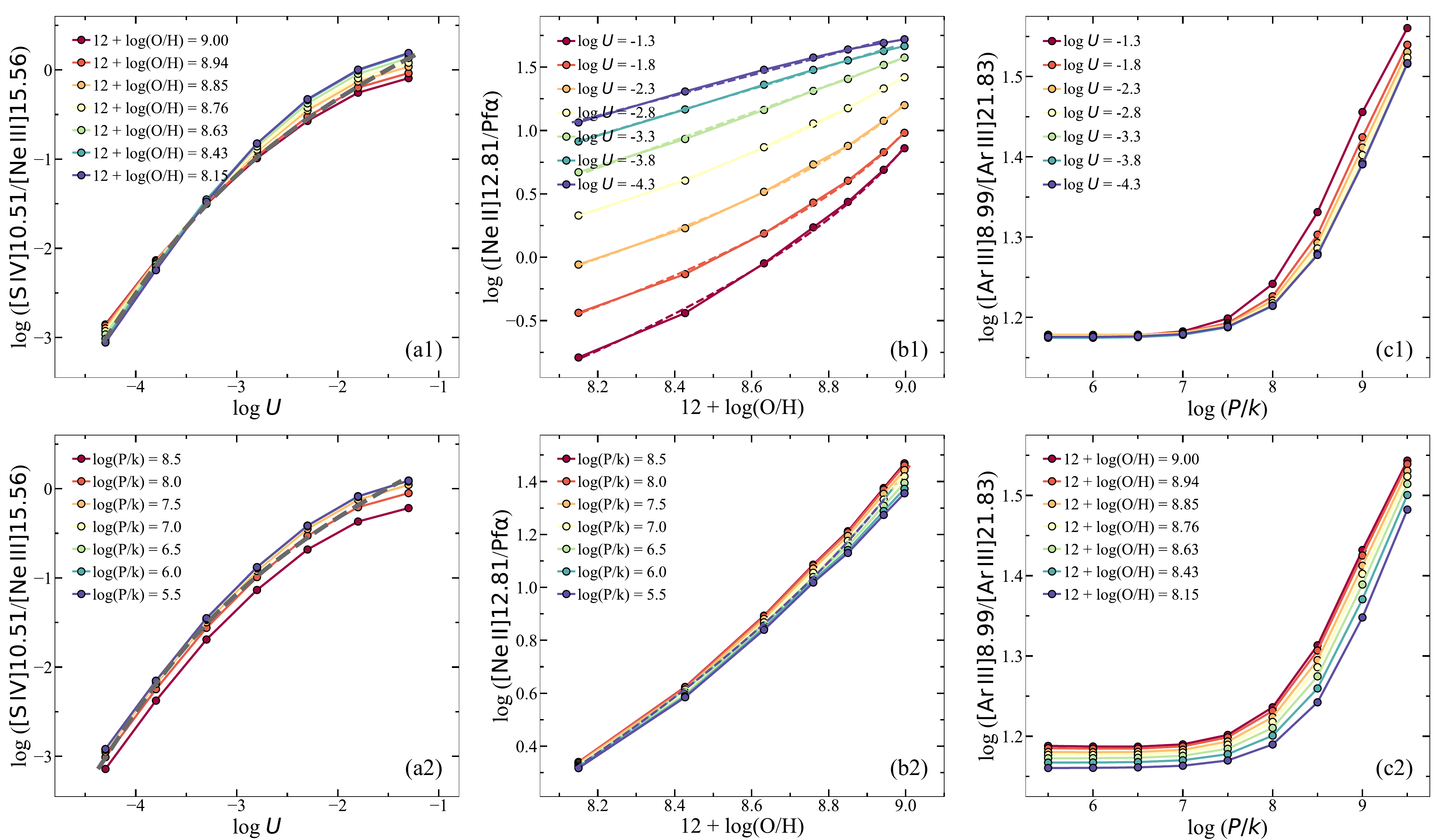}}
\caption{Dependence of (a1\&a2) [S~{\small IV}]10.51/[Ne~{\small III}]15.56 on the ionization parameter ($U$), (b1\&b2) [Ne~{\small II}]12.81/Pf$\alpha$ on the metallicity (12 + log (O/H)), and (c1\&c2) [Ar~{\small III}]8.99/[Ar~{\small III}]21.83 on the gas pressure $(P/k)$, as predicted by SF photoionization models (colored points). Each panel displays model results under specific parameter constraints: (a1\&b1) log $(P/k)$ = 7.0; (c1\&a2) 12 + log (O/H) = 8.76; (b2\&c2) log $U$ = $-2.8$. The gray-dashed lines in panel a1\&a2 represent the best-fit correlation (Equation~\ref{equ6}) derived from all SF models discussed in this section, while the colored-dashed lines in panel b1\&b2 represent the best-fit correlations (Equation~\ref{equ7}--\ref{equ7_3}) derived from all SF models with the given log $U$.}\label{Fig_SF}
\end{figure*}

The coevolution of galaxies and AGN has become an widely accepted notion in the contemporary cosmology (see review \citealt{Kormendy&Ho2013, Heckman&Best2014}), and circumnuclear starbursts are prevalent in AGN (e.g., \citealt{CidFernandes.etal.2001, Watabe.etal.2008, Imanishi.etal.2011, Esquej.etal.2014, Garcia-Burillo.etal.2014, Garcia-Bernete.etal.2022, Zhang&Ho2023}). Similar to shock-dominated regions, star-formation-dominated regions in AGN need to be distinguished and discussed regarding the diagnostics of the physical conditions specifically as well. Therefore, in this subsection we discuss the theoretical infrared line ratio diagnostics in star-forming (SF) regions. The {\sc MAPPINGS V} code is again used to calculate the theoretical models of SF regions using the same settings, i.e., metallicity, dust depletion, geometry, ionization parameter, gas pressure, stop criterion, etc, as the AGN models, but with different ionizing radiation fields (see Table~\ref{tabRes3} for the modeling results).

The ionizing radiation fields for the SF models are adopted from \cite{Thomas.etal.2018} derived using the SLUG2 (\citealt{Krumholz.etal.2015}) stellar population synthesis code with the settings summarized as follows. The ionizing radiation fields are calculated for ``galaxy'' (continuous star formation) mode with the default Starburst99 spectral synthesis and Padova stellar tracks with thermally pulsing AGB stars, given an age of 10 million years and a star formation rate of 0.001 $M_{\odot}~\rm yr^{-1}$. The ionizing spectra were firstly calculated under five available metallicities $Z$ = 0.0004, 0.004, 0.008, 0.02, and 0.05, then interpolated for the ionizing spectra of the oxygen abundances as used in the AGN models. Unlike for the AGN models, the input ionizing spectra of the SF models are determined by the metallicity. As shown in Figure~\ref{SpecSF}, the low-metallicity SF models have harder ionizing spectra, but still lacking high energy photons for the excitation of high ionization potential lines such as [Ar~{\small V}] and [Ne~{\small V}].\footnote{[Ar~{\scriptsize V}] emission is strong enough only in SF models of hard and intensive radiation fields with low metallicities and high ionization parameters.}

We repeat the same feature extraction process as for the AGN models for the line ratios predicted by the SF models. As opposed to the AGN models (see Section~\ref{sec3.1}), we find that the optimal line ratios for $U$ diagnostics for the SF models are those involving [S~{\small IV}]10.51, e.g., the [S~{\small IV}]10.51/[Ne~{\small III}]15.56 line ratio as shown in Figure~\ref{Fig_SF}(a). This result is consist with the previous finding by \cite{Pereira-Santaella.etal.2017}. We adopt a cubic function, i.e., Equation~\ref{equ6},
\noindent
\begin{align}\label{equ6}
\begin{aligned}
{\rm log}\,U = \,\,&c_3\times({\rm log}\,R)^{3} + c_2\times({\rm log}\,R)^{2} + \\&c_1\times{\rm log}\,R + c_0.
\end{aligned}
\end{align}
\noindent
to calibrate $U$ in SF regions, and the best-fit coefficients for $R =$ [S~{\small IV}]10.51/[Ne~{\small III}]15.56 are listed in Table~\ref{tabEqu6} in Appendix~\ref{secA}. Quantitatively, this calibration has a RMS scatter of 0.17 dex.

Contrary to the AGN models, the optimal metallicity diagnostics in the SF models are the line ratios of low excitation neon and argon lines relative to hydrogen emission. These low excitation lines in the SF models also have a strong dependence on the ionization parameter as those high excitation lines in the AGN models. We therefore adopt the strategy as described by Equations~\ref{equ7}$-$\ref{equ7_2} to calibrate the metallicity in SF regions.
\noindent
\begin{align}\label{equ7}
\begin{aligned}
12+{\rm log (O/H)} = a\times({\rm log}\,R - b)^{2} + c,
\end{aligned}
\end{align}
\noindent
where
\noindent
\begin{align}\label{equ7_1}
\begin{aligned}
a = a_{U,2}\times({\rm log}\,U)^{2} + a_{U,1}\times{\rm log}\,U + a_0,
\end{aligned}
\end{align}
\begin{align}\label{equ7_2}
\noindent
\begin{aligned}
b = b_{U,2}\times({\rm log}\,U)^{2} + b_{U,1}\times{\rm log}\,U + b_0,
\end{aligned}
\end{align}
\noindent
\begin{align}\label{equ7_3}
\begin{aligned}
c = c_{U,2}\times({\rm log}\,U)^{2} + c_{U,1}\times{\rm log}\,U + c_0.
\end{aligned}
\end{align}
\noindent

\begin{figure*}[!ht]
\center{\includegraphics[width=0.9\linewidth]{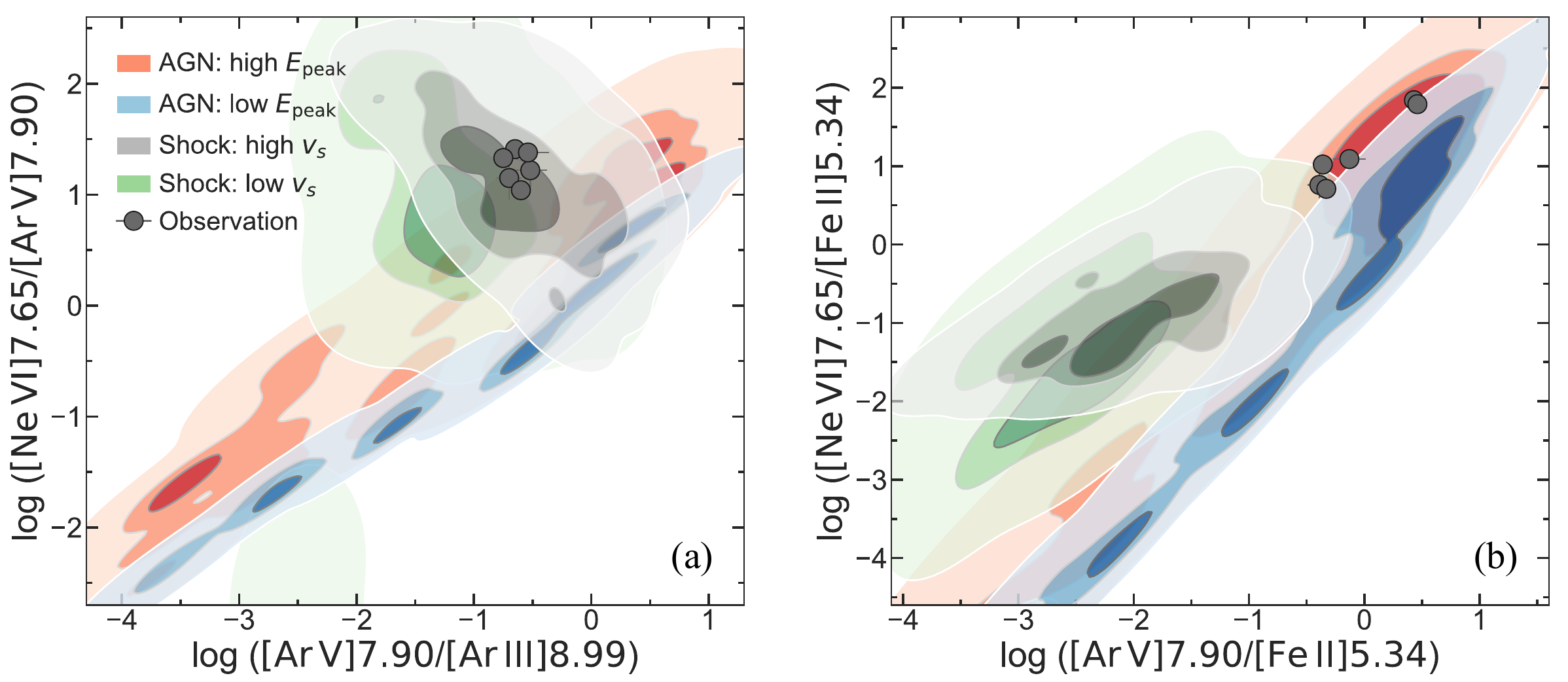}}
\caption{Diagrams consisting of (a) [Ne~{\small VI}]7.65/[Ar~{\small V}]7.90 versus [Ar~{\small V}]7.90/[Ar~{\small III}]8.99 and (b) [Ne~{\small VI}]7.65/[Fe~{\small II}]5.34 versus [Ar~{\small V}]7.90/[Fe~{\small II}]5.34 to distinguish between photoionization models excited by AGN and radiative shocks (i.e., including the shock precursors). Specifically, the reddish, blueish, grayish, and greenish contours (from dark to light covering 0-30\%, 30-70\%, 70-100\% of the corresponding models involved in this work) represent AGN models with $-2.0 \leq {\rm log} E_{\rm peak} \leq -0.95$ (high $E_{\rm peak} $) and $-3.25 \leq {\rm log} E_{\rm peak} \leq -2.25$ (low $E_{\rm peak} $), and radiative shock models with $v_{s} \geq 500\rm\,km\,s^{-1}$ (high $v_{s}$) and $v_{s} \leq 500\rm\,km\,s^{-1}$ (low $v_{s}$), respectively. The six gray points represent the measurements in Table~\ref{tabTest} of the six Seyferts discussed in Section~\ref{sec3.5}.}\label{Fig_diagram}
\end{figure*}

We propose $R =$ [Ne~{\small II}]12.81/Pf$\alpha$, as shown by Figure~\ref{Fig_SF}(b), for the metallicity calibration since the [Ar~{\small II}]6.99 emission line is weaker than [Ne~{\small II}]12.81 emission line by $\sim 1 - 2$ orders of magnitude in the SF models. The best-fit coefficients for $R =$ [Ne~{\small II}]12.81/Pf$\alpha$, as well as [Ar~{\small II}]6.99/Pf$\alpha$ with comparable RMS scatters of $\sim 0.02$ dex, are listed in Table~\ref{tabEqu7}.

The ionizing spectra of the SF models do not have a meaningful definition of $E_{\rm peak}$ as that of the AGN models. However, since the hardness of the SF ionizing radiation field is determined by metallicity, we can still qualitatively study some proposed radiation field hardness diagnostics for SF regions in the literature with the SF models calculated here. For example,  [Ne~{\small III}]15.56/[Ne~{\small II}]12.81, [Ar~{\small III}]8.99/[Ar~{\small II}]6.99, and [S~{\small IV}]10.51/[S~{\small III}]18.71 have long been proposed to be radiation field hardness diagnostics in SF regions (e.g., \citealt{Thornley.etal.2000, ForsterSchreiber.etal.2001,  Giveon.etal.2002, Verma.etal.2003, Snijders.etal.2007}), we therefore check the correlations between the predicted values of these line ratios and the metallicity of the SF models. We find that for a given $U$, [Ne~{\small III}]15.56/[Ne~{\small II}]12.81 and [Ar~{\small III}]8.99/[Ar~{\small II}]6.99 show tight monotonic correlations with metallicity and hence the radiation field hardness of the SF models, while [S~{\small IV}]10.51/[S~{\small III}]18.71 also shows a secondary dependence on gas pressure (see Figure~\ref{Fig_SF_H} in Appendix~\ref{secA}). All these line ratios have a strong dependence on $U$, with the predicted [Ne~{\small III}]15.56/[Ne~{\small II}]12.81 and [Ar~{\small III}]8.99/[Ar~{\small II}]6.99 values enhanced by 2 orders of magnitude, and the larger for [S~{\small IV}]10.51/[S~{\small III}]18.71, when log~$U$ changes from $-4.3$ to $-1.3$ (see Figure~\ref{Fig_SF_H}).

Finally, two line ratio candidates for diagnosing the gas pressure in SF regions within JWST/MRS bands are [Ar~{\small III}]8.99/[Ar~{\small III}]21.83 and [Fe~{\small II}]25.99/[Fe~{\small II}]5.34, since the SF ionizing spectra do not have enough high energy photons to excite the high ionization potential lines (i.e., [Fe~{\small VII}], [Ne~{\small V}], and [Ar~{\small V}]). However, the  [Ar~{\small III}]8.99/[Ar~{\small III}]21.83 and the same for [Fe~{\small II}]25.99/[Fe~{\small II}]5.34 are insensitive to gas pressure until log~$(P/k) \gtrsim 7.5$, which is beyond the gas pressure range of typical H~{\small II} regions in observations (e.g., \citealt{Pathak.etal.2025}).  We therefore do not provide the infrared-line-ratio-based gas pressure calibration for SF regions.

\subsection{Diagrams to Distinguish Excitation Sources}\label{sec4.3}

From the seminal work by \cite{Baldwin.etal.1981} and the extended studies (e.g., \citealt{Keel1983, Veilleux&Osterbrock1987}), two-dimensional diagrams based on optical emission line ratios are widely used to classify galaxies governed by different excitation sources, i.e., AGN, shocks, and SF activities (e.g., \citealt{Kewley.etal.2001, Kewley.etal.2006, Kauffmann.etal.2003}). Emission-line-ratio diagrams can be used to distinguish different types of galaxies because different excitation mechanisms operating on the ISM gas result in different line ratios in observation, which can infer the hidden physical properties of galactic environments (see review \citealt{Kewley.etal.2019}).

Recently, \cite{Feltre.etal.2023} advanced the application of such emission-line-ratio diagrams to the mid-IR band based on photoionization models and joint optical and mid-IR spectral observations of 42 Seyfert galaxies. They specifically focused on [Ne~{\small II}]12.81, [Ne~{\small III}]15.56, [Ne~{\small V}]14.32, [S~{\small III}]18.71, [S~{\small IV}]10.51, and [O~{\small IV}]25.89 measured from Spitzer/IRS spectra and strong optical emission lines (i.e.,  H$\alpha$, H$\beta$, [O~{\small I}]$\lambda6300$, [O~{\small III}]$\lambda5007$, [N~{\small II}]$\lambda6584$, and [S~{\small II}]$\lambda\lambda6716,31$) obtained with the Southern Africa Large Telescope. They found that although the mid-IR line ratios involving [Ne~{\small V}]14.32, [O~{\small IV}]25.89, or [S~{\small IV}]10.51 are good tracers of the AGN activity versus star formation as found in previous work (e.g., \citealt{Spinoglio&Malkan1992, Genzel.etal.1998, Dale.etal.2006, Armus.etal.2007}), such line ratios also show a dependence on the AGN contribution to the mid-IR continuum. In addition, they found that star formation or radiative shocks, especially the latter, can contribute to [Ne~{\small V}]14.32, [O~{\small IV}]25.89, and [S~{\small IV}]10.51 as well.  \cite{Feltre.etal.2023} therefore suggested that diagrams involving only [Ne~{\small V}]14.32, [S~{\small IV}]10.51, or [O~{\small IV}]25.89 are not sufficient to simultaneously reveal the relative roles of star formation, AGN, and shocks. This point is further demonstrated by the significant overlap in the distributions of SF, AGN, and shock models, especially the latter two, in Figure~\ref{Fig_Diags} in Appendix~\ref{secA} based on the photoionization models discussed in above sections.

As discussed in Section~\ref{sec3}, the mid-IR emission lines of highly ionized argon (i.e., [Ar~{\small V}]) and neon (i.e., [Ne~{\small VI}]) observable by JWST/MRS are effective as diagnostics of the ionization parameter and AGN spectral peak energy. In addition, as illustrated in Section~\ref{sec4.1.2}, [Fe~{\small II}] emission lines are significantly enhanced in shocked regions. We therefore expect that a diagram consisting of these emission lines would be able to pick out the AGN ionized regions well. Figure~\ref{Fig_diagram} presents two of such diagrams consisting of (a) [Ne~{\small VI}]7.65/[Ar~{\small V}]7.90 versus [Ar~{\small V}]7.90/[Ar~{\small III}]8.99 and (b) [Ne~{\small VI}]7.65/[Fe~{\small II}]5.34 versus [Ar~{\small V}]7.90/[Fe~{\small II}]5.34, respectively. The colored contours in Figure~\ref{Fig_diagram} (as well as those in Figure~\ref{Fig_Diags}) represent the line ratio distributions predicted by the photoionization models discussed in above sections. The discrete distribution peaks are mainly due to the sampling effect of model grids, not necessarily reflect actual distribution peaks in observations. Compared with the diagrams in Figure~\ref{Fig_Diags} involving [Ne~{\small V}]14.32, [O~{\small IV}]25.89, or [S~{\small IV}]10.51, the overall distributions of the radiative shock and AGN model predictions in Figure~\ref{Fig_diagram} shows that it is more feasible to use diagrams involving infrared [Ne~{\small VI}] and [Ar~{\small V}] emission lines to distinguish galactic regions dominated by different excitation sources. Figure~\ref{Fig_diagram} does not include SF models but can still be used to indirectly distinguish SF regions because SF regions can barely produce [Ne~{\small VI}]7.65 emission line due to its very high ionization potential (126.2~eV).\footnote{For all SF models involved in this work, their [Ne~{\scriptsize VI}]7.65 emission intensity are essentially less than one millionth of the H$\beta$ emission intensity.}

Although better than the diagrams in Figure~\ref{Fig_Diags}, the diagrams in Figure~\ref{Fig_diagram} still show overlapping areas between the radiative shock and AGN model results. This result further demonstrates that shocked regions can produce diagnostic emission line ratios the same as some AGN-excited regions. Therefore, in order to better distinguish the dominant excitation sources in different galaxies, the two diagrams in Figure~\ref{Fig_diagram} should be used synergistically. Specifically, Figure~\ref{Fig_diagram}(a) can be used to distinguish between AGN with different $E_{\rm peak}$ given the overall lower [Ne~{\small VI}]7.65/[Ar~{\small V}]7.90 ratios in AGN with low $E_{\rm peak}$. High $E_{\rm peak}$ and low $E_{\rm peak}$ AGN correspond to high and low accretion rate AGN, respectively, and can be empirically regraded as high luminosity AGN (e.g., Seyferts) and low luminosity AGN (e.g., LINERs). Figure~\ref{Fig_diagram}(a) also shows that radiative shock models with $v_{s} \geq 500\rm\,km\,s^{-1}$ have overall higher [Ne~{\small VI}]7.65/[Ar~{\small V}]7.90 as well as [Ar~{\small V}]7.90/[Ar~{\small III}]8.99 than radiative shock models with $v_{s} \leq 500\rm\,km\,s^{-1}$, indicating harder radiation fields and higher ionization parameters in radiative shocks with higher velocities. This trend is even more pronounced within radiative shock precursors. Specifically, pure shock precursors (not shown here) associated with shocks of $v_{s} \leq 500\,\rm km\,s^{-1}$ consistently lie just below the model predictions for low $E_{\rm peak}$ AGN, while those with $v_{s} \geq 500\,\rm km\,s^{-1}$ intersect the upper end of the model predictions for high $E_{\rm peak}$ AGN. Contrary to Figure~\ref{Fig_diagram}(a), Figure~\ref{Fig_diagram}(b) does not show a clear difference in the distributions of AGN with high $E_{\rm peak}$ and low $E_{\rm peak}$, as well as radiative shocks with high $v_{s}$ and low $v_{s}$, whereas it better distinguishes radiative shocks from AGN given their overall different distributions. Combing the two diagrams in Figure~\ref{Fig_diagram}, we find that the nuclear regions of the six targets discussed in Section~\ref{sec3.5} are governed by high $E_{\rm peak}$ AGN, which is consistent with their nature as Seyferts. 

Nevertheless, we cannot entirely rule out a scenario involving a composite contribution from both AGN and fast radiative shocks. Diagnostic diagrams involving the kinematic information of the emission lines can shed more light on the existence of shocked regions (\citealt{DAgostino.etal.2019, Zhu.etal.2025}). However, quantifying the contribution of shocked regions remains challenging, and they can contribute up to 80\% of the coronal emission in AGN (\citealt{Rodriguez-Ardila.etal.2025}). This is particularly true for shocks dominated by radiation from their radiative precursors, which closely mimic the ionizing radiation fields of AGN. The approaches of constructing starburst-AGN mixing sequences--either by anchoring empirical points (e.g., \citealt{Kewley.etal.2001, Wild.etal.2010}) or by fitting multiple line ratios with model results (e.g., \citealt{Davies.etal.2016, Thomas.etal.2018, Marconi.etal.2024})--are worth exploring to develop similar shock-AGN mixing sequences to help quantify the contribution of shocked regions to gas excitation in AGN. However, due to the fundamental differences in the underlying physics of shock and AGN models, constructing shock-AGN mixing sequences is inherently complex than building starburst-AGN mixing sequences, and thus requires dedicated investigation beyond the scope of this study.

\section{Summary and Conclusions}\label{sec5}

With excellent spectral and angular resolutions and, especially, sensitivity, JWST allows us to observe a variety of infrared emission lines that were previously inaccessible or barely accessible (e.g., [Ar~{\small V}]7.90, [Ne~{\small VI}]7.65). Covering a wide range of critical densities and ionization potentials (see Table~\ref{tablines}), these infrared emission lines provide us with an unprecedented opportunity to better diagnose the physical conditions in AGN, as well as in shocks and star-forming regions. Based on the {\sc MAPPINGS V} theoretical photoionization modeling, we systematically explore the feasibility of different infrared line combinations within JWST/MRS $\sim 5-28\,\mum$ spectral range in diagnosing the ionization parameter ($U$), the peak energy of AGN ionizing spectrum ($E_{\rm peak}$), metallicity (12 + log (O/H)), and gas pressure ($P/k$) in NLRs of AGN, as well as in star-forming regions. To find out the optimal prescriptions for the above diagnostics, we adopt a feature extraction method based on the Pearson correlation coefficient combining some practical criteria.

The main results about the infrared diagnostics can be summarized as follows:

\begin{enumerate}

\item Overall, the line ratios involving [Ar~{\small V}] (e.g., [Ar~{\small V}]7.90/[Ar~{\small III}]8.99 and [Ar~{\small V}]13.10/[Ne~{\small III}]15.56 as recommended) are the best $U$ diagnostics for NLRs in AGN, followed by line ratios involving [Ne~{\small V}], and finally line ratios involving [O~{\small IV}] with larger scatters (see Section~\ref{sec3.1}). In the meanwhile, the best $U$ diagnostics in star-forming regions are line ratios involving [S~{\small IV}] (e.g., [S~{\small IV}]10.51/[Ne~{\small III}]15.56 as recommended, see Section~\ref{sec4.2}).

\item High ionization potential lines (e.g., [Ne~{\small VI}]7.65) are recommended to diagnose $E_{\rm peak}$ (essentially the radiation field hardness) in AGN, since they have, among the infrared emission lines covered by JWST/MRS spectra, the strongest dependence on $E_{\rm peak}$ besides the primary dependence on $U$. In particular, the line ratio composed of emission lines from two noble gases, i.e., [Ne~{\small VI}]7.65/[Ar~{\small V}]7.90, is considered to be an effective diagnostic of the AGN radiation field hardness (see Section~\ref{sec3.2}).

\item The intensities of highly ionized argon emission (i.e., [Ar~{\small III}]8.99+[Ar~{\small V}]7.90) and singly ionized neon emission ([Ne~{\small II}]12.81) after normalization by hydrogen emission (i.e., Pf$\alpha$) are proposed to diagnose metallicity in NLRs of AGN and in SF regions, respectively, despite they also have a primary dependence on $U$. The line ratios involving singly ionized neon/argon and iron (i.e., [Ar~{\small II}]6.99/[Fe~{\small II}]5.34, and [Ne~{\small II}]12.81/[Fe~{\small II}]5.34) could be a complementary metallicity diagnostics in AGN, but subject to the dust depletion, as well as the X-ray and shock excitation (see Section~\ref{sec3.3} and Section~\ref{sec4.2}).

\item The ratios of infrared emission lines of the same ion species but different critical densities are explored to diagnose gas pressure in AGN and SF regions. While [Ne~{\small V}]14.32/[Ne~{\small V}]24.32 is a good choice in diagnosing gas pressure in AGN, there is no such line ratio that can be used in SF regions (see Section~\ref{sec3.4}). The two candidates of gas pressure diagnostic within JWST/MRS bands for SF regions, i.e., [Ar~{\small III}]8.99/[Ar~{\small III}]21.83 and [Fe~{\small II}]25.99/[Fe~{\small II}]5.34, are insensitive to gas pressure until log~$(P/k) \gtrsim 7.5$, which is beyond the gas pressure range of typical H~{\small II} regions in observations (see Section~\ref{sec4.2}).

\end{enumerate}

The proposed diagnostics are applicable to both high- and low-luminosity AGN and are applied to galaxies from the GATOS program. Specifically, the proposed diagnostics indicate log~$U$ values of $\sim -2.6$ to $-2.4$, log~$E_{\rm peak}$ values of $\sim -1.0$ to $-0.5$, metallicities of $\sim 0.2 - 0.9\,Z_{\odot}$, and log~$(P/k)$ values of $\sim 7.8 - 8.3$ in the central $\sim 200 - 300$ regions of the six Seyferts (see Section~\ref{sec3.5}). The derived log~$U$ values are overall consistent with previous measurements based on optical [O~{\small III}]/H$\beta$ and [N~{\small II}]/H$\alpha$ line ratios. The derived metallicities are statistically consistent with the sub-solar metallicities measured for large samples of type 2 AGN. However, the derived $E_{\rm peak}$ values are larger by of $\sim$ 1 dex than what would be expected from the correlation between $E_{\rm peak}$ and $\lambda_{\rm Edd}$. The elevated $E_{\rm peak}$ is plausibly due to extra heating from the circumnuclear hot gas associated with fast radiative shocks.

Additional attention is required especially when applying these emission-line-ratio diagnostics to low-luminosity AGN (see Section~\ref{sec4.1}), where the diagnostics could be affected by the hard X-ray and potential shocks. Although the line ratio diagnostics proposed in this study for AGN are essentially based on emission lines of highly charged ions in NLRs (i.e., [Ar~{\small V}], [Ne~{\small V}], and Ne~{\small VI}]) and thus largely avoid the impact of XRDs (see Section~\ref{sec4.1.1}), it is still crucial to identify and exclude shock-dominated regions when diagnosing the physical conditions of NLRs in AGN (see Section~\ref{sec4.1.2}). Last but not least, further studies based on the results here are deserved to help develop new diagrams (e.g., [Ne~{\small VI}]7.65/[Ar~{\small V}]7.90 versus [Ar~{\small V}]7.90/[Ar~{\small III}]8.99 and [Ne~{\small VI}]7.65/[Fe~{\small II}]5.34 versus [Ar~{\small V}]7.90/[Fe~{\small II}]5.34) for distinguishing galactic regions governed by different excitation sources, i.e., AGN, shocks, and star-forming activities (see Section~\ref{sec4.3}).
\acknowledgments
We thank the anonymous referee for constructive comments. L.Z. thanks Peixin Zhu (CfA) for helpful discussions at the beginning of this project and acknowledges grant support from the Space Telescope Science Institute (ID: JWST-GO-01670; JWST-GO-03535; JWST-GO-04972). C.P., E.K.S.H., and D. D. acknowledge grant support from the Space Telescope Science Institute (ID: JWST-GO-01670).  M.P.S. acknowledges support under grants RYC2021-033094-I, CNS2023-145506, and PID2023-146667NB-I00 funded by MCIN/AEI/10.13039/501100011033 and the European Union NextGenerationEU/PRTR. L.H.M. and A.A.-H. acknowledge support from grant PID2021-124665NB-I00  funded by MCIN/AEI/10.13039/501100011033 and by ERDF A way of making Europe. I.G.B is supported by the Programa Atracci\'on de Talento Investigador ``C\'esar Nombela'' via grant 2023-T1/TEC-29030 funded by the Community of Madrid. C.R. acknowledges support from Fondecyt Regular grant 1230345, ANID BASAL project FB210003 and the China-Chile joint research fund. E.B. acknowledges support from the Spanish grants PID2022-138621NB-I00 and PID2021-123417OB-I00, funded by MCIN/AEI/10.13039/501100011033/FEDER, EU. C.R.A. and A.A. acknowledge support from the Agencia Estatal de Investigaci\'on of the Ministerio de Ciencia, Innovaci\'on y Universidades (MCIU/AEI) under the grant ``Tracking active galactic nuclei feedback from parsec to kiloparsec scales'', with reference PID2022$-$141105NB$-$I00 and the European Regional Development Fund (ERDF). O.G.M. is supported by the UNAM PAPIIT project IN109123 and the SECIHTI Ciencia de Frontera project CF-2023-G100. T.D.-S. acknowledges the research project was supported by the Hellenic Foundation for Research and Innovation (HFRI) under the ``2nd Call for HFRI Research Projects to support Faculty Members \& Researchers" (Project Number: 03382)
\newline

\appendix

\section{Tables for Model Results, Tables for Coefficients in Best-fit Correlations, and Figures for Low-luminosity AGN}\label{secA}

\begin{figure*}[!ht]
\figurenum{B1}
\center{\includegraphics[width=1\textwidth]{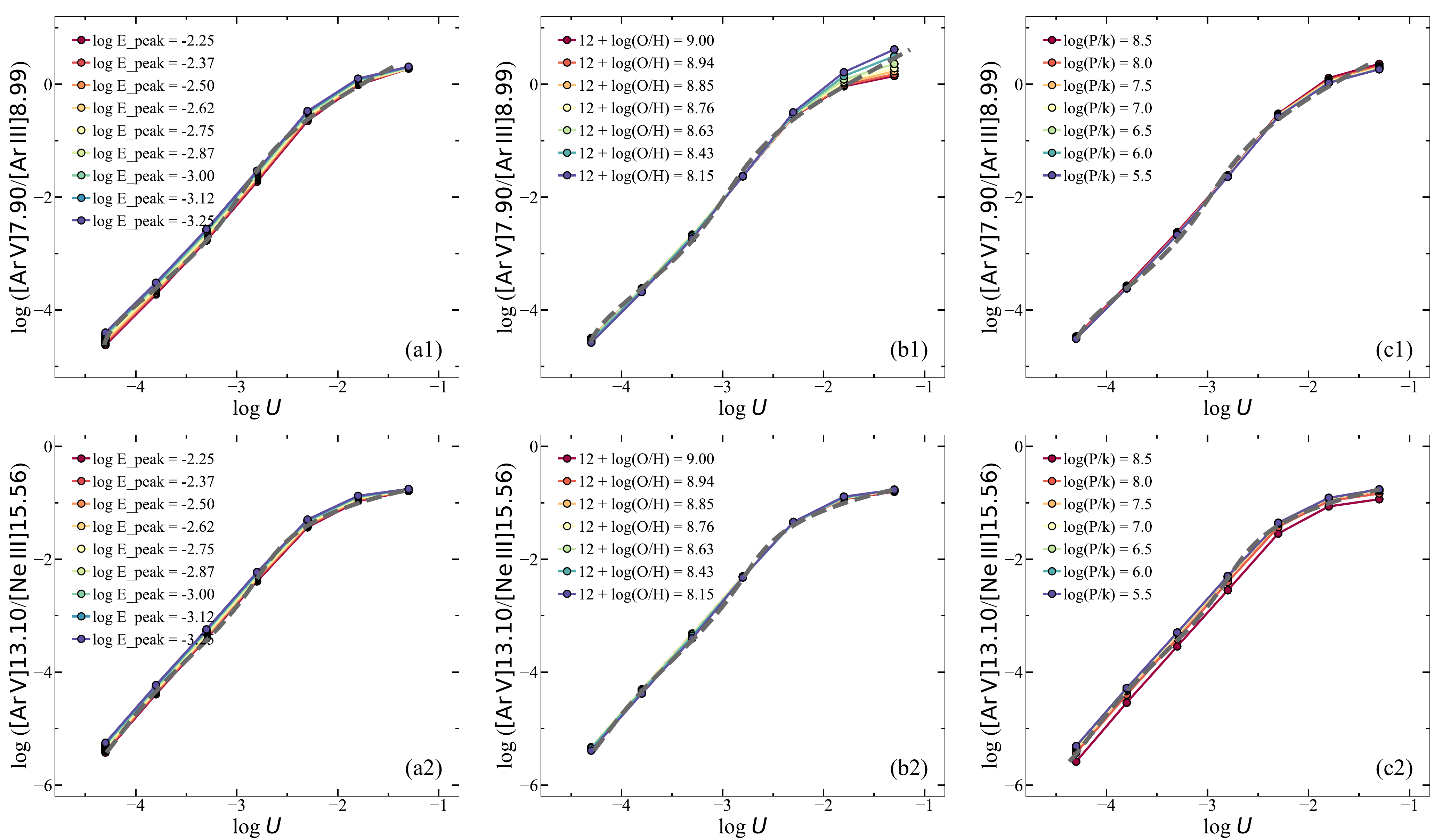}}
\caption{Same as Figure~\ref{Fig_U_AGN} but for low-luminosity AGN. Each panel, if not specifically marked, displays model results under the same parameter constraints as in Figure~\ref{Fig_U_AGN} for log $U$ (i.e., $-2.8$), 12 + log (O/H) (i.e., 8.76), and log $(P/k)$ (i.e., 7.0), but with a different value of log $E_{\rm peak}$ ($-2.75$ here).}\label{Fig_U_AGN_II}
\end{figure*}

\begin{figure*}[!ht]
\figurenum{B2}
\center{\includegraphics[width=1\textwidth]{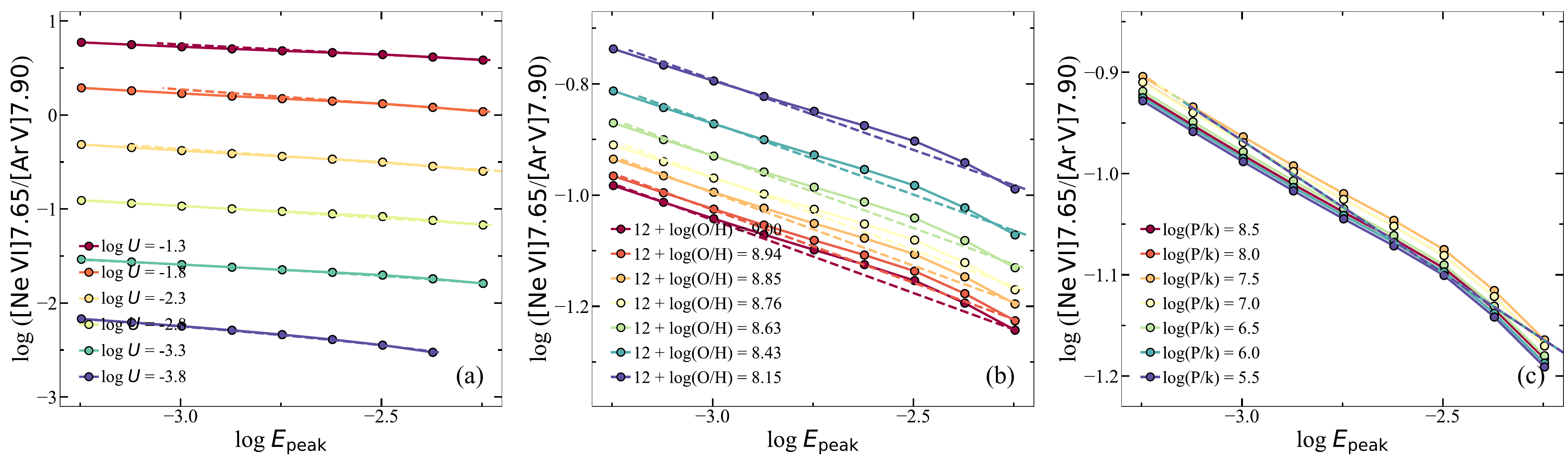}}
\caption{Same as Figure~\ref{Fig_Ep_AGN} but for low-luminosity AGN. Each panel, if not specifically marked, displays model results under the same parameter constraints as in Figure~\ref{Fig_Ep_AGN} for log $U$ (i.e, $-2.8$), 12 + log (O/H) (i.e., 8.76), and log $(P/k)$ (i.e., 7.0), but with a different value of log $E_{\rm peak}$ ($-2.75$ here).}\label{Fig_Ep_AGN_II}
\end{figure*}

\begin{figure*}[!ht]
\figurenum{B3}
\center{\includegraphics[width=1\textwidth]{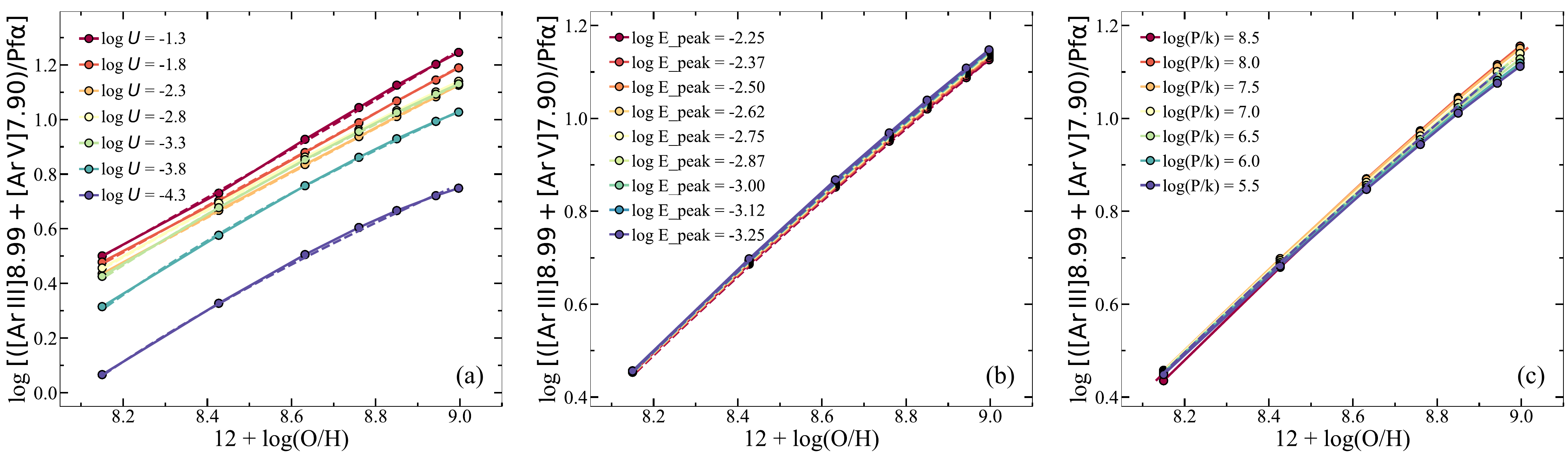}}
\caption{Same as Figure~\ref{Fig_A_AGN} but for low-luminosity AGN. Each panel, if not specifically marked, displays model results under the same parameter constraints as in Figure~\ref{Fig_A_AGN} for log $U$ (i.e., $-2.8$), 12 + log (O/H) (i.e., 8.76), and log $(P/k)$ (i.e., 7.0), but with a different value of log $E_{\rm peak}$ ($-2.75$ here).}\label{Fig_A_AGN_II}
\end{figure*}

\begin{figure*}[!ht]
\figurenum{B4}
\center{\includegraphics[width=1\textwidth]{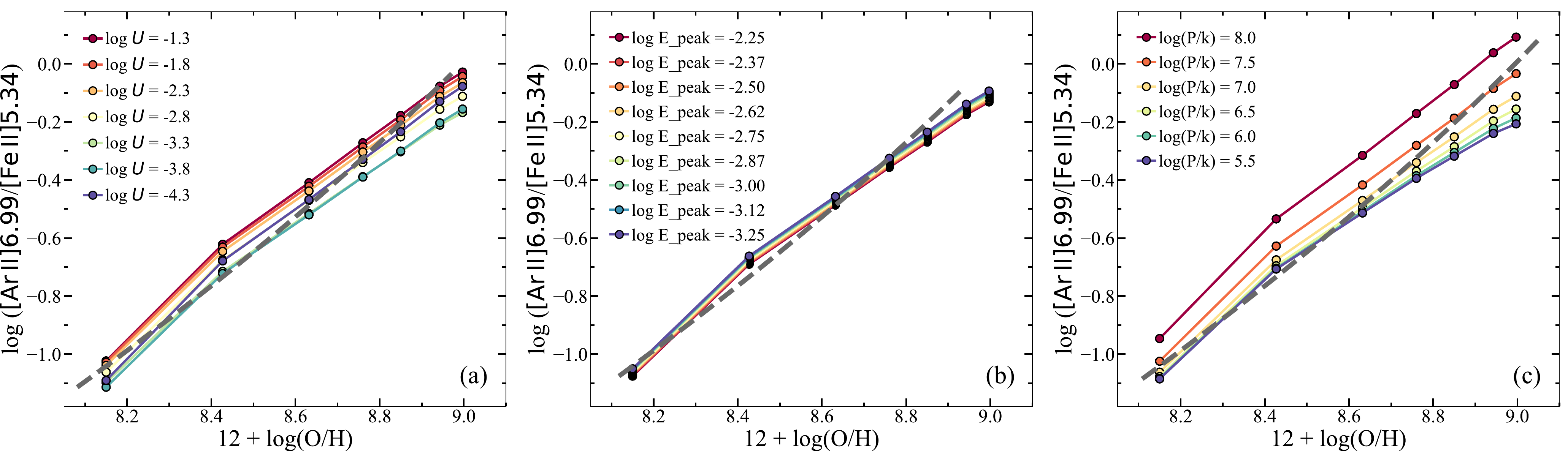}}
\caption{Same as Figure~\ref{Fig_A_II_AGN} but for low-luminosity AGN. Each panel, if not specifically marked, displays model results under the same parameter constraints as in Figure~\ref{Fig_A_II_AGN} for log $U$ (i.e., $-2.8$), 12 + log (O/H) (i.e., 8.76), and log $(P/k)$ (i.e., 7.0), but with a different value of log $E_{\rm peak}$ ($-2.75$ here).}\label{Fig_A_II_AGN_II}
\end{figure*}

\begin{figure*}[!ht]
\figurenum{B5}
\center{\includegraphics[width=1\textwidth]{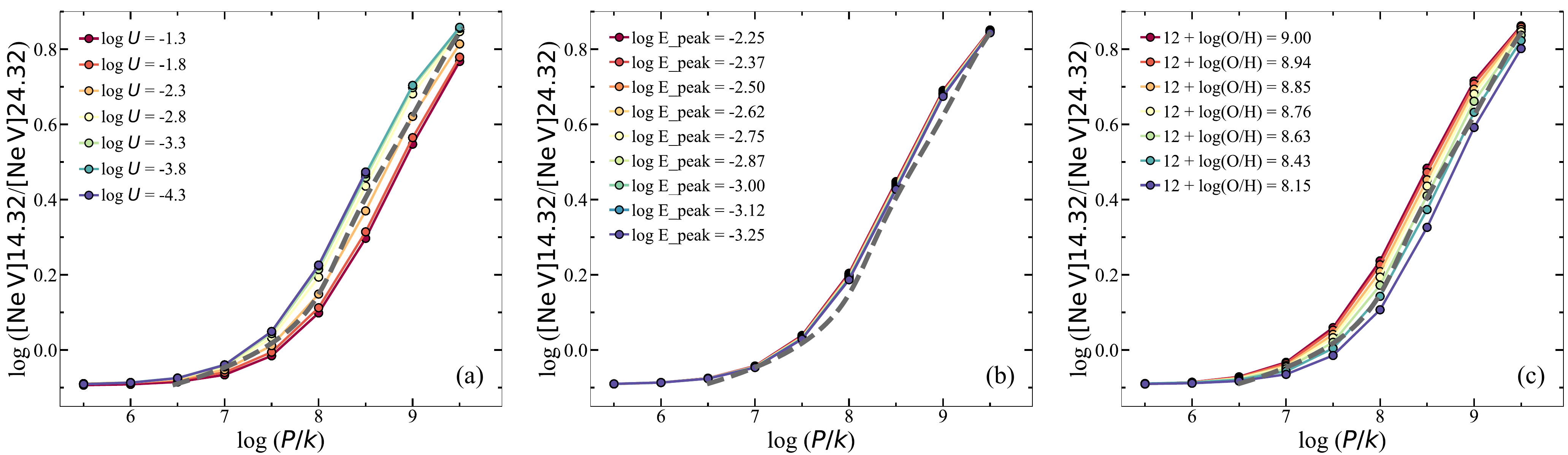}}
\caption{Same as Figure~\ref{Fig_P_AGN} but for low-luminosity AGN. Each panel, if not specifically marked, displays model results under the same parameter constraints as in Figure~\ref{Fig_P_AGN} for log $U$ (i.e., $-2.8$), 12 + log (O/H) (i.e., 8.76), and log $(P/k)$ (i.e., 7.0), but with a different value of log $E_{\rm peak}$ ($-2.75$ here).}\label{Fig_P_AGN_II}
\end{figure*}

\begin{figure*}[!ht]
\figurenum{B6}
\center{\includegraphics[width=1\textwidth]{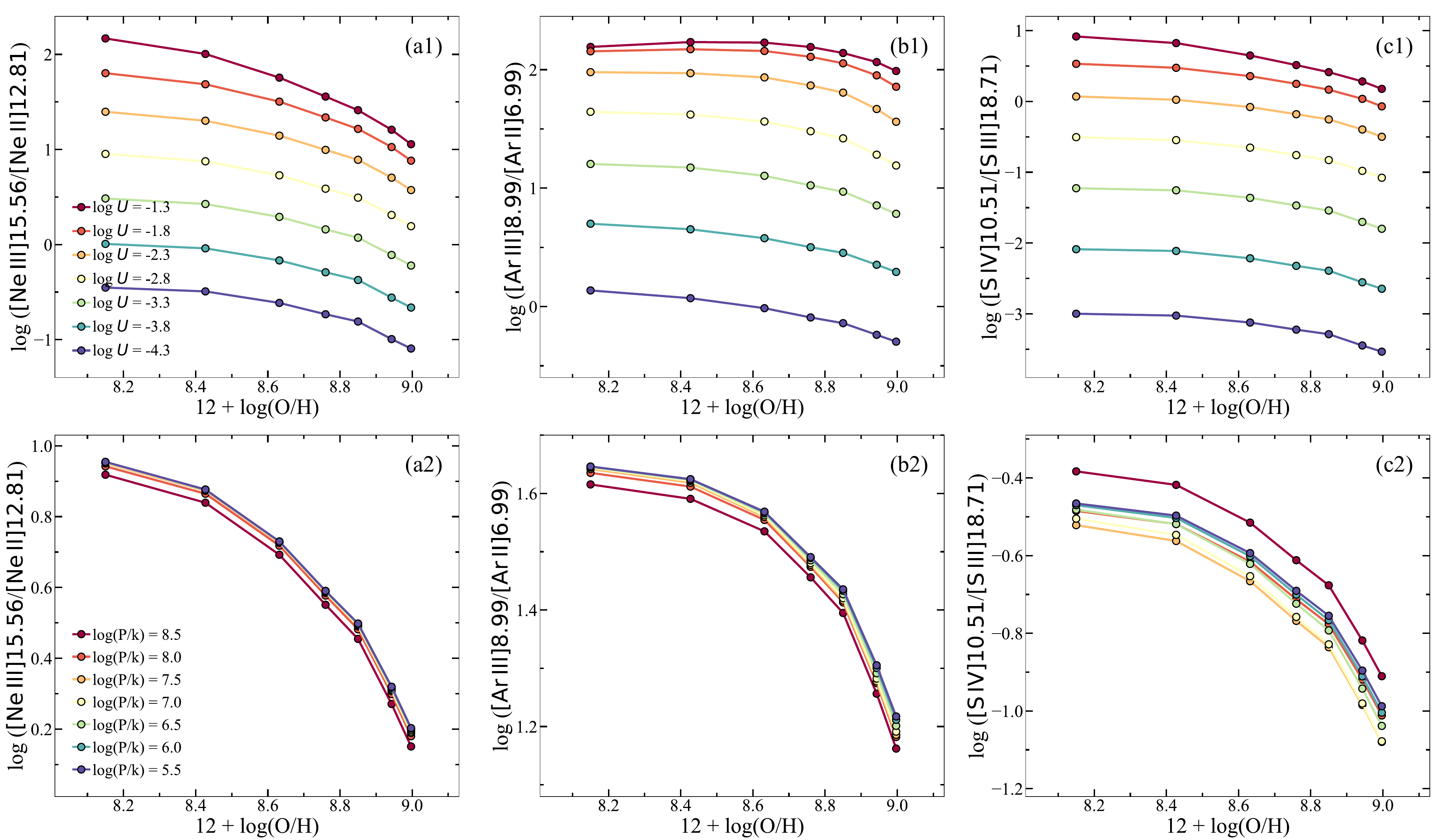}}
\caption{Dependence of (a1\&a2) [Ne~{\small III}]15.56/[Ne~{\small II}]12.81, (b1\&b2) [Ar~{\small III}]8.99/[Ar~{\small II}]6.99, and (c1\&c2) [S~{\small IV}]10.51/[S~{\small III}]18.71 on the metallicity (12 + log (O/H)), as predicted by SF photoionization models (colored points) discussed in Section~\ref{sec4.2}. The top and bottom panels display model results under specific constraints on log $(P/k)$ (i.e., 7.0) and log $U$ (i.e., $-2.8$), respectively.}\label{Fig_SF_H}
\end{figure*}

\begin{figure*}[!ht]
\figurenum{B7}
\center{\includegraphics[width=1\textwidth]{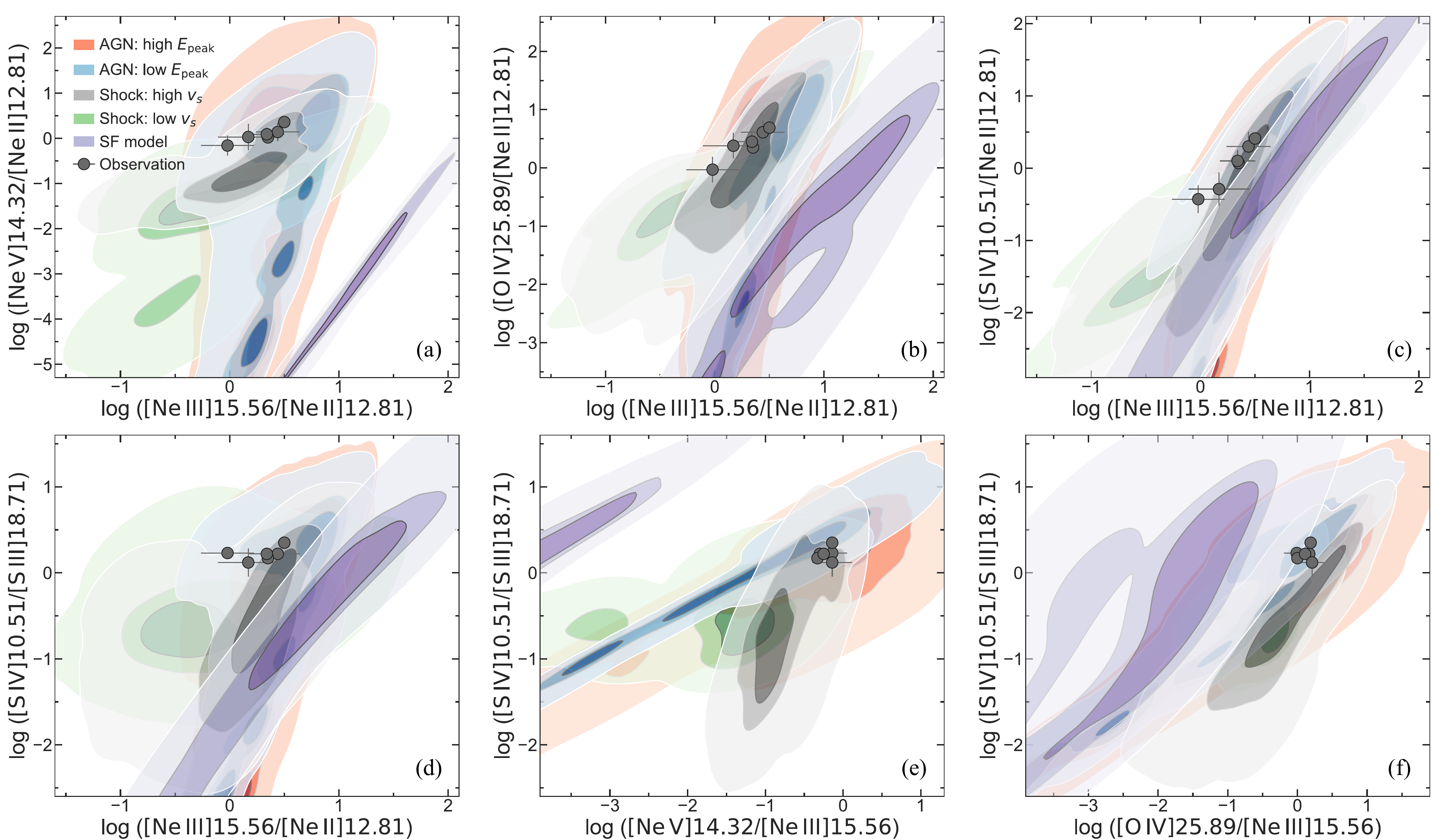}}
\caption{The same as Figure~\ref{Fig_diagram}, but with added contours for SF photoionization models discussed in Section~\ref{sec4.2}. Measurements for the central $r = 0\farcs75$ regions of six Seyferts represented by the six gray points are taken from \cite{Zhang.etal.2024a}.}\label{Fig_Diags}
\end{figure*}

Tables~\ref{tabRes1}, \ref{tabRes2}, and \ref{tabRes3} provide the model results calculated for the AGN, radiative shocks, and SF regions discussed in this paper, respectively. Figures~\ref{Fig_U_AGN_II}-\ref{Fig_P_AGN_II} are the same as Figures~\ref{Fig_U_AGN}-\ref{Fig_P_AGN} in the main text but for low-luminosity AGN. Figure~\ref{Fig_SF_H} is related to the discussion in Section~\ref{sec4.2} on the radiation field hardness diagnostics in SF regions. Figure~\ref{Fig_Diags} is related to the discussion in Section~\ref{sec4.3} on the mid-IR line ratio diagrams involving [Ne~{\small V}]14.32, [S~{\small IV}]10.51, or [O~{\small IV}]25.89. Tables~\ref{tabEqu1}-\ref{tabEqu7} summarize the coefficients in all best-fit correlations proposed in the main text (a more user-friendly compilation of all these coefficients is available online via this \href{https://github.com/LuluZhang-Astro/LuluZhang-Astro.github.io/blob/master/data/Coefficients_Functions.py}{link}).

\newpage

\newpage

\begin{longrotatetable}
\tablenum{A1}
\setlength{\tabcolsep}{2pt}
\begin{deluxetable}{ccccccccccccccccccccccccccccc}
\tabletypesize{\tiny}
\tablecolumns{29}
\tablecaption{The Modeling Results for AGN}
\tablehead{
\colhead{$M_{\rm BH}$} & \colhead{$\lambda_{\rm Edd}$} & \colhead{$E_{\rm peak}$} & \colhead{$U$} & \colhead{abund} & \colhead{$P/k$} & \colhead{FeII5.34} & \colhead{FeVIII5.45} & \colhead{MgVII5.50} & \colhead{MgV5.61} & \colhead{ArII6.99} & \colhead{NaIII7.32} & \colhead{Pfa} & \colhead{NeVI7.65} & \colhead{FeVII7.82} & \colhead{ArV7.90} & \colhead{ArIII8.99} & \colhead{FeVII9.53} & \colhead{SIV10.51} & \colhead{Hua} & \colhead{NeII12.81} & \colhead{ArV13.10} & \colhead{NeV14.32} & \colhead{NeIII15.56} & \colhead{SIII18.71} & \colhead{ArIII21.83} & \colhead{NeV24.32} & \colhead{OIV25.89} & \colhead{FeII25.99} \\
\colhead{(1)} & \colhead{(2)} & \colhead{(3)} & \colhead{(4)} & \colhead{(5)} & \colhead{(6)} & \colhead{(7)} & \colhead{(8)} & \colhead{(9)} & \colhead{(10)} & \colhead{(11)} & \colhead{(12)} & \colhead{(13)} & \colhead{(14)} & \colhead{(15)} & \colhead{(16)} & \colhead{(17)} & \colhead{(18)} & \colhead{(19)} & \colhead{(20)} & \colhead{(21)} & \colhead{(22)} & \colhead{(23)} & \colhead{(24)} & \colhead{(25)} & \colhead{(26)} & \colhead{(27)} & \colhead{(28)} & \colhead{(29)}}
\startdata
6.0 & -2.5 & -1.997 & -3.8 & 8.632 & 6.0 & 1.90e-01 & nan & nan & 4.42e-07 & 4.86e-02 & 5.06e-03 & 2.52e-02 & nan & nan & 2.58e-05 & 1.59e-01 & nan & 1.26e-02 & 9.63e-03 & 5.99e-01 & 4.45e-05 & 1.89e-05 & 1.33e+00 & 6.25e-01 & 1.09e-02 & 2.31e-05 & 4.49e-03 & 6.75e-02 \\
6.0 & -5.0 & -2.622 & -3.3 & 8.76 & 5.5 & 9.72e-02 & nan & nan & 2.40e-05 & 3.40e-02 & 5.30e-03 & 2.61e-02 & 9.62e-06 & 3.25e-06 & 4.62e-04 & 2.26e-01 & 1.20e-05 & 1.18e-01 & 1.00e-02 & 4.51e-01 & 8.12e-04 & 1.17e-03 & 1.70e+00 & 9.41e-01 & 1.53e-02 & 1.44e-03 & 1.44e-01 & 3.44e-02 \\
6.0 & -1.0 & -1.2685 & -3.8 & 8.76 & 5.5 & 3.86e-01 & nan & nan & 2.24e-06 & 1.62e-01 & 1.20e-02 & 2.46e-02 & 1.97e-06 & 4.54e-07 & 6.93e-05 & 2.07e-01 & 1.65e-06 & 2.53e-02 & 9.40e-03 & 1.34e+00 & 1.19e-04 & 1.11e-04 & 3.02e+00 & 1.81e+00 & 1.42e-02 & 1.37e-04 & 6.44e-03 & 1.37e-01 \\
8.5 & -2.5 & -2.622 & -1.8 & 8.632 & 7.0 & 4.98e-02 & 1.36e-03 & 9.95e-04 & 4.63e-02 & 1.88e-02 & 3.96e-03 & 2.33e-02 & 1.49e-01 & 4.85e-03 & 9.35e-02 & 7.96e-02 & 1.70e-02 & 1.74e+00 & 8.82e-03 & 1.80e-01 & 1.50e-01 & 7.82e-01 & 1.26e+00 & 5.59e-01 & 5.43e-03 & 9.12e-01 & 3.34e+00 & 1.89e-02 \\
6.0 & 0.0 & -0.94915 & -2.3 & 8.85 & 7.5 & 2.43e-01 & 1.24e-02 & 1.42e-02 & 4.34e-02 & 1.63e-01 & 8.46e-03 & 1.99e-02 & 1.44e+00 & 1.51e-02 & 1.13e-01 & 1.98e-01 & 5.21e-02 & 1.15e+00 & 7.49e-03 & 1.37e+00 & 1.73e-01 & 2.02e+00 & 2.34e+00 & 2.21e+00 & 1.35e-02 & 2.17e+00 & 4.74e+00 & 9.87e-02 \\
8.0 & -3.5 & -2.747 & -1.3 & 8.76 & 8.0 & 7.56e-02 & 5.74e-03 & 8.00e-03 & 5.83e-02 & 6.17e-02 & 5.17e-03 & 2.24e-02 & 8.88e-01 & 8.92e-03 & 1.75e-01 & 8.32e-02 & 3.03e-02 & 1.70e+00 & 8.44e-03 & 5.65e-01 & 2.25e-01 & 1.69e+00 & 1.55e+00 & 6.64e-01 & 5.05e-03 & 1.35e+00 & 2.32e+00 & 4.03e-02 \\
8.5 & -1.0 & -2.247 & -2.3 & 8.943 & 7.0 & 3.75e-02 & 1.24e-04 & 1.03e-05 & 6.73e-03 & 2.81e-02 & 5.63e-03 & 2.60e-02 & 1.19e-02 & 1.01e-03 & 5.38e-02 & 2.45e-01 & 3.67e-03 & 2.07e+00 & 9.94e-03 & 3.21e-01 & 9.05e-02 & 3.02e-01 & 2.44e+00 & 1.47e+00 & 1.64e-02 & 3.32e-01 & 2.80e+00 & 1.50e-02 \\
6.0 & -4.5 & -2.497 & -2.3 & 8.15 & 6.0 & 9.09e-02 & 2.67e-04 & 5.22e-05 & 1.64e-02 & 7.90e-03 & 1.88e-03 & 2.16e-02 & 6.31e-03 & 2.71e-03 & 1.27e-02 & 4.47e-02 & 9.37e-03 & 3.58e-01 & 8.15e-03 & 8.02e-02 & 2.04e-02 & 7.98e-02 & 4.80e-01 & 2.19e-01 & 3.13e-03 & 9.95e-02 & 1.16e+00 & 3.26e-02 \\
9.0 & -3.5 & -2.997 & -4.3 & 8.15 & 8.5 & 2.70e-01 & nan & nan & nan & 4.35e-02 & 2.21e-03 & 2.39e-02 & nan & nan & 8.78e-07 & 2.64e-02 & nan & 2.05e-04 & 9.04e-03 & 5.14e-01 & 8.09e-07 & 1.85e-07 & 3.03e-01 & 1.31e-01 & 1.47e-03 & nan & 6.67e-06 & 1.36e-01 \\
6.0 & -2.5 & -1.997 & -3.8 & 8.85 & 5.5 & 1.66e-01 & nan & nan & 4.51e-07 & 6.69e-02 & 6.34e-03 & 2.65e-02 & 1.42e-07 & nan & 4.23e-05 & 2.51e-01 & nan & 2.14e-02 & 1.02e-02 & 8.28e-01 & 7.47e-05 & 3.10e-05 & 2.07e+00 & 1.01e+00 & 1.70e-02 & 3.82e-05 & 5.54e-03 & 5.92e-02 \\
\enddata
\tablecomments{\footnotesize Columns (1) -- (6): The input parameters for {\sc MAPPINGS V} models of AGN (see Section~\ref{sec2}). Columns (7) -- (29): The predicted emission line intensities (normalized by the corresponding H$\beta$ emission) for the different AGN models that assume a homogeneous distribution of ionized gas. (This table is available in its entirety in machine-readable form in the online article).}
\label{tabRes1}
\end{deluxetable}
\end{longrotatetable}

\begin{longrotatetable}
\tablenum{A2}
\setlength{\tabcolsep}{2pt}
\begin{deluxetable}{ccccccccccccccccccccccccccc}
\tabletypesize{\tiny}
\tablecolumns{27}
\tablecaption{The Modeling Results for Radiative Shocks}
\tablehead{
\colhead{$n_{\rm H}$} & \colhead{$v_{\rm s}$} & \colhead{$\eta_{\rm M}$} & \colhead{abund} & \colhead{FeII5.34} & \colhead{FeVIII5.45} & \colhead{MgVII5.50} & \colhead{MgV5.61} & \colhead{ArII6.99} & \colhead{NaIII7.32} & \colhead{Pfa} & \colhead{NeVI7.65} & \colhead{FeVII7.82} & \colhead{ArV7.90} & \colhead{ArIII8.99} & \colhead{FeVII9.53} & \colhead{SIV10.51} & \colhead{Hua} & \colhead{NeII12.81} & \colhead{ArV13.10} & \colhead{NeV14.32} & \colhead{NeIII15.56} & \colhead{SIII18.71} & \colhead{ArIII21.83} & \colhead{NeV24.32} & \colhead{OIV25.89} & \colhead{FeII25.99} \\
\colhead{(1)} & \colhead{(2)} & \colhead{(3)} & \colhead{(4)} & \colhead{(5)} & \colhead{(6)} & \colhead{(7)} & \colhead{(8)} & \colhead{(9)} & \colhead{(10)} & \colhead{(11)} & \colhead{(12)} & \colhead{(13)} & \colhead{(14)} & \colhead{(15)} & \colhead{(16)} & \colhead{(17)} & \colhead{(18)} & \colhead{(19)} & \colhead{(20)} & \colhead{(21)} & \colhead{(22)} & \colhead{(23)} & \colhead{(24)} & \colhead{(25)} & \colhead{(26)} & \colhead{(27)} }
\startdata
10000.0 & 1000.0 & 0.0001 & 8.15 & 4.32e+02 & 1.17e+02 & 2.18e+02 & 5.19e+01 & 2.65e+02 & 2.59e+01 & 1.14e+02 & 1.28e+03 & 3.90e+01 & 1.49e+01 & 1.01e+02 & 1.17e+02 & 3.00e+01 & 4.18e+01 & 1.95e+03 & 1.58e+01 & 4.03e+02 & 1.82e+03 & 4.93e+02 & 3.54e+00 & 2.44e+02 & 5.12e+02 & 2.19e+02 \\
10000.0 & 100.0 & 0.0001 & 8.15 & 1.43e+00 & 0.00e+00 & 0.00e+00 & 2.24e-05 & 2.58e-01 & 4.15e-03 & 2.32e-01 & 0.00e+00 & 6.59e-05 & 4.86e-04 & 7.37e-02 & 1.63e-04 & 3.55e-02 & 8.54e-02 & 3.11e+00 & 2.99e-04 & 3.19e-04 & 3.33e-01 & 7.06e-02 & 3.23e-03 & 9.97e-05 & 2.99e-02 & 1.24e+00 \\
10000.0 & 150.0 & 0.0001 & 8.15 & 2.88e+00 & 3.86e-02 & 1.97e-04 & 1.37e-01 & 8.76e-01 & 2.05e-02 & 9.10e-01 & 3.01e-01 & 3.54e-02 & 4.71e-02 & 2.88e-01 & 9.67e-02 & 1.51e-01 & 3.40e-01 & 8.88e+00 & 3.39e-02 & 4.63e-01 & 1.97e+00 & 3.66e-01 & 1.41e-02 & 2.37e-01 & 5.58e-01 & 2.56e+00 \\
10000.0 & 200.0 & 0.0001 & 8.15 & 6.65e+00 & 1.69e+00 & 5.66e-01 & 7.17e-01 & 2.12e+00 & 1.12e-01 & 2.22e+00 & 1.11e+01 & 1.54e-01 & 1.04e-01 & 1.05e+00 & 4.12e-01 & 5.58e-01 & 8.29e-01 & 2.55e+01 & 6.38e-02 & 1.86e+00 & 8.11e+00 & 1.44e+00 & 4.86e-02 & 9.38e-01 & 1.17e+00 & 6.50e+00 \\
10000.0 & 25.0 & 0.0001 & 8.15 & 1.50e+00 & 0.00e+00 & 0.00e+00 & 0.00e+00 & 8.11e-04 & 0.00e+00 & 2.14e-03 & 0.00e+00 & 0.00e+00 & 0.00e+00 & 4.58e-08 & 0.00e+00 & 0.00e+00 & 7.98e-04 & 9.66e-03 & 0.00e+00 & 0.00e+00 & 0.00e+00 & 2.50e-05 & 0.00e+00 & 0.00e+00 & 0.00e+00 & 8.77e-01 \\
10000.0 & 300.0 & 0.0001 & 8.15 & 2.51e+01 & 5.99e+00 & 3.89e+00 & 1.92e+00 & 1.13e+01 & 7.99e-01 & 7.14e+00 & 2.17e+01 & 4.29e-01 & 7.12e-01 & 3.90e+00 & 1.16e+00 & 1.07e+01 & 2.66e+00 & 1.05e+02 & 6.49e-01 & 5.84e+00 & 4.62e+01 & 1.21e+01 & 1.57e-01 & 2.47e+00 & 2.19e+01 & 2.95e+01 \\
10000.0 & 400.0 & 0.0001 & 8.15 & 1.10e+02 & 8.84e+00 & 5.82e+00 & 8.48e+00 & 5.33e+01 & 2.34e+00 & 1.60e+01 & 3.80e+01 & 1.83e+00 & 4.33e+00 & 8.02e+00 & 5.50e+00 & 3.61e+01 & 5.97e+00 & 3.67e+02 & 4.43e+00 & 3.51e+01 & 1.21e+02 & 5.62e+01 & 2.94e-01 & 1.78e+01 & 8.35e+01 & 1.33e+02 \\
10000.0 & 50.0 & 0.0001 & 8.15 & 5.56e-01 & 0.00e+00 & 0.00e+00 & 0.00e+00 & 1.30e-02 & 0.00e+00 & 1.53e-02 & 0.00e+00 & 0.00e+00 & 0.00e+00 & 7.00e-06 & 0.00e+00 & 0.00e+00 & 5.40e-03 & 1.41e-02 & 0.00e+00 & 0.00e+00 & 2.98e-07 & 6.45e-04 & 2.82e-07 & 0.00e+00 & 0.00e+00 & 4.64e-01 \\
10000.0 & 600.0 & 0.0001 & 8.15 & 4.51e+02 & 1.86e+01 & 2.57e+01 & 2.79e+01 & 2.72e+02 & 9.04e+00 & 4.40e+01 & 2.56e+02 & 9.05e+00 & 1.80e+01 & 2.32e+01 & 2.82e+01 & 4.02e+01 & 1.64e+01 & 2.30e+03 & 1.92e+01 & 1.77e+02 & 6.32e+02 & 2.68e+02 & 8.61e-01 & 1.02e+02 & 2.16e+02 & 5.18e+02 \\
10000.0 & 75.0 & 0.0001 & 8.15 & 6.11e-01 & 0.00e+00 & 0.00e+00 & 0.00e+00 & 4.94e-02 & 1.27e-05 & 8.48e-02 & 0.00e+00 & 0.00e+00 & 0.00e+00 & 8.74e-04 & 0.00e+00 & 1.31e-05 & 3.06e-02 & 1.81e-01 & 0.00e+00 & 0.00e+00 & 1.13e-04 & 1.76e-03 & 2.46e-05 & 0.00e+00 & 0.00e+00 & 5.66e-01 \\
\enddata
\tablecomments{\footnotesize Columns (1) -- (4): The input parameters for {\sc MAPPINGS V} models of radiative shocks (see Section~\ref{sec4.1.2}). Columns (5) -- (27): The predicted emission line intensities (normalized by the corresponding H$\beta$ emission) for the different radiative shock models that assume a homogeneous distribution of ionized gas in the radiative shock precursor. (This table is available in its entirety in machine-readable form in the online article).}
\label{tabRes2}
\end{deluxetable}
\end{longrotatetable}

\begin{longrotatetable}
\tablenum{A3}
\setlength{\tabcolsep}{2pt}
\begin{deluxetable}{cccccccccccccccccccccccccc}
\tabletypesize{\tiny}
\tablecolumns{27}
\tablecaption{The Modeling Results for SF Regions}
\tablehead{
\colhead{$U$} & \colhead{abund} & \colhead{$P/k$} & \colhead{FeII5.34} & \colhead{FeVIII5.45} & \colhead{MgVII5.50} & \colhead{MgV5.61} & \colhead{ArII6.99} & \colhead{NaIII7.32} & \colhead{Pfa} & \colhead{NeVI7.65} & \colhead{FeVII7.82} & \colhead{ArV7.90} & \colhead{ArIII8.99} & \colhead{FeVII9.53} & \colhead{SIV10.51} & \colhead{Hua} & \colhead{NeII12.81} & \colhead{ArV13.10} & \colhead{NeV14.32} & \colhead{NeIII15.56} & \colhead{SIII18.71} & \colhead{ArIII21.83} & \colhead{NeV24.32} & \colhead{OIV25.89} & \colhead{FeII25.99} \\
\colhead{(1)} & \colhead{(2)} & \colhead{(3)} & \colhead{(4)} & \colhead{(5)} & \colhead{(6)} & \colhead{(7)} & \colhead{(8)} & \colhead{(9)} & \colhead{(10)} & \colhead{(11)} & \colhead{(12)} & \colhead{(13)} & \colhead{(14)} & \colhead{(15)} & \colhead{(16)} & \colhead{(17)} & \colhead{(18)} & \colhead{(19)} & \colhead{(20)} & \colhead{(21)} & \colhead{(22)} & \colhead{(23)} & \colhead{(24)} & \colhead{(25)} & \colhead{(26)} }
\startdata
-2.3 & 8.997 & 6.0 & 7.92e-03 & nan & nan & nan & 7.73e-03 & 2.52e-03 & 3.21e-02 & nan & nan & 1.20e-07 & 2.90e-01 & nan & 4.52e-01 & 1.25e-02 & 4.44e-01 & 2.41e-07 & nan & 1.65e+00 & 1.26e+00 & 1.87e-02 & nan & 1.29e-03 & 3.00e-03 \\
-2.8 & 8.632 & 7.5 & 1.19e-02 & nan & nan & 1.63e-07 & 4.60e-03 & 2.64e-03 & 2.55e-02 & nan & nan & 1.31e-05 & 1.67e-01 & nan & 1.35e-01 & 9.71e-03 & 1.94e-01 & 2.05e-05 & 7.33e-06 & 1.03e+00 & 6.27e-01 & 1.09e-02 & 6.35e-06 & 2.00e-02 & 4.95e-03 \\
-3.8 & 8.943 & 6.0 & 9.54e-02 & nan & nan & nan & 6.87e-02 & 3.81e-04 & 2.96e-02 & nan & nan & nan & 1.56e-01 & nan & 2.45e-03 & 1.14e-02 & 1.18e+00 & nan & nan & 3.27e-01 & 7.69e-01 & 1.02e-02 & nan & 2.39e-06 & 3.59e-02 \\
-2.3 & 8.427 & 6.0 & 7.36e-03 & nan & nan & 6.73e-06 & 1.04e-03 & 2.94e-03 & 2.54e-02 & nan & nan & 2.00e-04 & 9.75e-02 & 3.51e-07 & 3.89e-01 & 9.73e-03 & 4.18e-02 & 3.47e-04 & 9.64e-05 & 8.33e-01 & 3.35e-01 & 6.63e-03 & 1.18e-04 & 6.94e-02 & 2.53e-03 \\
-1.8 & 8.427 & 5.5 & 3.40e-03 & nan & nan & 4.98e-05 & 4.50e-04 & 3.20e-03 & 2.57e-02 & 2.44e-07 & 1.19e-06 & 8.32e-04 & 6.75e-02 & 4.29e-06 & 8.73e-01 & 9.84e-03 & 1.83e-02 & 1.43e-03 & 4.98e-04 & 8.71e-01 & 2.63e-01 & 4.60e-03 & 6.12e-04 & 1.00e-01 & 1.17e-03 \\
-1.3 & 8.632 & 7.0 & 1.81e-03 & nan & nan & 8.83e-05 & 4.82e-04 & 4.04e-03 & 2.60e-02 & 1.20e-06 & 3.57e-06 & 1.96e-03 & 8.15e-02 & 1.30e-05 & 1.80e+00 & 9.93e-03 & 2.33e-02 & 3.29e-03 & 1.64e-03 & 1.32e+00 & 4.05e-01 & 5.42e-03 & 1.81e-03 & 1.23e-01 & 6.85e-04 \\
-4.3 & 8.76 & 8.5 & 2.58e-02 & nan & nan & nan & 8.27e-02 & 4.03e-04 & 2.52e-02 & nan & nan & nan & 6.33e-02 & nan & 1.25e-04 & 9.58e-03 & 1.05e+00 & nan & nan & 1.74e-01 & 1.57e-01 & 3.34e-03 & nan & 2.37e-07 & 1.78e-02 \\
-3.3 & 8.76 & 6.0 & 4.80e-02 & nan & nan & nan & 1.76e-02 & 1.38e-03 & 2.76e-02 & nan & nan & 2.09e-07 & 1.89e-01 & nan & 2.78e-02 & 1.06e-02 & 5.38e-01 & 3.77e-07 & 1.24e-07 & 7.88e-01 & 7.12e-01 & 1.27e-02 & 1.51e-07 & 2.68e-03 & 1.71e-02 \\
-4.3 & 8.427 & 8.0 & 7.33e-02 & nan & nan & nan & 3.84e-02 & 5.42e-04 & 2.49e-02 & nan & nan & nan & 4.39e-02 & nan & 1.31e-04 & 9.46e-03 & 5.32e-01 & nan & nan & 1.63e-01 & 1.34e-01 & 2.74e-03 & nan & 1.04e-06 & 3.62e-02 \\
-2.3 & 8.632 & 8.5 & 1.12e-03 & nan & nan & 2.38e-06 & 1.80e-03 & 3.57e-03 & 2.44e-02 & nan & nan & 1.85e-04 & 1.41e-01 & nan & 2.58e-01 & 9.22e-03 & 8.55e-02 & 1.56e-04 & 3.85e-05 & 1.07e+00 & 2.29e-01 & 7.40e-03 & 1.23e-05 & 1.21e-02 & 7.36e-04 \\
\enddata
\tablecomments{\footnotesize Columns (1) -- (3): The input parameters for {\sc MAPPINGS V} models of SF regions (see Section~\ref{sec4.2}). Columns (4) -- (26): The predicted emission line intensities (normalized by the corresponding H$\beta$ emission) for the different SF models that assume a homogeneous distribution of ionized gas. (This table is available in its entirety in machine-readable form in the online article).}
\label{tabRes3}
\end{deluxetable}
\end{longrotatetable}

\startlongtable
\tablenum{A4}
\setlength{\tabcolsep}{3pt}
\begin{deluxetable*}{ccccccc}
\tabletypesize{\tiny}
\tablecolumns{7}
\tablecaption{Coefficients in Best-fit Correlations for $U$ of AGN with $R$ = [Ar~V]7.90/[Ar~III]8.99, [Ar~V]13.10/[Ne~III]15.56, and [Ar~V]7.90/[Ne~III]15.56}
\tablehead{
\colhead{$R$} & \colhead{$\rm[Ar~V]7.90/[Ar~III]8.99$} & \colhead{$\rm[Ar~V]13.10/[Ne~III]15.56$} & \colhead{$\rm[Ar~V]7.90/[Ne~III]15.56$} & \colhead{$\rm[Ar~V]7.90/[Ar~III]8.99$} & \colhead{$\rm[Ar~V]13.10/[Ne~III]15.56$} & \colhead{$\rm[Ar~V]7.90/[Ne~III]15.56$} \\
\colhead{log $E_{\rm peak}$} & \colhead{[$-2.0, -0.95$]} & \colhead{[$-2.0, -0.95$]} & \colhead{[$-2.0, -0.95$]} & \colhead{[$-3.25, -2.25$]} & \colhead{[$-3.25, -2.25$]} & \colhead{[$-3.25, -2.25$]} }
\startdata
$c_{5}$ & $-4.525e-03$ & $3.576e-03$ & $7.195e-03$ & $-5.677e-03$ & $8.764e-03$ & $1.533e-02$ \\
$c_{4}$ & $-3.193e-02$ & $7.535e-02$ & $1.373e-01$ & $-5.021e-02$ & $1.642e-01$ & $2.772e-01$ \\
$c_{3}$ & $-3.924e-03$ & $5.996e-01$ & $1.009e+00$ & $-1.013e-01$ & $1.173e+00$ & $1.922e+00$ \\
$c_{2}$ & $3.291e-01$ & $2.216e+00$ & $3.524e+00$ & $1.350e-01$ & $3.958e+00$ & $6.353e+00$ \\
$c_{1}$ & $1.022e+00$ & $4.165e+00$ & $6.193e+00$ & $9.735e-01$ & $3.958e+00$ & $1.043e+01$ \\
$c_{0}$ & $-1.993e+00$ & $3.979e-01$ & $1.685e+00$ & $-1.762e+00$ & $3.958e+00$ & $4.295e+00$ \\
RMS scatter & 0.16 dex & 0.17 dex & 0.16 dex & 0.10 dex & 0.10 dex & 0.07 dex \\
\enddata
\tablecomments{\footnotesize Coefficients in Equation~\ref{equ1}.}
\label{tabEqu1}
\end{deluxetable*}

\startlongtable
\tablenum{A5}
\setlength{\tabcolsep}{3pt}
\begin{deluxetable*}{crrrrrrrrrr}
\tabletypesize{\tiny}
\tablecolumns{11}
\tablecaption{Coefficients in Best-fit Correlations for $E_{\rm peak}$ of AGN with $R$ = [Ne~VI]7.65/[Ar~V]7.90}
\tablehead{
\colhead{log $E_{\rm peak}$} & \colhead{[$-2.0, -0.95$]} & \colhead{ } & \colhead{ } & \colhead{ } & \colhead{ } & \colhead{[$-3.25, -2.25$]} & \colhead{ } & \colhead{ } & \colhead{ } & \colhead{ } \\
\colhead{log $U$} & \colhead{[$-3.8, -3.3$]} & \colhead{[$-3.3, -2.8$]} & \colhead{[$-2.8, -2.3$]} & \colhead{[$-2.3, -1.8$]} & \colhead{[$-1.8, -1.3$]} & \colhead{[$-3.8, -3.3$]} & \colhead{[$-3.3, -2.8$]} & \colhead{[$-2.8, -2.3$]} & \colhead{[$-2.3, -1.8$]} & \colhead{[$-1.8, -1.3$]} }
\startdata
$a_{Z,2}$ & $2.328e-02$ & $-6.770e-03$ & $3.188e-03$ & $2.589e-02$ & $1.201e-01$ & $-3.461e+00$ & $2.693e-04$ & $-6.189e-06$ & $1.904e-04$ & $9.298e-02$ \\
$a_{Z,1}$ & $-3.564e-01$ & $1.201e-01$ & $-8.139e-02$ & $-5.474e-01$ & $-2.344e+00$ & $6.260e+01$ & $-4.837e-03$ & $-3.189e-03$ & $-5.130e-03$ & $-2.174e+00$ \\
$a_{U,1}$ & $-6.036e-01$ & $-2.726e-01$ & $1.671e-02$ & $2.721e-02$ & $3.613e-02$ & $-5.197e+00$ & $-3.382e-04$ & $-5.607e-03$ & $-9.945e-05$ & $1.452e-01$ \\
$a_{0}$ & $-3.804e-01$ & $-1.163e+00$ & $6.438e-01$ & $3.052e+00$ & $1.175e+01$ & $-3.047e+02$ & $2.612e-02$ & $3.733e-02$ & $3.331e-02$ & $1.249e+01$ \\
$b_{Z,2}$ & $-1.814e-01$ & $-2.021e-01$ & $-2.144e-01$ & $-3.855e-01$ & $-3.609e-01$ & $-1.523e-01$ & $8.256e-02$ & $-2.629e-01$ & $3.154e+00$ & $-6.706e-01$ \\
$b_{Z,1}$ & $3.207e+00$ & $3.450e+00$ & $3.203e+00$ & $5.701e+00$ & $5.470e+00$ & $2.186e+00$ & $-2.568e+00$ & $4.522e+00$ & $6.398e+00$ & $1.392e+01$ \\
$b_{U,1}$ & $8.213e-01$ & $-5.776e-01$ & $1.167e+00$ & $4.948e-01$ & $1.353e-01$ & $1.696e+00$ & $6.858e+00$ & $8.318e-01$ & $-2.817e+00$ & $2.158e+00$ \\
$b_{0}$ & $-1.381e+01$ & $-1.900e+01$ & $-1.106e+01$ & $-2.076e+01$ & $-2.027e+01$ & $-3.953e+00$ & $3.908e+02$ & $5.906e+01$ & $2.121e+02$ & $-6.115e+01$ \\
$c_{Z,2}$ & $-7.294e-02$ & $-2.031e-02$ & $-6.264e-02$ & $-9.451e-02$ & $-3.524e-02$ & $2.312e-01$ & $-3.484e+01$ & $1.071e+00$ & $4.126e+01$ & $1.183e+01$ \\
$c_{Z,1}$ & $1.337e+00$ & $4.443e-01$ & $1.068e+00$ & $1.483e+00$ & $5.172e-01$ & $-3.530e+00$ & $6.288e+02$ & $2.125e-02$ & $-4.495e+02$ & $-2.008e+02$ \\
$c_{U,1}$ & $-1.940e-01$ & $-4.896e-01$ & $-7.116e-02$ & $-4.659e-01$ & $-5.835e-01$ & $3.316e-01$ & $2.094e+01$ & $3.466e+01$ & $3.992e+01$ & $-9.069e+00$ \\
$c_{0}$ & $-8.895e+00$ & $-6.098e+00$ & $-7.158e+00$ & $-9.052e+00$ & $-4.978e+00$ & $1.244e+01$ & $-3.459e+03$ & $-1.348e+02$ & $-6.133e-02$ & $8.230e+02$ \\
RMS scatter & 0.05 dex & 0.03 dex & 0.02 dex & 0.02 dex & 0.03 dex & 0.02 dex & 0.04 dex & 0.06 dex & 0.10 dex & 0.13 dex \\
\enddata
\tablecomments{\footnotesize Coefficients in Equation~\ref{equ2} for different ranges of log $U$.}
\label{tabEqu2}
\end{deluxetable*}

\startlongtable
\tablenum{A6}
\setlength{\tabcolsep}{1pt}
\begin{deluxetable*}{crrrrrrrrrrrr}
\tabletypesize{\tiny}
\tablecolumns{13}
\tablecaption{Coefficients in Best-fit Correlations for 12 + log (O/H) of AGN with $R$ = ([Ar~III]8.99+[Ar~V]7.90)/Pf$\alpha$}
\tablehead{
\colhead{log $E_{\rm peak}$} & \colhead{[$-2.0, -0.95$]} & \colhead{ } & \colhead{ } & \colhead{ } & \colhead{ } & \colhead{ } & \colhead{[$-3.25, -2.25$]} & \colhead{ } & \colhead{ } & \colhead{ } & \colhead{ } & \colhead{ } \\
\colhead{log $U$} & \colhead{[$-4.3, -3.8$]} & \colhead{[$-3.8, -3.3$]} & \colhead{[$-3.3, -2.8$]} & \colhead{[$-2.8, -2.3$]} & \colhead{[$-2.3, -1.8$]} & \colhead{[$-1.8, -1.3$]} & \colhead{[$-4.3, -3.8$]} & \colhead{[$-3.8, -3.3$]} & \colhead{[$-3.3, -2.8$]} & \colhead{[$-2.8, -2.3$]} & \colhead{[$-2.3, -1.8$]} & \colhead{[$-1.8, -1.3$]} }
\startdata
$a_{E,3}$ & $-1.179e-01$ & $-1.615e-02$ & $-4.535e-03$ & $-4.274e-02$ & $-4.723e-04$ & $-7.667e-03$ & $-4.523e-02$ & $-2.561e-02$ & $-1.327e-03$ & $5.427e-05$ & $-9.078e-07$ & $-1.111e-02$ \\
$a_{E,2}$ & $-4.574e-01$ & $-4.859e-02$ & $-3.221e-03$ & $-2.237e-01$ & $-2.183e-03$ & $-3.959e-02$ & $-2.732e-01$ & $-1.594e-01$ & $-1.143e-02$ & $4.498e-04$ & $-1.428e-05$ & $-1.223e-01$ \\
$a_{E,1}$ & $-5.220e-01$ & $-2.267e-02$ & $2.951e-02$ & $-3.476e-01$ & $-5.951e-06$ & $-6.117e-02$ & $-3.843e-01$ & $-2.474e-01$ & $-3.264e-02$ & $1.361e-03$ & $-4.716e-05$ & $-4.725e-01$ \\
$a_{U,1}$ & $3.616e-02$ & $4.227e-02$ & $4.429e-02$ & $1.898e-02$ & $3.553e-05$ & $-9.674e-03$ & $3.101e-02$ & $1.375e-02$ & $-1.684e-01$ & $-2.234e-01$ & $4.427e-04$ & $-3.569e-04$ \\
$a_{0}$ & $1.804e-01$ & $3.178e-01$ & $3.141e-01$ & $1.571e-02$ & $9.976e-03$ & $9.590e-03$ & $5.625e-01$ & $3.154e-01$ & $-3.461e-01$ & $-4.717e-01$ & $3.840e-03$ & $-6.767e-01$ \\
$b_{E,3}$ & $-1.482e+00$ & $-1.597e-02$ & $1.435e-01$ & $-8.123e-01$ & $4.264e+00$ & $-1.074e+00$ & $-5.291e-01$ & $-4.076e-01$ & $-1.391e-01$ & $-1.549e-01$ & $3.267e-01$ & $3.691e-01$ \\
$b_{E,2}$ & $-6.125e+00$ & $2.944e-02$ & $6.347e-01$ & $-5.457e+00$ & $-3.372e+00$ & $-5.066e+00$ & $-4.069e+00$ & $-3.051e+00$ & $-1.192e+00$ & $-1.326e+00$ & $1.508e+00$ & $4.293e+00$ \\
$b_{E,1}$ & $-7.469e+00$ & $7.895e-01$ & $1.560e+00$ & $-9.141e+00$ & $1.286e+01$ & $-6.051e+00$ & $-1.001e+01$ & $-7.203e+00$ & $-3.453e+00$ & $-3.862e+00$ & $3.514e+00$ & $8.758e+00$ \\
$b_{U,1}$ & $1.122e+00$ & $1.085e+00$ & $1.001e+00$ & $6.683e-01$ & $4.292e+00$ & $-1.833e-01$ & $7.654e-01$ & $2.453e-01$ & $-2.903e+00$ & $-2.203e+01$ & $4.718e+01$ & $-7.299e-01$ \\
$b_{0}$ & $-1.360e-01$ & $2.056e+00$ & $1.454e+00$ & $-5.911e+00$ & $-3.485e+01$ & $-1.087e+01$ & $-5.733e+00$ & $-5.770e+00$ & $-1.466e+01$ & $-6.875e+01$ & $-9.986e+01$ & $7.193e+00$ \\
$c_{E,3}$ & $-7.811e-01$ & $1.094e-01$ & $1.644e-01$ & $-3.626e-01$ & $1.475e+00$ & $-7.442e-01$ & $-3.195e-01$ & $-2.500e-01$ & $-1.247e-01$ & $-1.685e-01$ & $4.815e-01$ & $1.715e-01$ \\
$c_{E,2}$ & $-3.303e+00$ & $3.700e-01$ & $4.818e-01$ & $-2.883e+00$ & $-6.043e+00$ & $-2.681e+00$ & $-2.456e+00$ & $-1.880e+00$ & $-1.064e+00$ & $-1.440e+00$ & $2.829e+00$ & $2.334e+00$ \\
$c_{E,1}$ & $-3.998e+00$ & $8.657e-01$ & $8.866e-01$ & $-5.002e+00$ & $5.641e+00$ & $-1.566e+00$ & $-6.055e+00$ & $-4.498e+00$ & $-3.084e+00$ & $-4.190e+00$ & $7.403e+00$ & $5.039e+00$ \\
$c_{U,1}$ & $4.894e-01$ & $4.223e-01$ & $4.056e-01$ & $2.927e-01$ & $4.397e+00$ & $8.821e-01$ & $2.132e-01$ & $-1.263e-02$ & $-1.898e+00$ & $-1.340e+01$ & $3.847e+01$ & $-9.073e-01$ \\
$c_{0}$ & $7.766e+00$ & $8.946e+00$ & $8.508e+00$ & $4.663e+00$ & $-3.983e+00$ & $5.523e+00$ & $3.710e+00$ & $3.877e+00$ & $-2.217e+00$ & $-3.560e+01$ & $-2.531e+01$ & $1.165e+01$ \\
RMS scatter & 0.02 dex & 0.01 dex & 0.01 dex & 0.01 dex & 0.01 dex & 0.01 dex & 0.02 dex & 0.01 dex & 0.01 dex & 0.02 dex & 0.02 dex & 0.02 dex \\
\enddata
\tablecomments{\footnotesize Coefficients in Equation~\ref{equ3} for different ranges of log $U$.}
\label{tabEqu3}
\end{deluxetable*}

\startlongtable
\tablenum{A7}
\setlength{\tabcolsep}{3pt}
\begin{deluxetable*}{ccccccccc}
\tabletypesize{\tiny}
\tablecolumns{9}
\tablecaption{Coefficients in Best-fit Correlations for 12 + log (O/H) of AGN with $R$ = [Ar~II]6.99/[Fe~II]5.34, [Ne~II]12.81/[Fe~II]5.34, [Ar~II]6.99/[Fe~II]25.99, and [Ne~II]12.81/[Fe~II]25.99}
\tablehead{
\colhead{$R$} & \colhead{$\rm[Ar~II]/[Fe~II]5.34$} & \colhead{$\rm[Ar~II]/[Fe~II]5.34$} & \colhead{$\rm[Ne~II]/[Fe~II]5.34$} & \colhead{$\rm[Ne~II]/[Fe~II]5.34$} & \colhead{$\rm[Ar~II]/[Fe~II]25.99$} & \colhead{$\rm[Ar~II]/[Fe~II]25.99$} & \colhead{$\rm[Ne~II]/[Fe~II]25.99$} & \colhead{$\rm[Ne~II]/[Fe~II]25.99$} \\
\colhead{log $E_{\rm peak}$} & \colhead{[$-2.0, -0.95$]} & \colhead{[$-3.25, -2.25$]} & \colhead{[$-2.0, -0.95$]} & \colhead{[$-3.25, -2.25$]} & \colhead{[$-2.0, -0.95$]} & \colhead{[$-3.25, -2.25$]} & \colhead{[$-2.0, -0.95$]} & \colhead{[$-3.25, -2.25$]} }
\startdata
$c_{2}$ & $-7.021e-02$ & $-1.217e-01$ & $-6.849e-02$ & $-4.032e-02$ & $1.910e-01$ & $1.794e-01$ & $1.973e-01$ & $2.056e-01$ \\
$c_{1}$ & $7.508e-01$ & $6.842e-01$ & $8.979e-01$ & $8.627e-01$ & $9.507e-01$ & $9.507e-01$ & $5.772e-01$ & $5.243e-01$ \\
$c_{0}$ & $9.007e+00$  & $8.994e+00$ & $8.198e+00$ & $8.154e+00$ & $8.661e+00$ & $8.668e+00$ & $7.906e+00$ & $7.894e+00$ \\
RMS scatter & 0.08 dex & 0.08 dex  & 0.07 dex & 0.07 dex & 0.05 dex & 0.05 dex & 0.05 dex & 0.04 dex\\
\enddata
\tablecomments{\footnotesize Coefficients in Equation~\ref{equ4}.}
\label{tabEqu4}
\end{deluxetable*}

\startlongtable
\tablenum{A8}
\setlength{\tabcolsep}{10pt}
\begin{deluxetable*}{ccc}
\tabletypesize{\tiny}
\tablecolumns{3}
\tablecaption{Coefficients in Best-fit Correlations for $P/k$ of AGN with $R$ =[Ne~V]14.32/[Ne~V]24.32}
\tablehead{
\colhead{log $E_{\rm peak}$} & \colhead{[$-2.0, -0.95$]} & \colhead{[$-3.25, -2.25$]} }
\startdata
$c_{5}$ & $3.340e+01$ & $2.768e+01$ \\
$c_{4}$ & $-8.183e+01$ & $-6.887e+01$ \\
$c_{3}$ & $7.367e+01$ & $6.377e+01$ \\
$c_{2}$ & $-2.947e+01$ & $-2.662e+01$ \\
$c_{1}$ & $7.065e+00$ & $6.909e+00$ \\
$c_{0}$ & $7.494e+00$ & $7.384e+00$ \\
RMS scatter & 0.19 dex & 0.20 dex \\
\enddata
\tablecomments{\footnotesize Coefficients in Equation~\ref{equ5}.}
\label{tabEqu5}
\end{deluxetable*}

\startlongtable
\tablenum{A9}
\setlength{\tabcolsep}{10pt}
\begin{deluxetable*}{cc}
\tabletypesize{\tiny}
\tablecolumns{2}
\tablecaption{Coefficients in Best-fit Correlations for log $U$ of SF regions with $R$ = [S~IV]10.51/[Ne~III]15.56}
\tablehead{
\colhead{$R$} & \colhead{$\rm[S~IV]10.51/[Ne~III]15.56$}}
\startdata
$c_{3}$ & $3.652e-02$  \\
$c_{2}$ & $3.357e-01$ \\
$c_{1}$ & $1.605e+00$ \\
$c_{0}$ & $-1.513e+00$  \\
RMS scatter & 0.17 dex  \\
\enddata
\tablecomments{\footnotesize Coefficients in Equation~\ref{equ6}.}
\label{tabEqu6}
\end{deluxetable*}

\startlongtable
\tablenum{A10}
\setlength{\tabcolsep}{4pt}
\begin{deluxetable*}{ccccccc}
\tabletypesize{\tiny}
\tablecolumns{4}
\tablecaption{Coefficients in Best-fit Correlations for 12 + log (O/H) of SF regions with $R$ = [Ne~II]15.56/Pf$\alpha$ and [Ar~II]6.99/Pf$\alpha$}
\tablehead{
\colhead{$R$} & \colhead{$\rm[Ne~II]15.56/Pf\alpha$} & \colhead{ } & \colhead{ } & \colhead{$\rm[Ar~II]6.99/Pf\alpha$} & \colhead{ } & \colhead{ } \\
\colhead{log $U$} & \colhead{[$-4.3, -3.3$]} & \colhead{[$-3.3, -2.3$]} & \colhead{[$-2.3, -1.3$]} & \colhead{[$-4.3, -3.3$]} & \colhead{[$-3.3, -2.3$]} & \colhead{[$-2.3, -1.3$]} }
\startdata
$a_{U,2}$ & $-1.894e-04$ & $1.338e-01$ & $6.611e-02$ & $-2.480e-01$ & $1.458e-01$ & $2.534e-01$ \\
$a_{U,1}$ & $-1.775e-03$ & $6.566e-01$ & $3.383e-01$ & $-2.191e+00$ & $5.631e-01$ & $1.173e+00$ \\
$a_{0}$ & $-1.827e-03$ & $5.575e-01$ & $1.836e-01$ & $-4.840e+00$ & $-3.860e-02$ & $7.948e-01$ \\
$b_{U,2}$ & $-2.126e+00$ & $2.059e+00$ & $7.611e-01$ & $3.641e+02$ & $7.973e-01$ & $7.452e-01$ \\
$b_{U,1}$ & $2.271e+01$ & $9.311e+00$ & $2.590e+00$ & $2.577e+03$ & $2.991e+00$ & $2.490e+00$ \\
$b_{0}$ & $-1.339e+02$ & $1.245e+01$ & $3.864e+00$ & $4.540e+03$ & $2.317e+00$ & $1.440e+00$ \\
$c_{U,2}$ & $-2.601e-03$ & $1.138e+00$ & $1.120e-01$ & $2.030e+02$ & $3.618e-01$ & $7.675e-02$ \\
$c_{U,1}$ & $6.382e+01$ & $5.502e+00$ & $3.958e-01$ & $1.438e+03$ & $1.755e+00$ & $3.242e-01$ \\
$c_{0}$ & $1.123e+02$ & $1.576e+01$ & $9.442e+00$ & $2.543e+03$ & $1.114e+01$ & $9.355e+00$ \\
RMS scatter & 0.03 dex & 0.02 dex & 0.01 dex & 0.03 dex & 0.02 dex & 0.02 dex \\
\enddata
\tablecomments{\footnotesize Coefficients in Equation~\ref{equ7} for different ranges of log $U$.}
\label{tabEqu7}
\end{deluxetable*}
\,\,\,

\end{document}